\documentclass[12pt,british]{iopart}
\usepackage[T1]{fontenc}
\usepackage[latin9]{inputenc}
\usepackage{babel}
\usepackage{textcomp}
\usepackage{amssymb}
\usepackage{graphicx}
\usepackage{color}
\usepackage{array}
\usepackage{tikz}
\usepackage{enumitem}
\usepackage{txfonts}
\usepackage{marginnote}
\usetikzlibrary{shapes, arrows, positioning}
\usepackage {hyperref}

\renewcommand{\vec}[1]{\mathbf{#1}}
\newcommand{\alphabar}{\bar{\alpha}}

\begin{document}

\title[The physics of streamers]{The physics of streamer discharge phenomena}

\author{Sander~Nijdam$^{1}$, Jannis Teunissen$^{2,3}$ and
Ute~Ebert$^{1,2}$}

\address{$^{1}$ Eindhoven University of Technology, Dept.\ Applied Physics\\
 P.O. Box 513, 5600 MB Eindhoven, The Netherlands\\}
\address{$^{2}$ Centrum Wiskunde \& Informatica (CWI), Amsterdam, The Netherlands}
\address{$^{3}$ KU Leuven, Centre for Mathematical Plasma-astrophysics, Leuven, Belgium}
\ead{s.nijdam@tue.nl}

\begin{abstract}

In this review we describe a transient type of gas discharge which is
commonly called a streamer discharge, as well as a few related phenomena in
pulsed discharges. Streamers are propagating ionization fronts with
self-organized field enhancement at their tips that can appear in gases at
(or close to) atmospheric pressure. They are the precursors of other
discharges like sparks and lightning, but they also occur in for example
corona reactors or plasma jets which are used for a variety of plasma
chemical purposes. When enough space is available, streamers can also form
at much lower pressures, like in the case of sprite discharges high up in
the atmosphere.

We explain the structure and basic underlying physics of streamer
discharges, and how they scale with gas density. We discuss the chemistry
and applications of streamers, and describe their two main stages in
detail: inception and propagation. We also look at some other topics, like
interaction with flow and heat, related pulsed discharges, and electron
runaway and high energy radiation. Finally, we discuss streamer simulations
and diagnostics in quite some detail.

This review is written with two purposes in mind: First, we describe recent
results on the physics of streamer discharges, with a focus on the work
performed in our groups. We also describe recent developments in
diagnostics and simulations of streamers. Second, we provide background
information on the above-mentioned aspects of streamers. This review can
therefore be used as a tutorial by researchers starting to work in the
field of streamer physics.
\end{abstract}

version of \today

\maketitle

\tableofcontents

\section{Introduction}
\begin{figure}
   \centering
   \includegraphics[width=8 cm]{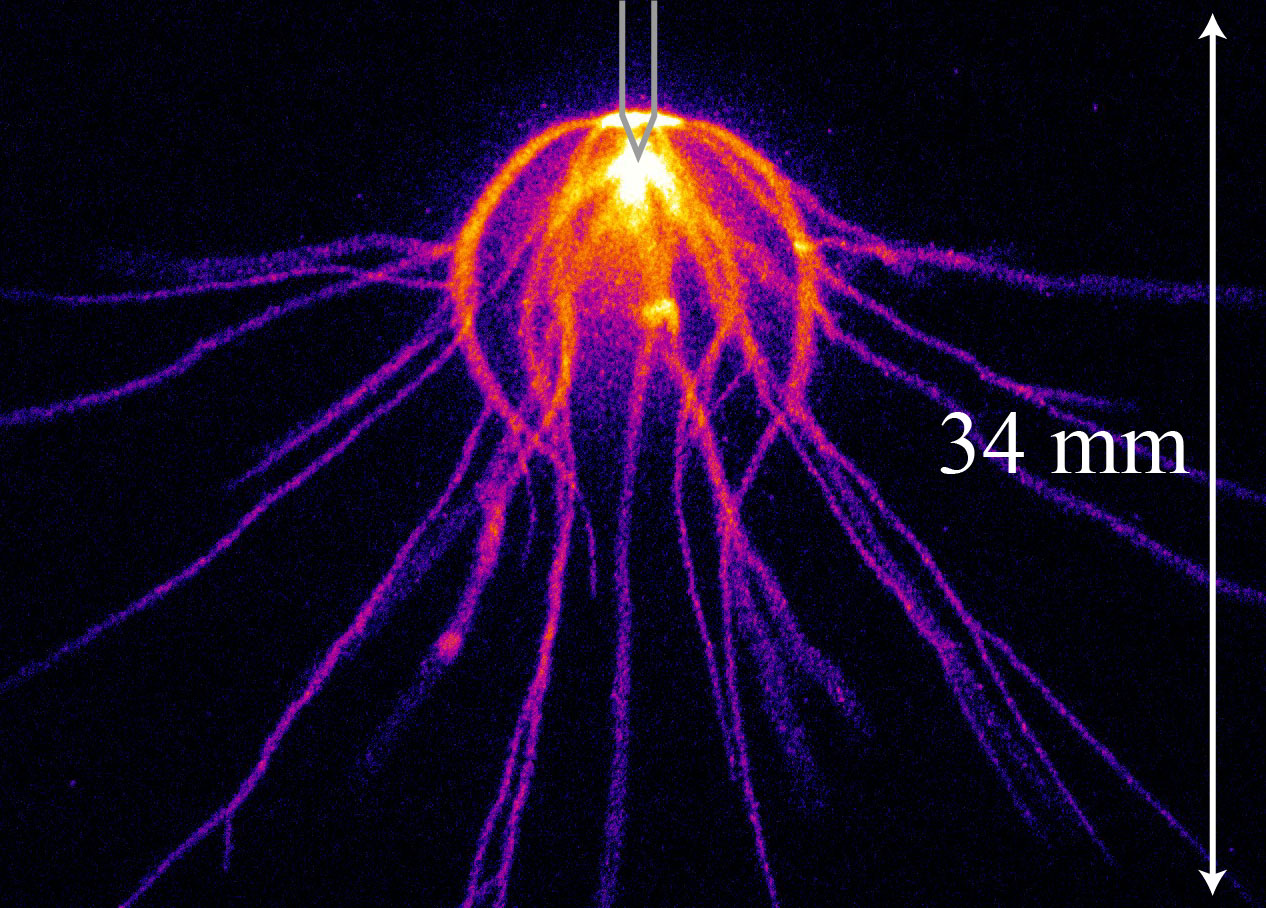}
   \caption{A long exposure, false colour, image of a peculiar streamer discharge
   caused by a complex voltage pulse. Image taken from~\cite{Nijdam2011}.}
   \label{fig:Onion}
\end{figure}

Streamers are fast-moving ionization fronts that can form complex tree-like structures
or other shapes, depending on conditions (see e.g. figure~\ref{fig:Onion}).
In this paper, we review our present understanding of streamer discharges. We start from the
basic physical mechanisms and concepts, aiming also at beginners in the field. We also touch
on related phenomena such as discharge inception, diffuse discharges, nanosecond pulsed
discharges, plasma jets, transient luminous events and lightning propagation, electron runaway
and high energy radiation.

The paper is organized as follows:
In the present introductory section we briefly review streamer phenomena in nature and
technology, we discuss the relevant physical mechanisms with their multiscale nature
and we have a first look at numerical models and streamers in laboratory experiments.
The following two sections are devoted to the details of the different temporal stages
of the discharge evolution: discharge inception (section~\ref{sec:inception}) and streamer propagation and branching  (section~\ref{sec:propagation}).
In these two sections we mainly concentrate on streamers in (ambient) air but
in section~\ref{sec:diff_media} we treat streamers in other media and pressures.
In section~\ref{sec:additional} this is followed by discussions on
streamer-relevant plasma theory and chemistry, interaction with flow and heat,
high energy phenomena, plasma jets and sprite discharges.
The final main sections treat the available methods in detail, first in modeling and simulations
(section~\ref{sec:simulations}),
and then in plasma diagnostics (section~\ref{sec:diagnostics}). We end with a short outlook and an overview of open
questions on the physics of streamer discharge phenomena (section~\ref{sec:outlook}).

\subsection{Streamer phenomena in nature and technology}\label{sec:applications}

The most common and well-known occurrence of streamers is as the precursor of
sparks where they create the first ionized path for the later heat-dominated
spark discharge. Streamers play a similar role in the inception and in the
propagation of lightning leaders.
Streamers are directly visible in our atmosphere as so-called sprites,
discharges far above active thunderstorms; they will be discussed in more
detail in section \ref{sec:sprites}.

Streamers are members of the cold atmospheric plasma (CAP) discharge family.
Most industrial applications of streamers and other CAPs (i.e., not as
precursors of discharges like sparks) utilize the unique chemical properties
of such discharges. The highly non-equilibrium character and the resulting
high electron energies enable CAPs to start high-temperature chemical
reactions close to room temperature. This leads to two major advantages
compared to thermal plasmas and other hot reactors: firstly it enables such
reactions in environments that cannot withstand high temperatures, and
secondly it can make the chemistry very efficient as no energy is lost on gas
heating.

The electrons that trigger the chemical reactions can have energies of the
order of 10\,eV or higher, something that cannot be achieved in any thermal
process (as 1~eV corresponds to 11600 Kelvin). In air and air-like gas
mixtures this leads to the production of OH, O and N radicals as well as of
excited species and ions like O$_2^-$, O$^-$, O$^+$, N$_4^+$ and O$_4^+$
after an initial production of N$_2^+$, O$_2^+$, see
section~\ref{sec:chemistry}. Each of these species can start other chemical
reactions, either within the bulk gas, on nearby surfaces or even in nearby
liquids when the species survives long enough and can be easily absorbed. The
initial energy distribution of the generated excited species is typically far
from thermal equilibrium.

Due to these properties of streamers and other CAPs they are used or
developed for a myriad of applications, most of which are described
extensively in the following review
papers~\cite{Becker2005,Fridman2005,Bruggeman2013,Bruggeman2017}. Popular
applications are plasma medicine~\cite{Fridman2008, Laroussi2014, Graves2014}
including cancer therapy~\cite{Keidar2013} and
sterilization~\cite{Klaempfl2012}, industrial surface
treatment~\cite{Bardos2010}, air treatment for cleaning or ozone
production~\cite{Kim2004, Heesch2008}, plasma assisted
combustion~\cite{Starikovskaia2014,Popov2016} and
propulsion~\cite{Leonov2016,Kotsonis2015} and liquid
treatment~\cite{Kolb2008,Bruggeman2016}. Two recent reviews on nanosecond
pulsed streamer generation, physics and applications are by
Huiskamp~\cite{Huiskamp2020} and Wang and Namihira~\cite{Wang2020}.

\begin{figure*}
  \tikzstyle{block} = [rectangle, draw, fill=blue!20,
  text width=7em, text centered, rounded corners, minimum height=2em]
  \tikzstyle{block_appl} = [rectangle, draw, fill=black!10,
  text width=7em, text centered, rounded corners, minimum height=2em,
  node distance=0.7cm]
  \tikzstyle{line} = [draw, -latex']
  \tikzstyle{cloud} = [draw, ellipse,fill=red!20, text width=7em, minimum height=2em]

  \centering
  \begin{tikzpicture}[thick, align=center, node distance = 1cm, auto]
    \footnotesize
    \node [block, text width=12em, fill=yellow!20] (power) {Pulsed electric power};
    \node [cloud, below=of power, node distance=0.7cm] (electrons) {Energetic electrons};
    \node [cloud, right=of electrons] (collisions) {Collisions with gas molecules};
    \node [block, below left= of collisions] (wind) {Corona wind};
    \node [block, left=of wind] (runaway) {Electron runaway};
    \node [block, right=of wind] (chemistry) {Plasma chemistry};
    \node [block, right=of chemistry] (breakdown) {Electric breakdown};
    \node [block_appl, below=of wind] (wind_appl) {Plasma actuators};
    \node [block_appl, below=of chemistry] (chemistry_appl)
    {Plasma: medicine, agriculture, combustion, disinfection, air cleaning};
    \node [block_appl, below=of breakdown] (breakdown_appl)
    {High-voltage switch gear, lightning protection};
    \path [line] (power) -- (electrons);
    \path [line] (electrons) -- (collisions);
    \path [line] (collisions) -- (electrons);
    \path [line] (electrons) -- (runaway);
    \path [line] (collisions) -- (wind);
    \path [line] (collisions) -- (chemistry);
    \path [line] (collisions) -- (breakdown);
    \path [line,dashed] (wind) -- (wind_appl);
    \path [line,dashed] (chemistry) -- (chemistry_appl);
    \path [line,dashed] (breakdown) -- (breakdown_appl);
  \end{tikzpicture}
  \caption{Energy conversion in pulsed atmospheric discharges with application fields.}
  \label{fig:Applications}
\end{figure*}
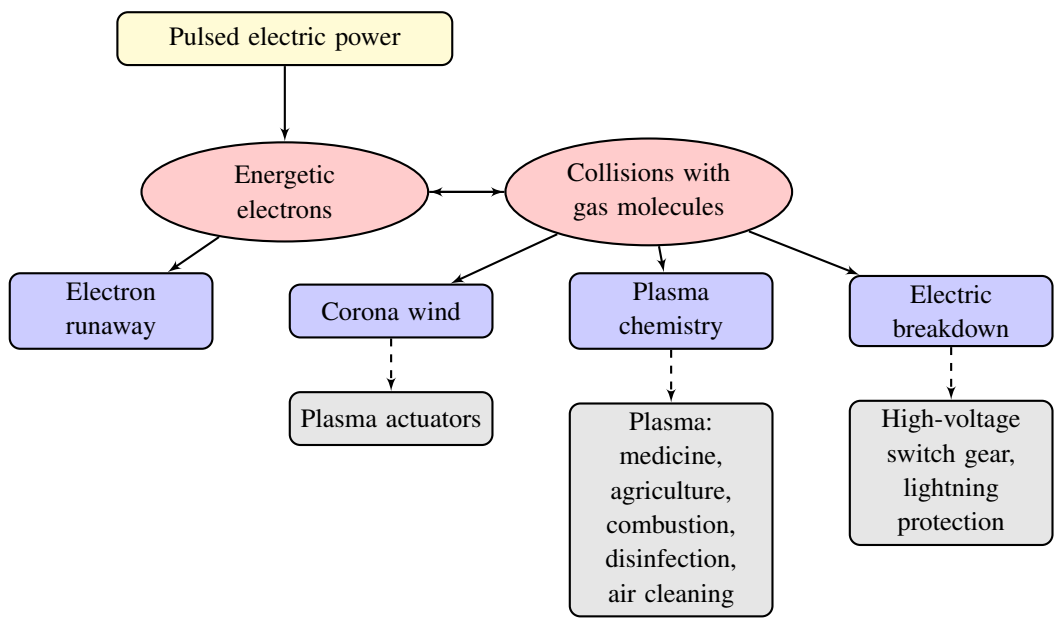

A fast pulsed discharge like a streamer has the advantage that the electric
field is not limited by the breakdown field. The electric field and thereby
the electron energy can transiently reach much higher values than in static
discharges. Pulsed discharges can be seen as energy conversion processes, as
sketched in Figure~\ref{fig:Applications}. First, pulsed electric power is
applied to gas at (close to) atmospheric pressure. When the gas discharge
starts to develop, this energy is converted to ionization and to free
electron energies in the eV range, far from thermal equilibrium. The further
plasma evolution can include different physical and chemical mechanisms. a)
If the local electric field is high enough, electrons can keep accelerating
up to electron runaway, and create Bremsstrahlung photons in collisions with
gas molecules; the photons can initiate other high-energy processes in the
gas, as is in particular seen in thunderstorms, see
section~\ref{sec:fastelectrons}. b) The drift of unbalanced charged particles
through the gas can create so-called corona wind, see
section~\ref{sec:flowheat}. c) Excitation, ionization and dissociation of
molecules by electron impact trigger plasmachemical reactions in the gas, see
section~\ref{sec:chemistry}. d) Electric breakdown means that the
conductivity increases further by ionization, heating and thermal gas
expansion; it is used in high voltage switch gear, and has to be controlled
in lightning protection.

Streamer discharges are often produced in ambient air. For this reason, we
and many other authors use the term Standard Temperature and Pressure (STP)
as a simple definition of ambient (air) conditions. Its exact definition
varies, but it always represents a temperature of either room temperature or
0$^{\circ}$C and a pressure close to 1\,atmosphere.

\subsection{A first view on the theory of streamers}
\label{sec:first_view_model}

In this section we will discuss the basics of streamer discharges. The theory
presented here is primarily based on streamers in atmospheric air, but most
of the concepts are also valid for (or can be generalized to) other gas
densities and/or compositions.

Streamer discharges can appear when a gas with low to vanishing electric
conductivity is suddenly exposed to a high electric field. Key for streamer
discharges are the acceleration of electrons in the local electric field and
the collisions between electrons and neutral gas molecules (for brevity, we
will use the term \emph{gas molecules} instead of writing \emph{gas atoms or
molecules}), which can be of the following type:
\begin{itemize}
  \item Elastic collisions, in which the total kinetic energy is conserved,
      although some of it is typically transferred from the electron to the
      gas molecule.
  \item Excitations, in which some of the electron's kinetic energy is used
      to excite the molecule. Depending on the gas molecule (or atom),
      there can be rotational, vibrational and electronic excitations.
  \item Ionization, in which the gas molecule is ionized.
  \item Attachment, in which the electron attaches to the gas molecule,
      forming a negative ion.
\end{itemize}
Data for the collisions of electrons with different types of atoms and
molecules can be found, e.g., on the community webpage
\href{https://www.lxcat.net}{www.lxcat.net}.

As explained below, streamer discharges can form where the electric field is
above the breakdown value. However, streamers can also enter into regions
where the electric field is below breakdown. This is due to the nonlinear
streamer mechanism, which is based on the following physical processes.

\subsubsection{Impact ionization. } Free electrons that are accelerated by a high local
electric field, can create new electron-ion pairs when they impact with
sufficient kinetic energy on gas molecules. If there is also an electron
attachment reaction, then the impact ionization rate must be larger than the
attachment rate for the plasma to grow; the local electric field is then said
to be above the breakdown value. In such fields, the chain reaction of
ionization growth leads to the creation of electrically conducting plasma
regions.

How many ionization and attachment events occur per electron per unit length
is described by the \emph{ionization and attachment coefficients} $\alpha$
and $\eta$. As discussed above, breakdown requires that $\alpha > \eta$, or
in other words, that the effective ionization coefficient
\begin{equation}\label{eq:bar-alpha}
\alphabar = \alpha - \eta
\end{equation}
is positive. The electric field $E_k$ where the effective ionization
coefficient vanishes $\alphabar(E_k)=0$, is called the classical breakdown
field; for $E>E_k$, the ionization density grows.

In electronegative gases like air, electron loss due to recombination
is negligible relative to attachment, because recombination is quadratic
in the degree of ionization, and the degree of ionization is small.

\subsubsection{Electron drift. } Electrons gain kinetic energy in the
local field and lose energy in collisions with gas molecules. This leads to
an average drift motion that can be described by $\vec{v}_\mathrm{drift} =
-\mu_e \vec{E}$, where $\mu_e$ is the \emph{electron mobility}. Only in very
high fields, electrons can overcome the friction barrier caused by
collisions; they then keep accelerating and become runaway electrons (for
further discussion see section~\ref{sec:fastelectrons}).

The drift motion of charged particles in the field leads to an electric
current that usually satisfies Ohm's law
\begin{equation}
  \label{eq:ohms-law} \vec{j} = \sigma \vec{E},
\end{equation} where $\vec{j}$ is the electric current density and $\sigma$
the conductivity of the plasma. Note that magnetic fields are not taken into
account here, as their effect is typically negligible, see section
\ref{sec:plasma-theory}.

As long as electron and ion densities are similar
(i.e., during and after the ionization process and before attachment depletes the electrons), the electron contribution
dominates the conductivity, hence $\sigma \approx e \mu_e n_e$, where $e$ is
the elementary charge and $n_e$ the electron density. Ions also drift in the
field, but as they carry the same electric charge and are much heavier, they
are much slower than the electrons. Furthermore, ions lose kinetic energy
more easily than electrons as they have a similar mass as the gas molecules
they collide with. (This is a consequence of the conservation of energy and
momentum in collisions.)

\subsubsection{Electric field enhancement. } Equation (\ref{eq:ohms-law})
shows that in an ionized medium or plasma with conductivity $\sigma$, an
electric field creates an electric current. Due to the conservation of
electric charge
\begin{equation}
  \label{eq:conservation} \partial_t\rho+\nabla\cdot\vec{j}=0,
\end{equation}
the charge density distribution $\rho$ changes in time due to a current
density $\vec{j}$, and the electric field $\vec{E}$ changes as well according
to Gauss' law of electrostatics (in vacuum or in not too dense gases, in the
absence of solid or liquid bodies)
\begin{equation}
  \label{eq:Gauss}
  \nabla\cdot\vec{E} = \rho/\epsilon_0,
\end{equation}
where $\epsilon_0$ is the dielectric constant.

\begin{figure}
    \centering
    \includegraphics[width=8cm]{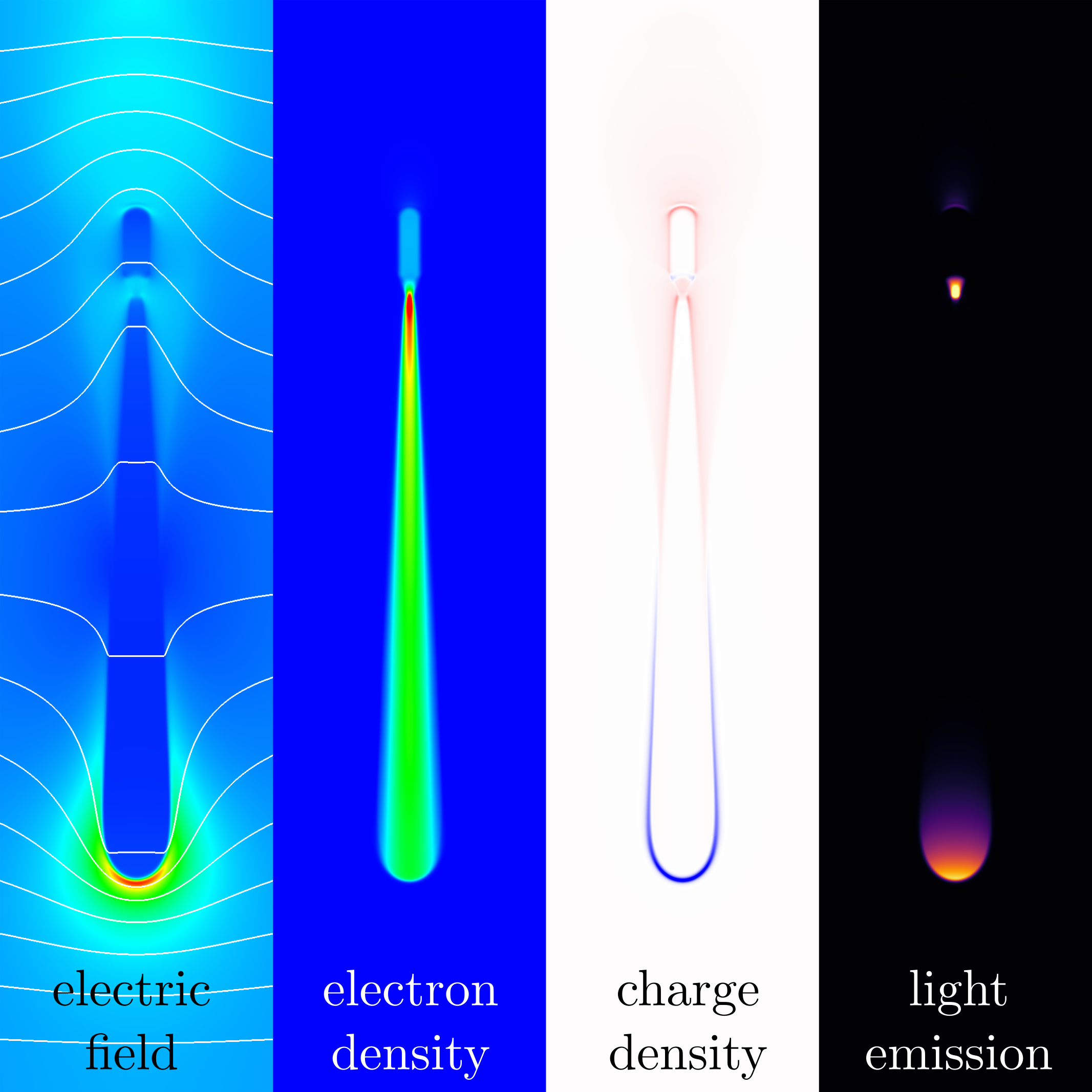}
    \caption{Simulation example showing a cross section of a positive streamer
      propagating downwards. A strong electric field is present at the streamer
      tip. A charge layer surrounds the streamer channel, with both positive
      charge (blue) and negative charge (red) present. A cross section of the instantaneous light emission is also shown, which is
      concentrated near the streamer head. The simulation was performed with an
      axisymmetric fluid model~\cite{Teunissen_2017} in air at
      $1 \, \textrm{bar}$, in a gap of $1.6 \, \textrm{cm}$ with an applied
      voltage of $32 \, \textrm{kV}$.}
    \label{fig:streamer_sim}
\end{figure}

Equations (\ref{eq:ohms-law})--(\ref{eq:Gauss}) imply that the interior of a
non-moving body with constant conductivity $\sigma$ is screened on the time
scale of the dielectric relaxation time
\begin{equation}
  \label{eq:tau-drt} \tau = \epsilon_0 / \sigma,
\end{equation}
while a surface charge builds up at the edges that screens the field in the
interior. If the shape of the conducting body is elongated in the direction
of the electric field, there is significant surface charge around the sharp
tips, and therefore a strong field enhancement ahead of these tips. If the
locally enhanced field at a tip exceeds the breakdown value $E_k$, a
conducting streamer body can grow at such a location, even if the background
field is below breakdown. This is illustrated with numerical modeling results
in figure~\ref{fig:streamer_sim}.

To be more precise, there are two important corrections to this simple
picture of a streamer: first, the ionization and hence the conductivity of a
streamer is not constant, but changes in space and time, and for a
generalization of the dielectric relaxation time to a reactive plasma with
$\alphabar>0$, we refer to~\cite{Teunissen_2014}. Second, the shape of the
conducting body changes in time, and therefore the electric field is
typically not completely screened from the interior.

\subsubsection{Electron source ahead of the ionization front.} \label{sec:electronsource}
The above mechanisms suffice to explain the propagation of negative (i.e.,
anode-directed) streamers in the direction of electron drift. However,
positive (i.e., cathode-directed) streamers frequently move with similar
velocity against the electron drift direction. They require an electron
source ahead of the ionization front. The dominant mechanism in air is
photo-ionization, a nonlocal mechanism. Photons are generated in the active
impact ionization region at the streamer tip, but create electron-ion pairs
at some characteristic distance determined by their absorption cross-section.
Other sources of free electrons ahead of a streamer ionization front can be
earlier discharges, external radiation sources like radioactivity or cosmic
rays, electron detachment from negative ions, or bremsstrahlung photons from
runaway electrons.

\subsubsection{Coherent structure.}
The nonlinear interaction of impact ionization, electron drift and field enhancement creates the streamer head, see Figure~\ref{fig:streamer_sim}. It can be considered as a coherent structure
that propagates with a dynamically stabilized shape.
Other examples of coherent structures are solitons or chemical or combustion fronts.

\subsection{The multiple scales in space, time and energy}
\label{sec:multiscale}

\begin{figure}
    \centering
    \includegraphics[width=8 cm]{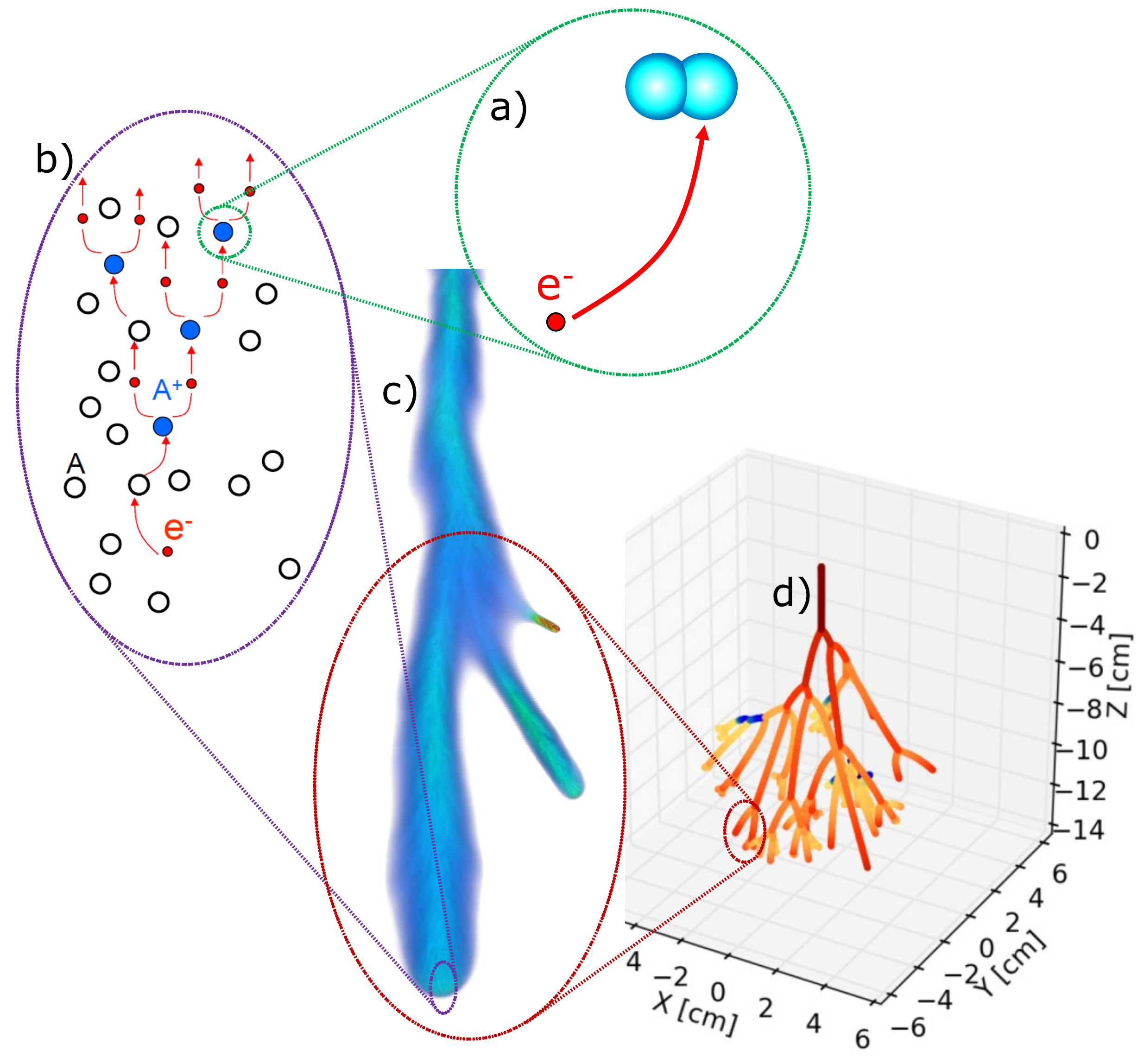}
    \caption{The multiple spatial scales in streamer discharges: a) collision of an
    electron with an atom or molecule, b) multiple electrons accelerate in a local electric
    field, collide with neutral gas molecules and form an ionization avalanche, c) a
    branching streamer discharge with field enhancement at the tips, d) a discharge tree
    with multiple streamer branches. Panel d is reproduced from a figure in \cite{Luque2014}.}
    \label{fig:MultiScale}
\end{figure}

The multiple spatial scales in a streamer discharge are illustrated in
Figure~\ref{fig:MultiScale}. From small to large, the following processes
take place:

{\bf Collisions:} On the most microscopic level (panel a), electrons that are
accelerated by the electric field collide with gas molecules. A proper
characterization of the collision processes is key to understanding the
electron energy distribution as well as the excitation, ionization and
dissociation of molecules.

{\bf Motion of an ensemble of electrons:} Panel b in
Figure~\ref{fig:MultiScale} shows an ensemble of ``individual'' electrons
moving in an electric field, colliding with gas molecules, and forming an
ionization avalanche. The modeling of such electrons with Monte Carlo
particle methods is described in sections~\ref{sec:first_view_PIC} and
\ref{sec:intro-particle}.

{\bf Field enhancement and streamer mechanism:} Panel c in
Figure~\ref{fig:MultiScale} illustrates a streamer discharge with local field
enhancement at the channel tips, as described above. The picture shows the
result of a 3D simulation~\cite{Teunissen_2017}. Such simulations are often
performed with fluid models, which use a density approximation for electrons
and ions, see sections~\ref{sec:intro-fluid-model} and \ref{sec:intro-fluid}.

{\bf Multi-streamer structures:} In most natural and technical processes,
streamers do not come alone, and they interact through their space charges
and their internal electric currents. A reduced model that approximates the
growing streamer channels as growing conductors with capacitance is shown in
panel d of Figure~\ref{fig:MultiScale} and discussed in more detail in
section \ref{sec:macroscopic-models}; such so-called fractal models are a key
to understanding processes with hundreds or more streamers.

{\bf Different scales in time and energy:} A pulsed discharge starts from
single electrons and avalanches, and eventually develops space charge effects
to form a streamer. Later, behind the streamer ionization front, the ion
motion, the deposited heat and consecutive gas expansion, and the initiated
plasma-chemistry become important. These mechanisms can cause a transition to
a discharge with a higher gas temperature and a higher degree of ionization.
Such discharges are known as leaders, sparks and arcs.

The electron energy scales depend on the local electric field and are much
higher at the streamer tip than elsewhere, but typically in the eV range.
However, electron runaway can accelerate electrons into the range of tens of
MeV in the streamer-leader phase of lightning, in a not yet fully understood
process.

In sections \ref{sec:inception}, \ref{sec:streamer} and \ref{sec:flowheat}, we will discuss the
temporal sequence of physical processes in a pulsed discharge in detail.

\subsection{Introduction to numerical models}
\label{sec:intro-models}

We now briefly introduce two types of models that are often used to simulate
streamer discharges: fluid and particle models. A more detailed description
of these models and their range of validity can be found in
 section \ref{sec:simulations}.

\subsubsection{Particle description of a discharge}
\label{sec:first_view_PIC}

Microscopically, the physics of a streamer discharge is determined by the
dynamics of particles: electrons, ions, neutral gas molecules (or atoms) and
photons. The electrons and ions interact electrostatically through the
collectively generated electric field. Their momentum $\vec{p}$ and energy
$\varepsilon$ change in time as
\begin{eqnarray*}
  \label{eq:momentum-eq}
  \partial_t \vec{p} &= q \vec{E},\\
  \partial_t \varepsilon &= q \vec{v} \cdot \vec{E},
\end{eqnarray*}
where $q$ is the particle's charge and $\vec{v}$ its velocity. The energy and
momentum gained from the field is however quickly lost in collisions with
neutral gas molecules. As the typical degree of ionization in streamers at up
to 1~bar is below $10^{-4}$ (see sections~\ref{sec:Conductivity} and
\ref{sec:scalinglaws}), charged particles predominantly collide with neutrals
rather than with other charged particles.
In a particle-in-cell (PIC) code for streamer discharges, it is therefore
common to describe the electrons as particles that move and collide with
neutrals, the slower ions as densities, and the neutrals as a background
density.
To reduce computational costs, each simulation particle typically represents
multiple physical electrons. The neutral gas is included only implicitly
through the collision rates for electron-neutral scattering, excitation,
ionization and attachment collisions.

In a PIC code, the electron and ion densities are used to compute the charge
density $\rho$ on a numerical mesh. The electric potential $\phi$ and the
electric field $\vec{E}=-\nabla\phi$ can then be computed by solving
Poisson's equation
\begin{equation}
  \label{eq:poisson-phi}
  \nabla \cdot (\varepsilon \nabla \phi) = -\rho,
\end{equation}
with suitable boundary conditions, where $\varepsilon$ is the dielectric
permittivity. Note that the electrostatic approximation is used here; its
validity is discussed in section \ref{sec:plasma-theory}.

Compared to fluid models, the main drawback of particle models is their
higher computational cost. Particle models have several important advantages,
however:
\begin{itemize}
  \item They can be used when there are few particles, so that a density
      approximation is not valid. This is for example relevant during the
      inception phase of a discharge, see section~\ref{sec:inception}.
  \item Stochastic processes can be described properly. Such processes
      include not only the electron-neutral collisions, but for example
      also the photo-ionization mechanism. If there are few photoionization
      events, their stochasticity can contribute to streamer branching, see
      section~\ref{sec:branching}.
  \item The distribution of electrons in physical and velocity space is
      directly approximated, whereas additional assumptions are required in
      a fluid model, which may not be valid.
\end{itemize}
For more details about particle models, see section \ref{sec:intro-particle}.

\subsubsection{Fluid models}
\label{sec:intro-fluid-model}

Fluid models employ a continuum description of a discharge, which means that
they describe the evolution of one or more densities in time. In the classic
drift-diffusion-reaction model, the electron density $n_e$ evolves as
\begin{equation}
\label{eq:classical_ne}
  \partial_t n_e = \nabla \cdot \left(n_e \mu_e \vec{E} + D_e \nabla n_e\right) + S_e +
  S_\mathrm{ph},\\
\end{equation}
where $D_e$ is the electron diffusion coefficient and $S_\mathrm{ph}$ is a
source term accounting for nonlocal photo-ionization. The source term $S_e$
corresponds to electron impact ionization $\alpha$ and attachment $\eta$, and
is usually given by
\begin{equation}
  \label{eq:source-term}
  S_e = \alphabar \mu_e E n_e,~~~\alphabar=\alpha-\eta,
\end{equation}
where $E = |\vec{E}|$. Depending on the gas composition, one or more ion
species can be generated. In the simplest case, no additional reactions for
these ions are included, and they are assumed to be immobile. A single
density $n_i$ that describes the sum of positive minus negative ion densities
can then be used, which changes in time as
\begin{equation}
  \label{eq:classical_ni}
  \partial_t n_i = S_e + S_\mathrm{ph}.
\end{equation}
Due to the conservation of electric charge, the source terms have to be equal
in equations (\ref{eq:classical_ne}) and (\ref{eq:classical_ni}).

The transport coefficients ($\mu_e$ and $D_e$) and the source term $S_e$ in
equation (\ref{eq:classical_ne}) depend on the electron velocity
distribution. They are often parameterized using the local electric field or
the local mean energy, see section \ref{sec:intro-fluid}. Details about the
computation of photo-ionization are given in section
\ref{sec:simulations-photoionization}. An example of a simulation of a
positive streamer discharge in atmospheric air with the classic fluid model
is shown in figure~\ref{fig:streamer_sim}.

It should be noted that the reactions in the classical discharge model only
contain interactions of discharge products (like electrons, ions or photons)
with neutrals, and not directly with each other, except through the electric
field.
The reason is that the degree of ionization in a streamer at up to
atmospheric pressure is typically below $10^{-4}$. Processes that are
quadratic or higher in the degree of ionization are therefore negligible.
This is discussed in more detail in section~\ref{sec:scalinglaws}.

\subsection{A first view on streamers in experiments}
\label{sec:first_view_experiment}

We have started with models, because they allow understanding how microscopic
mechanisms interact to create the inner nonlinear structure of a single
streamer. The challenge for modeling lies in covering multi-streamer
processes and discharge phenomena on earlier and later time scales (that will
be addressed in later sections) based on proper micro-physics input.

For experiments, the situation is quite the opposite: It is easier to observe
phenomena with many streamers over longer times than to zoom into the inner
structure of single streamer tips on the intrinsic (nanosecond) time scale.
Therefore, all streamer experimental images shown here are of complete
discharges containing one or more streamer channels.

\begin{figure*}
  \centering
  \includegraphics[width=17 cm]{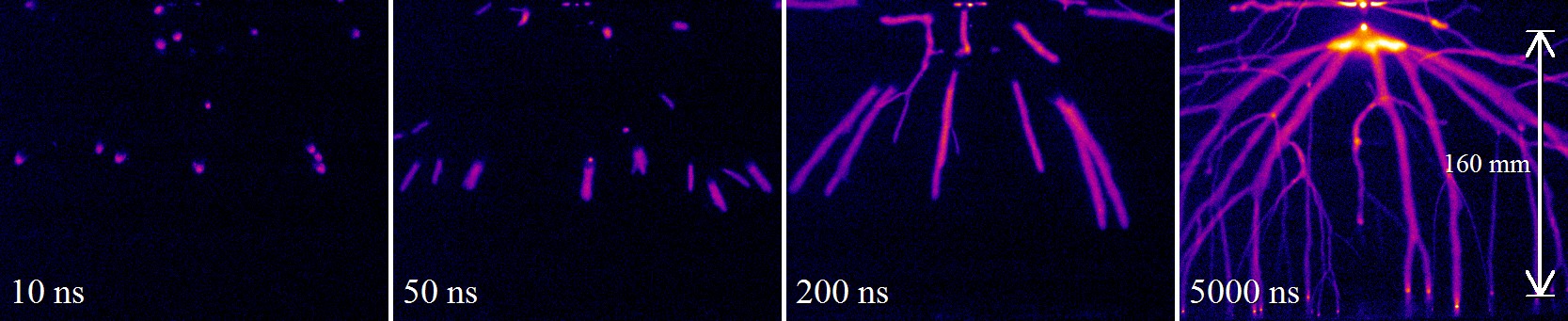}\\
  \caption{Example of ICCD
images for positive streamer discharges under the same conditions using different gate
(exposure) times, as indicated on the images. The camera delay has
been varied so that the streamers are roughly in the centre of the
image. The discharges were captured in artificial air at 200~mbar with
a voltage pulse of about 24.5~kV.
  Image from~\cite{NijdamThesis}.
  \label{fig:velocity-example}}
\end{figure*}

The easiest to acquire, and therefore the most often shown quantity in
streamer experiments is the light emission. Light can easily be imaged by
ICCD or other cameras (see sections \ref{sec:imaging} and
\ref{sec:othergasses} for limitations). In air, a camera will only image the
actively growing regions of a streamer discharge, i.e., the tips, while the
current carrying channels mostly stay dark, as the electric fields and hence
the electron energies are too low in the channels to excite the molecules to
emissions in the optical range. This effect is demonstrated in
figure~\ref{fig:velocity-example} where for short exposures only small dots
are visible.

\begin{figure*}
  \centering
  \includegraphics[width=16 cm]{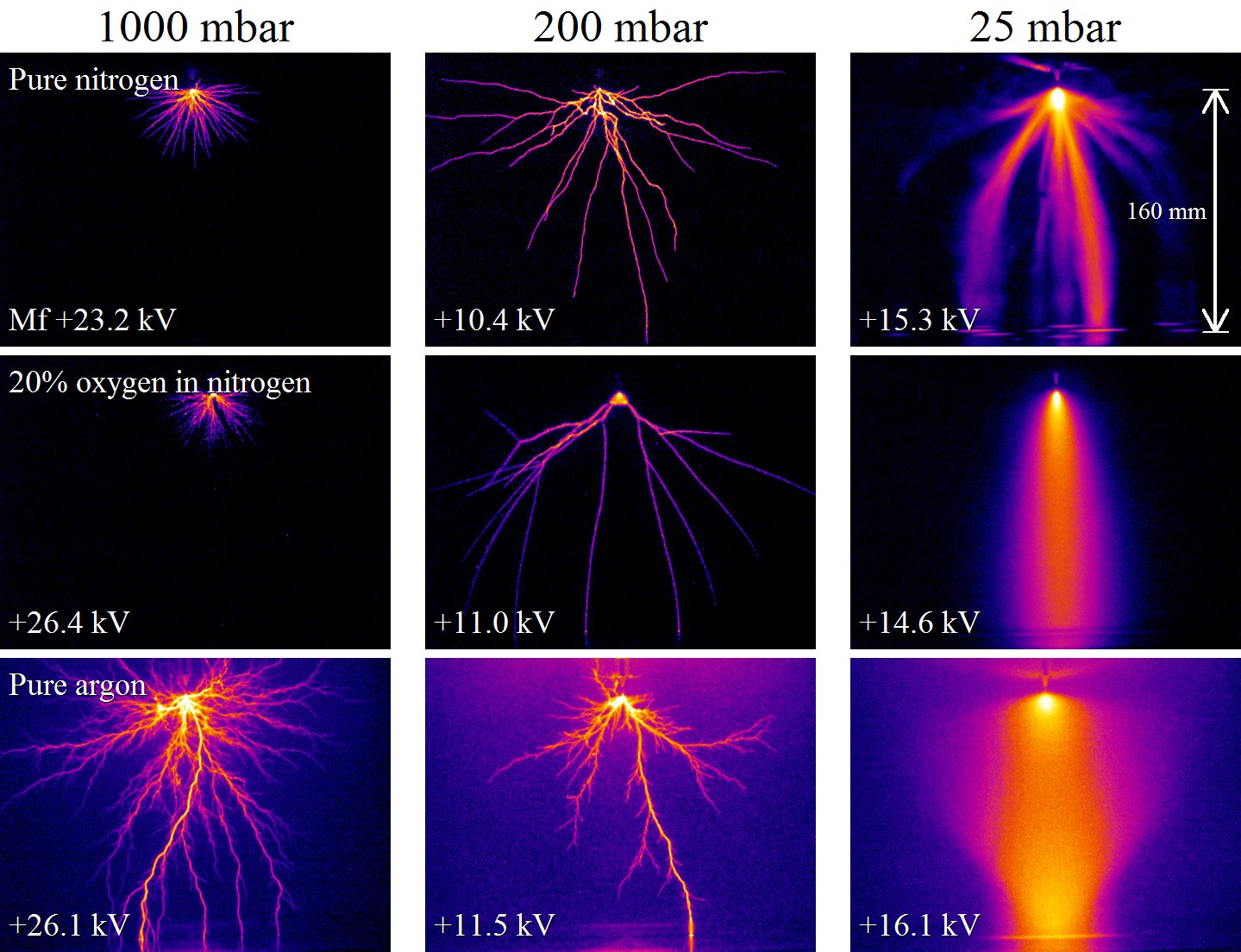}\\
  \caption{Overview of positive streamer discharges produced
in three different gas mixtures (rows), at 1000, 200
and 25~mbar (columns). All measurements have a long exposure time
and therefore show the whole discharge, including transition to glow
for 25~mbar.
  Image adapted from~\cite{Nijdam2010}. 
  \label{fig:gas-pressure-dependence}}
\end{figure*}

Figure~\ref{fig:gas-pressure-dependence} shows long exposure images of
streamers in different gases and pressures. It showcases the wide variety of
shapes and sizes of streamers, ranging from single channels to complex
streamer trees at higher pressures. It also shows the variability in streamer
width and branching behaviour between the different conditions.

\begin{figure*}
  \centering
  \includegraphics[width=16 cm]{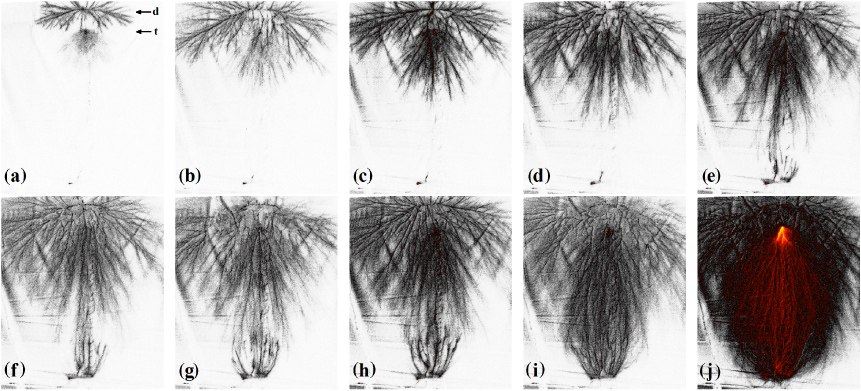}\\
  \vspace{3mm}
  \includegraphics[width=16 cm]{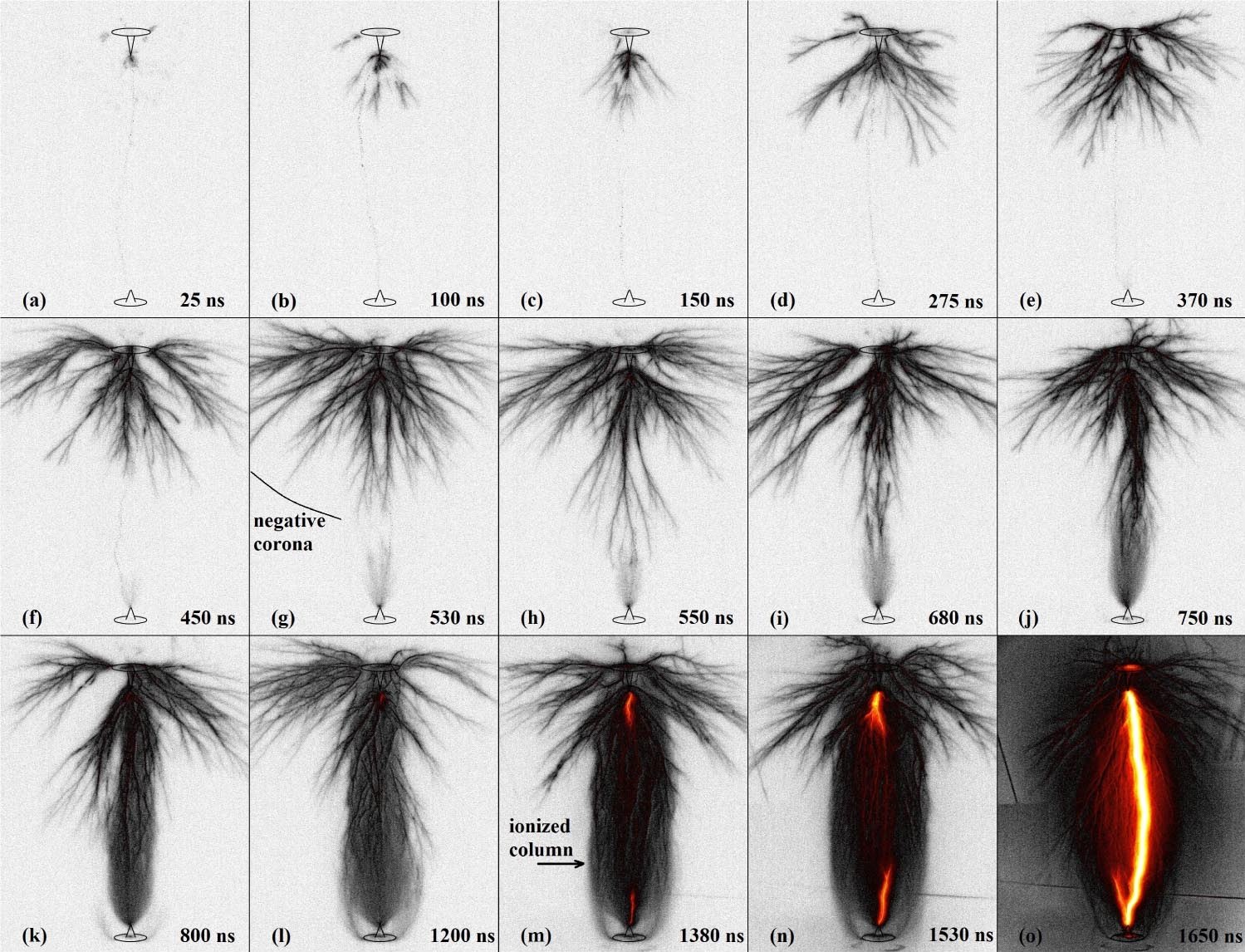}\\
  \caption{Development of positive (top panel) and negative
  (bottom panel) streamers creating
  a high-voltage spark in gap lengths
  of 100 and 127\,cm respectively
  at applied voltages of 1.0 and 1.1\,MV respectively, both
  with a voltage rise time of 1.2\,${\rm \mu}$s in atmospheric air.
  Each picture shows a different discharge under the same conditions
  with increasing exposure time from discharge inception.
  In the top panel these times are (for a-j):
  70, 160, 190, 250, 320, 340, 370, 410, 460 and 610\,ns.
  In the bottom panel they are indicated on the images.
  Images from~\cite{Kochkin2012} and \cite{Kochkin2014}.
  \label{fig:kochkin-full-discharge}}
\end{figure*}

Two examples of the development of a streamer discharge at an applied voltage
of 1~MV over a distance of 1~m in ambient air can be seen in
figure~\ref{fig:kochkin-full-discharge}. The top panel shows positive
streamers propagating smoothly from the top (HV) electrode to the ground
bottom electrode, which are, in the end, met by short negative
counter-propagating streamers and then grow into a hot, spark-like channel.
The bottom panel shows that negative streamer expansion from the top
electrode instead happens in bursts, likely related to the microsecond
voltage rise time (see section~\ref{sec:stabilityfield}). Almost
simultaneously, positive streamers are growing from the elevated bottom
electrode. These meet each other after around 550\,ns, again forming a
spark-like channel.

\section{The initial stage: Discharge inception} \label{sec:inception}
The formation of a discharge requires two conditions: First, a sufficiently high electric field should be present in a sufficiently large
region. Second, free electrons should be present in this region. If no or few of these electrons are present, the discharge may form with a significant delay or not at all.
On the other hand, a sufficient supply of free electrons can reduce the inception delay and jitter, and also the required electric field to start a discharge within a given time.

Below, we will first discuss possible sources of free electrons, and then the conditions on the
electric field to start a discharge, both in the bulk and near a surface. Finally, we discuss
\emph{inception clouds}, a stage immediately before streamer emergence near a pointed electrode in air.

\subsection{Sources of free electrons}\label{sec:incept-sources}
In repetitive discharges, one discharge can serve as an electron source
for the next discharge. Depending on the time span between them,
some electrons can still be present, or they can detach from negative ions
like O$_2^-$ or O$^-$ in air, or they can be liberated through
Penning ionization. Another possibility is storage on solid surfaces.

For the first discharge in a non-ionized gas, possible electron sources
are the decay of radioactive elements within the gas or external radiation.
The actual mechanisms depend on local circumstances. E.g., in the lab, the materials used for the vessel and the lab itself, together with possible radioactive
gas admixtures, determine the local radiation level. UV light can supply
electrons as well, especially from surfaces which can emit for much lower photon energies
than gases.

In the Earth's atmosphere, the availability of free electrons strongly
depends on altitude; we discuss it here in descending order. Above about
85~km at night time or about 40~km at day time, the D and the E layer of the
ionosphere contain free electrons. In fact, the lower edge of the E layer at
night time can sharpen under the action of electric fields from active
thunderstorms, and launch sprite discharges downward which are upscaled
versions of streamers at very low air
densities~\cite{luqueEmergenceSpriteStreamers2009,
luqueMesosphericElectricBreakdown2012, liuSpriteStreamerInitiation2015}, see
also section~\ref{sec:sprites}. On the other hand, electrons are scarce at
lower altitudes, as they easily attach to oxygen molecules. In particular, in
wet air, water clusters grow around these ions and electron detachment is
very unlikely~\cite{Gallimberti1979}. On the other hand, when a high energy
cosmic particle enters our atmosphere, it can liberate large electron numbers
in extensive air showers which could be a mechanism for lightning
inception~\cite{rutjesGenerationSeedElectrons2019}. Up to 3\,km altitude, the
radioactive decay of radon from the ground is the main source of free
electrons~\cite{pancheshnyiRoleElectronegativeGas2005}, except for specific
local soil conditions.

\subsection{Avalanche-to-streamer transition far from boundaries}\label{sec:avalanche-to-streamer}

\subsubsection{Starting with a single free electron. }

The simplest case to consider is a single free electron in a gas in a homogeneous field.
According to equations (\ref{eq:classical_ne}) and (\ref{eq:source-term}),
the ionization avalanche grows if the effective Townsend ionization coefficient $\alphabar$ in a given
electric field strength $E$ is positive, i.e., if $E>E_k$.
During a time $t$, the centre of an avalanche drifts a distance $d = \mu_e E t$ in the electric field, and the number of electrons is multiplied by a factor $\exp\left(\alphabar(E) \, d \right)$.

Eventually, the space charge density of the avalanche creates an electric field comparable to
the external field. At this moment, space charge effects have to be included, and the discharge transitions into the streamer phase. In
ambient
air, this happens when $\alphabar(E)d \approx 18$; this is known as the Meek criterion.
The avalanche to streamer transition is analyzed in~\cite{Montijn2006}.
In particular, it was found that electron diffusion yields a small correction to the Meek number, and that it determines the width of avalanches.
(In contrast, Raizer \cite{Raizer1991} relates the width of avalanches to electrostatic repulsion which is not consistent with the concept that their space charge is negligible.)

When a single electron develops an avalanche in an inhomogeneous electric field $\vec{E}(\vec{r})$, the local multiplication rates $\alphabar(E)$ add up over the electron trajectory $L$ like
$\int_L \alphabar(E(s))\,ds$. The Meek criterion for the avalanche to streamer transition
in air at standard temperature and pressure is then
\begin{equation} \label{eq:Meek}
    \int_L \alphabar(E(s))\,ds \approx 18.
\end{equation}
The Meek number gets a logarithmic correction in the gas number density when it deviates from
atmospheric conditions~\cite{Montijn2006}. This follows from the scaling laws discussed in section
\ref{sec:scalinglaws}.

If there are $N_e$ electrons starting together from about the same location, the required
electron multiplication for an avalanche to streamer transition decreases with $\log N_e$,
since the criterion becomes $N_e\,\exp\left[\int_L \alphabar(E(s))\,ds\right]\approx \exp(18)$.

\subsubsection{Starting with many free or detachable electrons. }

\begin{figure*}
  \centering
  \includegraphics[width=12cm]{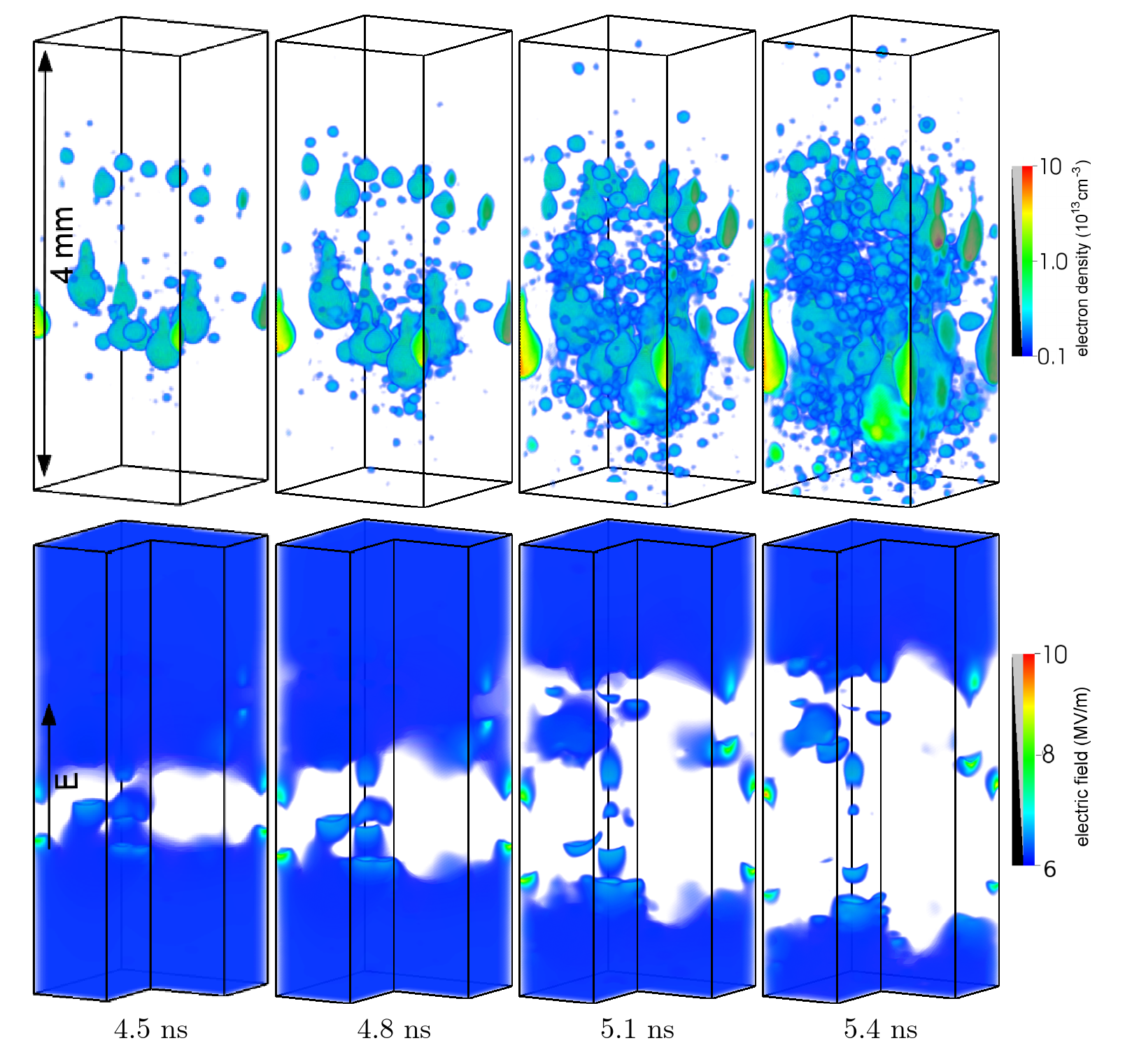}
  \caption{Simulation of discharge inception in atmospheric air in a field of
    twice the breakdown value, taken from \cite{Sun_2014a}. Shown are the
    electron density (top) and electric field (bottom). Initially, a layer of
    O$_2^-$ ions with a density of $10^4 \, \textrm{cm}^{-3}$ was present.
    Electrons detach from these ions and form multiple overlapping avalanches.}
  \label{fig:above-breakdown}
\end{figure*}

When the initial condition is a wide spatial distribution of electrons in an {\em electric
field above breakdown}, streamer formation competes with a more homogeneous breakdown due to
many overlapping ionization avalanches. Such a situation can arise when there is still a
substantial electron density from a previous discharge, or when electrons detach from ions in
the applied electric field. The dynamics
of a pre-ionized layer developing into an ionized and screened region
through a multi-avalanche process
are shown in figure \ref{fig:above-breakdown}. While the Meek number characterizes the critical propagation length of an avalanche
for space charge effects to set in, the ionization screening time \cite{Teunissen_2014}
\begin{equation}
  \label{eq:ionization-screening}
    \tau_{is}=\ln\left(1+\frac{\bar\alpha\epsilon_0E}{e n_0}\right)/(\bar\alpha\mu_eE)
\end{equation}
is the temporal equivalent for a multi-avalanche process, where $n_0$ is the initial electron density and $E$ the applied electric field.
The ionization screening time can be seen as the generalization
of the dielectric relaxation time (\ref{eq:tau-drt}) to an electron density that changes in time due to
the effective impact ionization $\bar\alpha$.

In the past, many authors have simulated streamers in electric fields
above the breakdown value. This was often done to reduce computational costs,
since such streamers can be simulated within shorter times in smaller computational domains.
However, the results of such simulations can change substantially if background ionization is added,
since streamer breakdown and the homogeneous breakdown mode of
Figure~\ref{fig:above-breakdown} are competing when the background field is above breakdown.

On the other hand, if the {\em electric field is below breakdown}, discharges would mostly not
start. However, if there is a sufficiently high and compact density of electrons and ions,
this ionized patch can screen the electric field from its interior and enhance it at its
edges. This leads to a local electron multiplication and drift only in the region above
breakdown, and to the emergence and growth of a positive streamer at one side of the initial
plasma, while the negative streamer on the opposite side is delayed if it grows at all.

The basic differences between discharge inception below and above the breakdown field are discussed in more detail in~\cite{Sun_2014a}.


\subsection{Streamer inception near surfaces}\label{sec:incept-surf}

Above, we have discussed discharge inception within the gas, far from any boundaries.
However, many discharges ignite near dielectric or conducting surfaces, such as electrode needles
or wires, water droplets or ice particles, because the electric field near such objects is enhanced.
For the same shape and material, positive discharges ignite more easily than negative
ones, at least in air.

The inception process again is determined by the availability of free electrons near the surface and by their
avalanche growth. As discussed above, the electron number in an avalanche
grows as the exponent of $\int_L\alphabar(E) ds$ where the integral is taken
over the avalanche path $L$ along an electric field line.
The Meek number is calculated on the path $L$ that has the largest value
of the integral and ends at the surface.
In electrical engineering, it is known from experiments
that a discharge near a strongly curved electrode can start
when the Meek number is as low as 9 or 10
\cite{nasserMathematicalPhysicalModel1974, zaenglApplicationStreamerBreakdown1994, lowkeOnsetCoronaFields2003a, naidisConditionsInceptionPositive2005, mikropoulosThresholdInceptionConditions2015},
but apparently this is not known to
geophysicists modeling lightning inception near ice particles in thunderclouds~\cite{babichPositiveStreamerInitiation2016, babichRoleChargedIce2017}
who use a Meek-number of 18 for their estimates.

In the lightning inception study \cite{Dubinova2015}, fluid simulations showed that a Meek
number of 10 is sufficient to start a streamer discharge from an elongated ice particle.
In their PhD theses \cite{Dubinova_Thesis, Rutjes_Thesis}, 
Dubinova and Rutjes argued that there is a the major difference between streamer inception far from or near a surface: a streamer forms from an avalanche
far from surfaces when a sufficient negative charge has accumulated in the
propagating electron patch, and the emergent streamer has negative polarity. 
(When photo-ionization is strong enough, a positive streamer can form 
at the other end of the ionized patch.) In contrast, a streamer near a 
conducting or dielectric surface forms when the approaching ionization
avalanches leave a sufficient density of (relatively immobile) 
positive ions behind near the surface, and the emerging streamer is positive. 
So there is no reason why the number of ionization events in both cases should be equal.

\subsection{Inception cloud or diffuse discharge or spherical streamer or wide ionization front} \label{sec:inception_cloud}
\begin{figure}
  \centering
  \includegraphics[width=8cm]{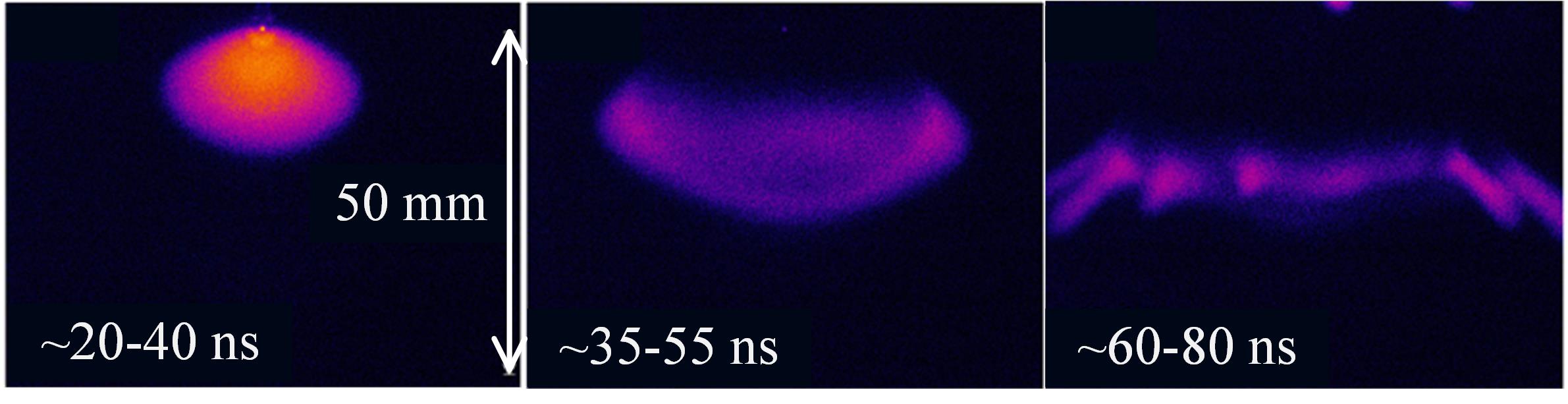}
  \caption{Inception cloud (left), shell (middle) and destabilization
  of the shell into streamer channels (right) of a streamer discharge
  in 200\,mbar artificial air.
A 130\,ns, +35\,kV voltage pulse is applied to 160\,mm point-plane gap.
Indicated times are from the start of the voltage pulse.
Figure from \cite{Chen2015}.}
  \label{fig:inception-cloud-chen}
\end{figure}

A positive discharge in air that starts from a needle electrode,
does not directly develop from an avalanche phase into an elongated
streamer, but there is a stage of evolution in between that has
been called \emph{inception cloud} in our experimental papers
\cite{Briels2008c,Briels2008b}.
The same phenomenon is also seen for negative polarity air
discharges~\cite{Nijdam2011} (see also figure~\ref{fig:Onion}).
An example of such an inception cloud is shown in
figure~\ref{fig:inception-cloud-chen}
but it can also be observed in figures~\ref{fig:doublepulse}
and~\ref{fig:Rep-freq-dep}.
These and other figures show that first light is emitted all
around the electrode, and that this cloud is
growing. In a second stage, the light is essentially emitted
from a thin expanding and later stagnating shell around the
previous cloud. And in a third stage, this shell breaks up into streamers.
Similar phenomena have also been discussed in literature
under the name of a diffuse discharge~\cite{Rep'ev2006,Tardiveau2009,
tarasenkoRunawayElectronsDiffuse2019} or recently
a spherical streamer~\cite{naidisSubnanosecondBreakdownHighpressure2018, tarasenkoFormationWideStreamers2018}
or an ionization wave~\cite{hoderEmergingExpandingStreamer2020}.

The shell phase is clearly a nonlinear structure with a propagating ionization front, while the
electric field is screened from the interior, almost like the streamer illustrated in
figure~\ref{fig:streamer_sim}, but not yet elongated,
but more semi-spherical. The localized light emission
indicates the location of the ionization front (just like in the streamers in Fig.~\ref{fig:velocity-example}), and the maximal radius fits reasonably well with the
assumption that the interior is electrically screened, and that the electric field on the
boundary is roughly the breakdown field $E_k$.
This is because the radius $R$ of an ideally conducting sphere on an electric
potential $U$ with a surface field $E$ is $R=U/E$; therefore the maximal
radius of the inception cloud is
\begin{equation} \label{eq:Rmax}
R_{\rm max}=U/E_k,
\end{equation}
where $U$ is the voltage applied to the electrode
and $E_k$ is the breakdown field~\cite{Nijdam2011}; and this radius
fits the experimental cloud radius quite well.
We mention that Tarasenko in his recent
review~\cite{tarasenkoRunawayElectronsDiffuse2019} attributes
the formation of inception clouds or diffuse discharges
to electron runaway; we will
discuss electron runaway in section~\ref{sec:fastelectrons},
but we stress here that the ionization dynamics and the
maximal radius $R_{\rm max}$ point to the radial expansion
of a streamer-like ionization wave with interior screening,
indeed a "spherical streamer", in the words
of Naidis {\it et al.}~\cite{naidisSubnanosecondBreakdownHighpressure2018}.

\begin{figure}
  \centering
  \includegraphics[width=8cm]{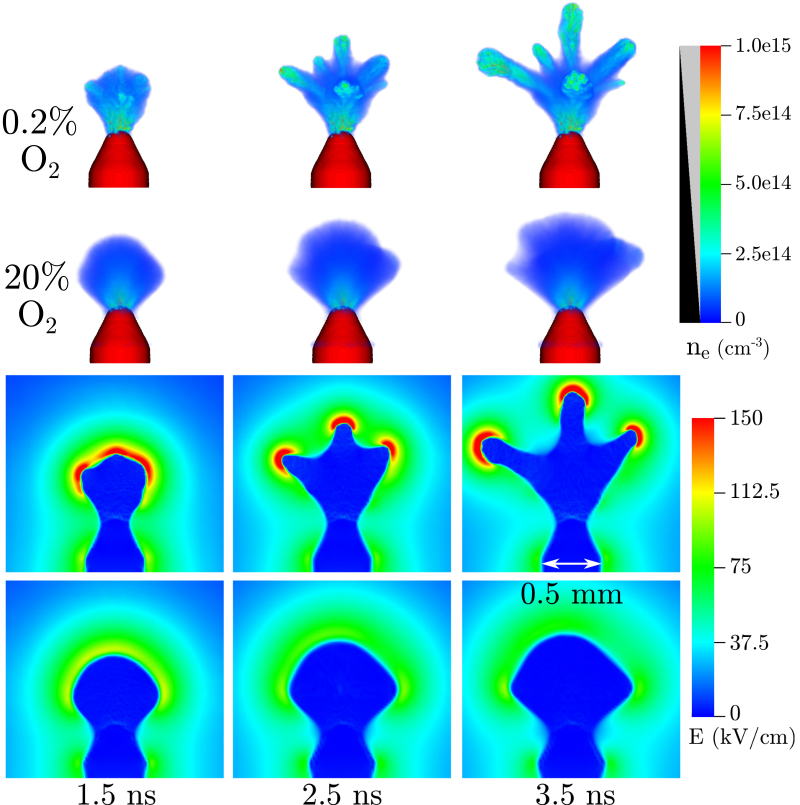}
  \caption{Particle-in-cell simulation of discharge inception around a needle
    electrode. Two gases are used: N$_2$ with 20\% and $0.2$\% O$_2$, both at 1
    bar. The electron density (top) and a cross section of the electric field
    (bottom) are shown. Figure adapted from \cite{Teunissen_2016}.}
  \label{fig:inception_cloud_simulation}
\end{figure}

The first estimates above were substantiated by further experimental and
simulation studies \cite{Clevis2013,Chen2015,Teunissen_2016}.
Figure~\ref{fig:inception_cloud_simulation} shows 3D simulations
of positive discharge inception near a pointed electrode in
nitrogen with 0.2\% or 20\% oxygen~\cite{Teunissen_2016}.
In the case of nitrogen with 20\% oxygen (artificial air),
the formation of an electrically screened, approximately spherical 
inception cloud can be seen in the plots for the electric field.

By varying nitrogen-oxygen ratios, Chen \emph{et al.}~\cite{Chen2015} showed that sufficient
photo-ionization is essential for the stable formation of an inception cloud, which was confirmed by the simulations in
\cite{Teunissen_2016}, see figure~\ref{fig:inception_cloud_simulation}.
At 100\,mbar, Chen {\it et al.}\ found that below 0.2\,\% oxygen, the size of the inception
cloud decreases significantly or breaks up almost immediately.
This is because photo-ionization has a stabilizing effect on the discharge front,
both in the phase of the nearly spherically expanding cloud, and later
in the streamer phase. This effect of photo-ionization
is seen similarly in streamer branching in different gas mixtures,
as discussed in section~\ref{sec:branching-other-gases}.

The applied voltage and the voltage rise time clearly determine the degree of ionization
within the cloud and the cloud radius. Diameters and velocities of the streamers that emerge
from the destabilization of the inception cloud, can vary largely as will be discussed in the
next section.
Understanding how the cloud properties determine the streamer properties is a task for the
future.

\section{Streamer propagation and branching}\label{sec:streamer}
\label{sec:propagation}

\subsection{Positive versus negative streamers}

\begin{figure*}
\centering
\includegraphics[width=16 cm]{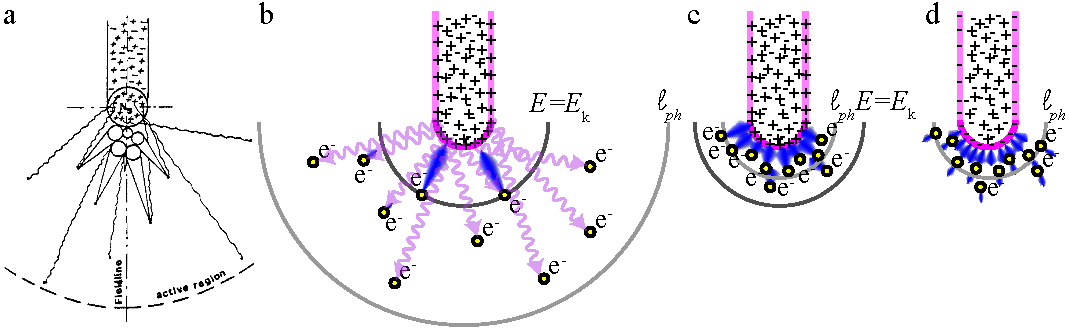}
\caption{\label{fig:avalanches-schematic}\label{fig:streamer-propagation-basics}
 Schematic depictions of streamer propagation. a)
 Illustration of positive streamer propagation in air based on the original concept
  of Raether~\cite{Raether1939}, published in English by
  Loeb and Meek~\cite{Loeb1940}. Picture taken
  from~\cite{Gallimberti1979}.
  Panels b-d) show an updated scheme for b) positive streamers with
  few photons with a long mean free path,
  c) positive streamers in air and d) negative streamers in air.
  Avalanches start from a yellow electron and are indicated in blue, $\ell_{\rm photo}$ indicates the photo-ionization range and $E=E_{\textrm{k}}$
  indicates the active region. Note also that panel a) shows a net positive charge in a spherical head region, while panels b-d) show have surfaces charges around the streamer head and along the lateral channel.
  }
\end{figure*}

\begin{figure}
    \centering
    \includegraphics[width=8cm]{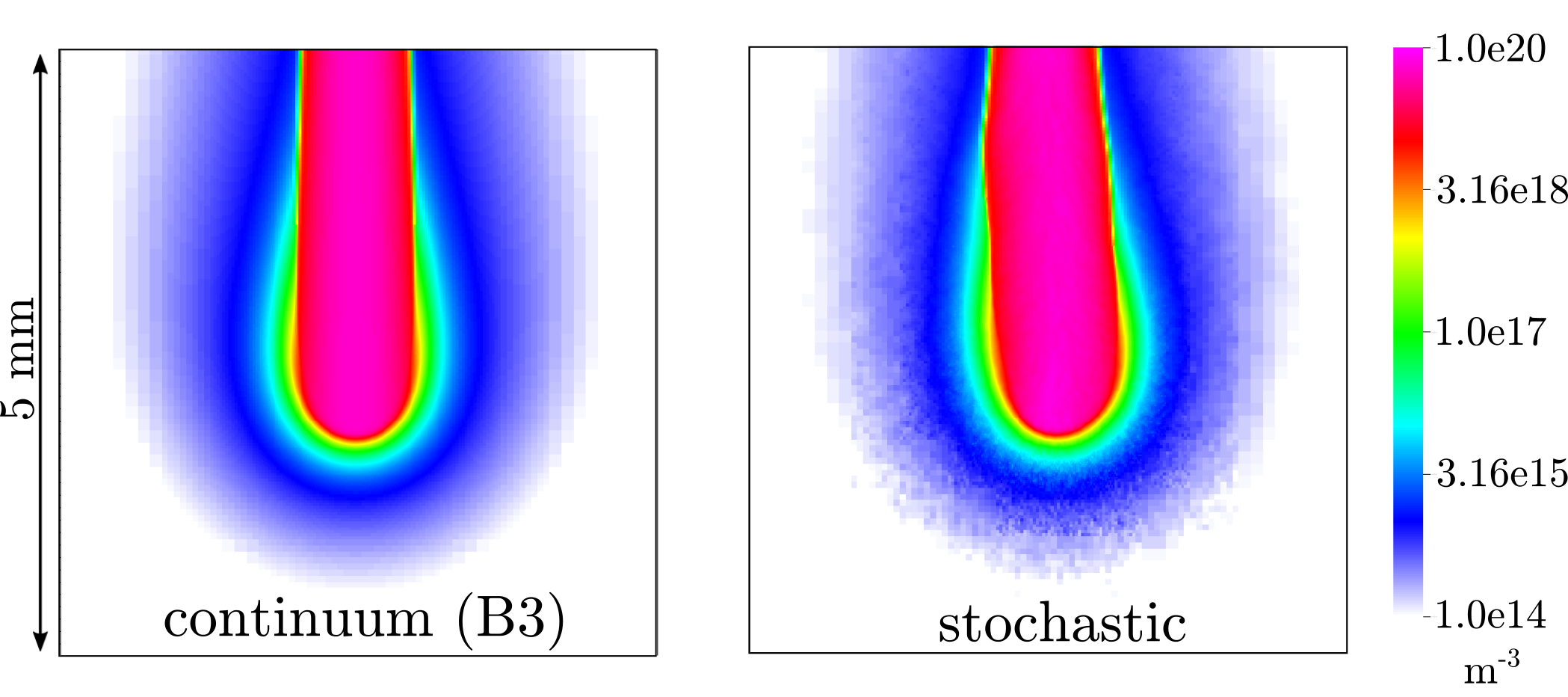}
    \caption{Cross sections through 3D simulations of positive streamers in air,
      showing the electron density on a logarithmic scale. Two photo-ionization
      models are used: a continuum approximation~\cite{Bourdon2007} (left) and a
      stochastic (Monte Carlo) model with discrete single photons (right). Due
      to the large number of ionizing photons in air, individual electron
      avalanches cannot be distinguished, and the continuum approximation works
      well. Figure adapted from~\cite{Bagheri_2019}.}
    \label{fig:photoelectrons}
\end{figure}

Streamer discharges can have positive or negative polarity.
See figure~\ref{fig:avalanches-schematic}c-d for a schematic comparison.
A positive streamer carries a positive charge surplus at its head and
typically propagates towards the cathode, i.e., against the electron
drift direction. A negative streamer propagates towards the anode.
While its propagation in the direction of the electron drift
seems to be the most natural motion, positive streamers in air are seen to start
more easily, and to propagate faster and further, as is discussed in more detail below. Luque~{\it et~al.}~\cite{Luque2008} explain the asymmetry between the polarities as follows. The space charge layer around a negative streamer is formed by an excess of electrons. These electrons drift outward from the streamer body, and their drift in the lateral direction decreases the focusing and enhancement of the electric field at the streamer tip. This process continues even when the lateral field is below the breakdown threshold.
On the contrary, a positive streamer grows essentially only where the field is high enough for
a substantial multiplication of approaching ionization avalanches. Their charge layers are formed by an excess of positive ions, and these ions hardly move.
(For the available free electrons to start these avalanches, see
section~\ref{sec:electronsource}.) Therefore the field enhancement is better maintained
ahead of positive streamers.

The traditional (but not fully correct) interpretation of a propagating positive streamer is reproduced in
figure~\ref{fig:avalanches-schematic}a. It shows the streamer head as a
sphere filled with positive charge, and 4 to 5 ionization
avalanches propagating towards it are shown.
The active region is the region
where the electric field is above the breakdown value.
Note that simulations in air (like in figure~\ref{fig:streamer_sim})
show a quite different picture: (i) the positive
charge is located in a thin layer around the head rather than in a sphere, and (ii) the
avalanches in air are so dense that they cannot be distinguished.
We have schematically depicted this in figure~\ref{fig:avalanches-schematic}c,
and a simulation example is shown in figure \ref{fig:photoelectrons}.

\subsection{Streamer diameter and velocity}
\label{sec:VeloRadField}

Streamer properties depend on gas composition and density.
The gas composition determines the transport and reaction coefficients and the strength and
properties of photo-ionization. The gas number density determines the mean free path of
the electrons between collisions with molecules, which is an important length scale for discharges, see section \ref{sec:scalinglaws}.
For the present section it suffices to know that for physically similar streamers at different gas number densities $N$, the length and time scales scale like $1/N$, electric fields with $N$, ionization degrees with $N^2$ and velocities and
voltages
are independent of $N$.

But even for one gas composition, density and polarity, there is not one
streamer diameter and velocity. A classical question in
streamer physics used to be: "What determines the radius of a streamer?"~\cite{Bazelyan_spark_discharge_1998},
as the radius is the input for so-called 1.5-dimensional
models~\cite{Morrow1985} that modelled streamer evolution
in one dimension on the streamer axis and included an electric field profile
based on the model input for the streamer radius.
But measurements show that streamer diameters and velocities can vary
by orders of magnitude in the same gas, as summarized below.

\subsubsection{Measurements. $\quad$}

Experimentally, streamer diameters and velocities can be measured relatively easily, although both have
their issues, as is explained in section~\ref{sec:diamvelodiagnostics}.
Experimental streamer diameters are always
optical diameters (usually full width at half maximum intensity), while
the natural radius evaluated in models is the radius of the space charge layer,
which is also called the electrodynamic radius; it is about twice the optical radius~\cite{nudnovaStreamerHeadStructure2008,luqueEmergenceSpriteStreamers2009}.

\begin{figure}
\centering
\includegraphics[width=7.5 cm]{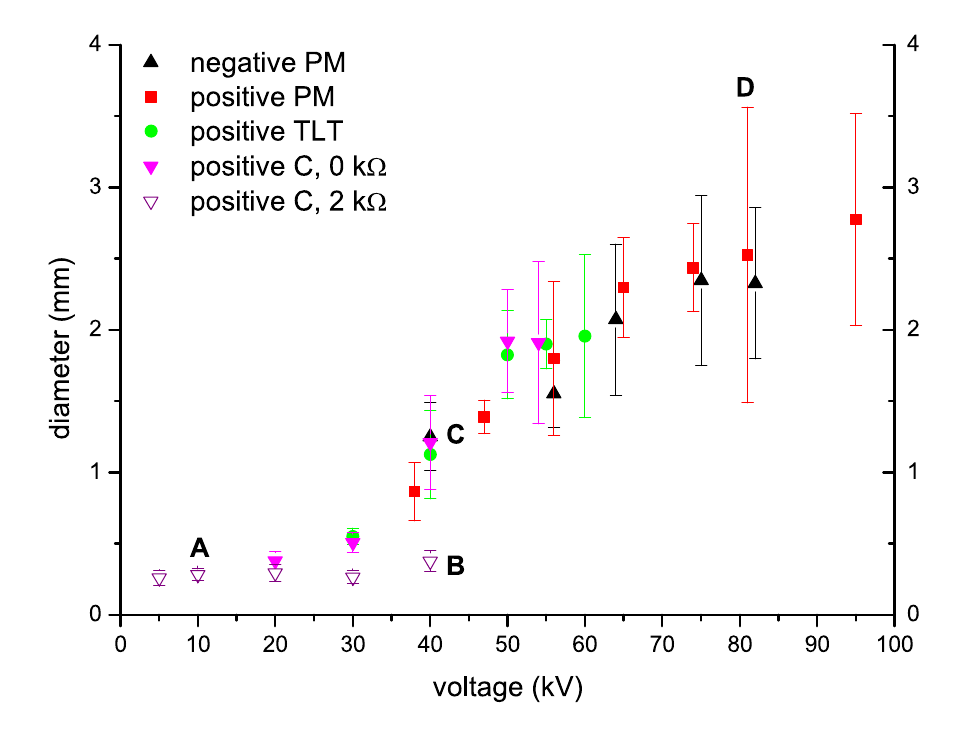}
\includegraphics[width=7.5 cm]{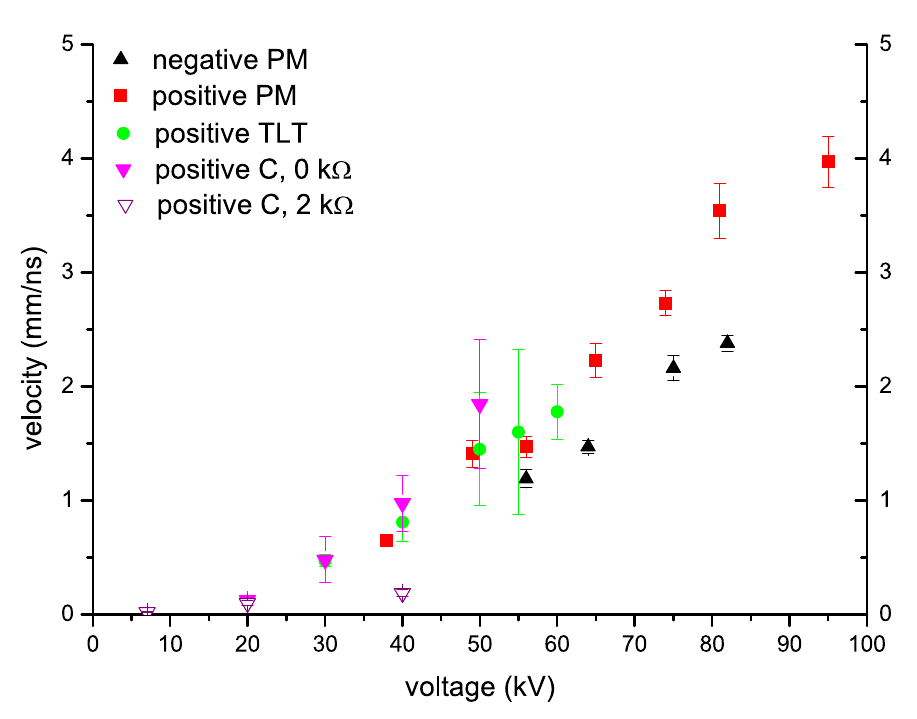}\\
\caption{\label{fig:TanjaPM} Diameter (left) and velocity (right) of streamers as a function
of applied voltage and polarity, reproduced from \cite{Briels2008}. The different voltage
sources and their voltage rise times are 15 ns for PM, 25 ns for TLT, 30 ns for C with 0~k$\Omega$, and 150
ns for C with 2~k$\Omega$.}
\end{figure}

{\bf Overview of diameters and velocities of positive and negative streamers in STP air. }
In air at standard temperature and pressure, Briels {\it et al.}\ \cite{Briels2008} found in a
study published in 2008, that streamer diameter and velocity depend strongly on voltage,
voltage rise time and polarity. Their results are reproduced in Fig.~\ref{fig:TanjaPM}. They
show for their needle plane set-up with a 40 mm gap that:
\begin{itemize}
    \item positive streamers appear for voltages above 5\,kV, but negative ones
        only above 40\,kV,
    \item velocities and diameters of positive streamers stay small and do not change with voltage, when the voltage rise time is as long as 150\,ns,
    \item for the faster rise times of 15, 25 and 30\,ns, positive streamer diameters
        grow by a factor 15 in the voltage range from 5 to 96\,kV, and their velocity grows
        by a factor of 40,
    \item for a rise time of 15\,ns and for voltages varying from 40 to 96\,kV, diameter and
        velocity of positive and negative streamers are getting more similar, but the positive
        streamers are always wider and faster,
    \item for any fixed set of conditions, a minimal streamer diameter $d_\mathrm{min}$
        could be identified and
        such minimal streamers do no longer branch, but they can still propagate for long distances.
\end{itemize}
It should be noted that in longer gaps with a point-plane (or similarly inhomogeneous)
geometry streamers can branch into thinner streamers or decrease in diameter and velocity
without branching. Examples of this are shown in the 200\,mbar images in
figure~\ref{fig:gas-pressure-dependence}.

{\bf Fits for velocities and diameters. } Briels {\it et al.}
\cite{Briels2008} also presented the empirical fit
$v=0.5d^2\,\textrm{mm}^{-1}\,\textrm{ns}^{-1}$ for the relation between velocity $v$ and
diameter $d$. A similar relation between diameter
and velocity was found for sprite discharges (see section~\ref{sec:sprites}) in~\cite{Kanmae2012},
but for larger reduced diameters and velocities on a similar curve.
Chen \textit{et al.}.~\cite{Chen2013} find that the relation of Briels {\it et al.} overestimates
velocities for higher voltages (they use up to 290\,kV in a 57\,cm gap)
and give the relation $v=(0.3\,\textrm{mm} + 0.59d)\,\textrm{ns}^{-1}$.
However, these discrepancies should be seen in
the perspective that a functional dependence was left out of these fits: as Naidis
\cite{Naidis2009} has pointed out, an evaluation of the classical fluid model shows that the
velocity depends not only on the diameter, but also on the maximal electric field at the
streamer head.

{\bf Range of measured velocities. } The lowest velocities reported for positive laboratory streamers in air
(and other nitrogen-oxygen mixtures) are around
$10^5$~m/s, or at a late stage of development even as low as $6\cdot 10^{4}$~m/s in air and $3\cdot 10^{4}$~m/s in nitrogen ~\cite{Briels2008b}.
Typical velocities range between $10^5$~m/s
and around $10^6$~m/s~\cite{Dawson1965,Allen1995,Tardiveau2002,Pancheshnyi_2005,Winands2008a,
Briels2008b,Nijdam2010,Meng2015}.
Maximum velocities are reported at $3-5\cdot10^6$~m/s~\cite{Yi2002,Namihira2003,Chen2013,Zeng2013}
for high applied voltages.
For sprite discharges (see section~\ref{sec:sprites}), velocities of up to $5\cdot10^7$~m/s
are commonly reported~\cite{Kanmae2012,Stenbaek-Nielsen2013}
with one exceptionally high observation of velocities up to
$1.4\cdot10^8$~m/s~\cite{McHarg2002}, but velocities of $10^5$~m/s
are also seen in sprites~\cite{Moudry2002,Ebert2011}.

\begin{figure}
\centering
\includegraphics[width=8.5 cm]{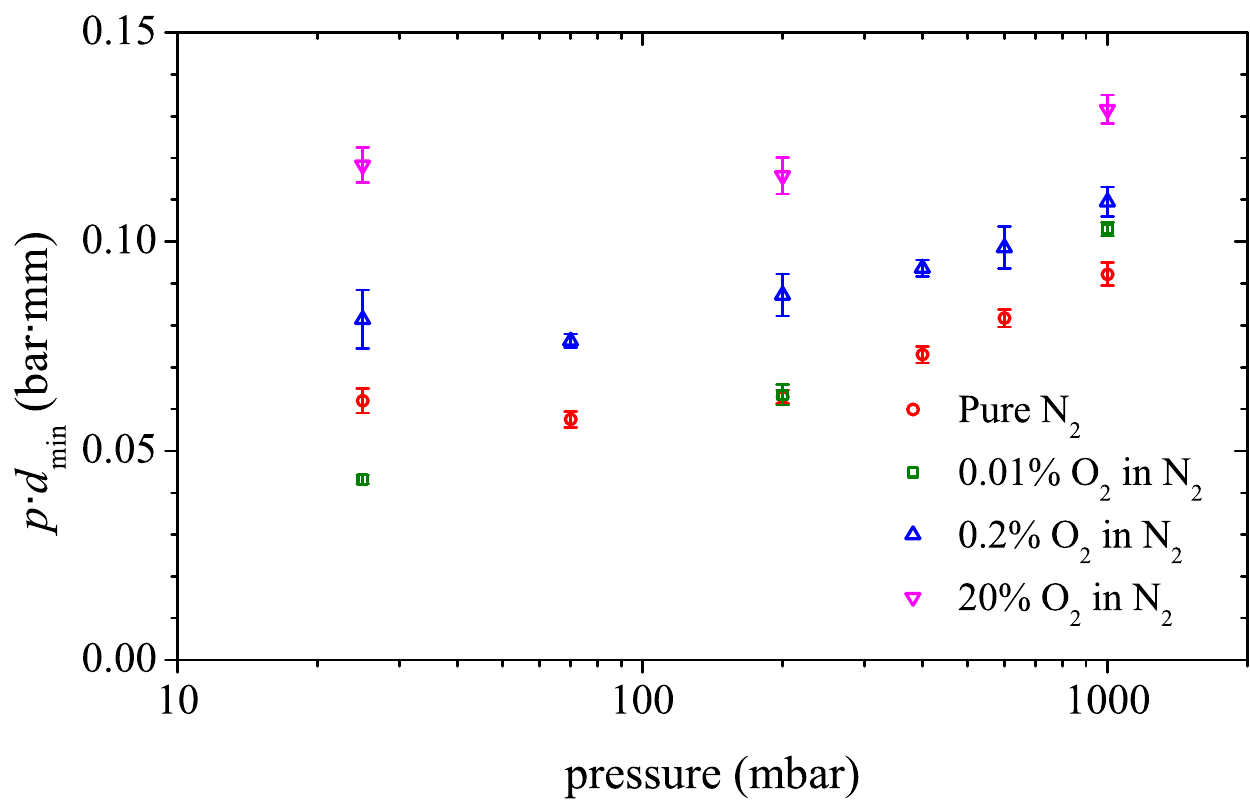}\\
\caption{\label{fig:pdminGraph} Scaling of the reduced minimal diameter ($p\cdot d_\mathrm{min}$)
with pressure ($p$) for the four different nitrogen oxygen mixtures.
Image from~\cite{Nijdam2010}.}
\end{figure}

{\bf Range of measured diameters. } In \cite{Briels2008b}, streamer discharges in air and in nitrogen of
unknown purity were compared. By using a slow voltage rise time of 100-180\,ns, the streamers
are intentionally kept thin. Here, minimal streamer diameters $d_\mathrm{min}$ in air
as function of pressure $p$ were found to scale well with inverse pressure
with values of $p \cdot d_\mathrm{min} = 0.20 \pm 0.02\,\textrm{mm\,bar}$.
(Support for the $d_\mathrm{min}$ concept is given in the next subsection.)
In nitrogen, streamers are thinner with minimal diameters
$p \cdot d_\mathrm{min} = 0.12 \pm 0.02\,\textrm{mm\,bar}$. These values
are consistent with reduced diameters of sprites for which
$p \cdot d_\mathrm{min} / T = 0.3 \pm 0.2\,\textrm{mm\,bar/(293\,K)}$ was
found in~\cite{Gerken2000}.
In \cite{Nijdam2010}, we improved gas purity and optical diagnostics and
studied more nitrogen oxygen mixtures. This led to similar trends but somewhat
lower values of $p \cdot d_\mathrm{min}$ as is shown in figure~\ref{fig:pdminGraph}.
Here $d_\mathrm{min}$ is the minimal streamer diameter observed experimentally.

\subsubsection{Theory. $\quad$}\label{sec:prop-theory}

{\bf The minimal streamer diameter $d_\mathrm{min}$.}
Figure \ref{fig:TanjaPM} shows that for low voltages
and/or large voltage rise times, the streamers have a fixed small diameter. Should one assume
that there is indeed a minimal streamer diameter, or could there be streamers with smaller
diameter that are just not detected?
The minimal streamer diameter can be estimated from the classical fluid model of
section \ref{sec:first_view_model},
as already argued in~\cite{Briels2006,Briels2008b,Ebert2010}. The key to streamer formation is the field enhancement
ahead of its tip, as illustrated in figure \ref{fig:streamer_sim}. This enhancement can only
take place if the thickness $\ell$ of the space charge layer is considerably smaller than the streamer
radius $R=d/2$. But $\ell$ has a lower limit as well. This is because a change $\Delta E$ of the electric field across
the layer requires a surface charge density $\epsilon_0 \Delta E$
according to electrostatics (\ref{eq:Gauss}). This surface charge is created by the
charge density $\rho$ within the layer integrated over its width:
\begin{equation}
    \epsilon_0 \Delta E = \int_\ell \rho(z)\;dz, \quad \mbox{where }\rho = e(n_i - n_e).
\end{equation}
The charge density is of the order of $e n_i$ where $n_i$ is the ionization density (\ref{eq:ne})
behind the front.
As a result, the width of the space charge layer is of the order of
\begin{equation}
    \frac1\ell \approx \frac{\int_{E_{\rm behind}}^{E_{\rm max}}\bar\alpha(E') dE'}{E_{\rm max} - E_{\rm behind}}\le\bar\alpha(E_{\rm max}),
\end{equation}
where $\Delta E = E_{\rm max}-E_{\rm behind}$ is the difference between
the maximum of the electric field $E_{\rm max}$ in the front and the
electric field $E_{\rm behind}$ immediately behind the ionization front.


{\bf A lower bound for the velocity $v_{min}$ of negative streamers} can be derived as
follows. A negative streamer ionization front moves with the electron drift velocity, augmented with
effects of diffusion, impact ionization and photo-ionization.
Therefore the electron drift velocity is a lower bound to the velocity of the streamer
ionization front. Furthermore the electric field at the streamer tip must have at least the
breakdown value.
If the drift velocity increases with electric field, then $v_{min}=\mu_e(E_k)\;E_k$ is a lower bound for the velocity of a negative streamer. In air at standard temperature (i.e., at $0^\circ$~C.), this velocity is approximately $1.3 \cdot 10^5$~m/s.
Within the range of validity of the scaling laws (see section~\ref{sec:scalinglaws}), this velocity is independent of air
density.

{\bf The inception cloud} was already discussed in section~\ref{sec:inception_cloud}.
When the cloud destabilizes into streamers, velocity and radius of the streamers
are determined by radius and inner ionization profile
of the cloud, and these in turn depend on the voltage characteristics
like rise time and maximal voltage. This dependence is clearly seen in
experiments~\cite{Briels2008c,Briels2008b,Nijdam2011,Clevis2013,Chen2015}.
Understanding the cloud destabilization is the key to understanding
how streamers of different diameter and velocity emerge.
Some first steps have been taken in \cite{Teunissen_2016}.

\subsection{Electric currents}
\label{sec:currents}

\subsubsection{Measurements. }
As the velocities and diameters of streamers vary widely, so do their electric currents.
Pancheshnyi {\it et al.}~\cite{Pancheshnyi_2005} measured streamer currents
of the order of 1~A or less in 2005, and Briels {\it et al.}~\cite{Briels2006}
explored a wider parameter range and measured streamer currents from 10~mA
up to 25~A in 2006 for streamers of different velocity and diameter.

\subsubsection{Theory. }
\label{sec:currents-theory}
The streamer current is typically maximal at the streamer head, and
dominated by the displacement of the streamer head charge.
This current can be estimated.
As argued above, the surface charge density within the screening layer
is approximated by $\varepsilon_0 \Delta E$, and hence an upper bound for the surface charge density around the streamer head
is $\varepsilon_0 E_\mathrm{max}$, where $E_\mathrm{max}$
is the streamer's maximal electric field.
Furthermore, we approximate this surface charge density as being present over an area $2 \pi R^2$,
i.e., over a semi-sphere, where $R$ is the streamer radius.
The streamer's head then has approximately a charge of
$Q = 2 \pi R^2 \varepsilon_0 E_\mathrm{max}$, distributed over
the head radius $R$. An approximation for the current at the head
is therefore $I_{\rm max}\approx Q\cdot v/R$, where $v$ is the streamer velocity.
For instance, for a wide and fast streamer in ambient air
with a maximal field of 20~MV/m,
an electrodynamic radius of 2,5~mm and a velocity of $3\cdot10^6$~m/s,
the electric current at the head is approximately 8~A. (We remark,
that the electrodynamic
radius characterizes the location of the space charge layer,
and it is approximately equal to the diameter (full width half maximum)
of light emission observed in experiments.)

It should be noted that the scaling laws that relate physically similar streamers
at different gas densities $N$ (see section~\ref{sec:scalinglaws}) imply
that the currents do not depend on density.

\subsection{Electron density and conductivity in a streamer}
\label{sec:Conductivity}

\subsubsection{Measurements. }

The conductivity of a streamer channel is dominated by electron
mobility times electron density, except if electron attachment has seriously depleted the electrons.
Electron densities in streamer channels in ambient air
are of the order of $10^{19}-4\cdot10^{21}$\,m$^{-3}$, see
e.g.~\cite{Hubner2013,Inada2017}, i.e, there is one free electron
per $(60~{\rm nm})^3$ to $(500~{\rm nm})^3$,
while the neutral density in ambient air is $2.5\cdot10^{25}$\,m$^{-3}$, so one per (3.4\,nm)$^3$.

\subsubsection{Theory. }
{\bf Electron density behind the ionization front. }
The ionization density $n_e\approx n_i$ in the neutral plasma
immediately behind the ionization front depends
on the electric field ahead and in the front.
For ionization fronts propagating with constant velocity,
the approximation
\begin{equation}\label{eq:ne}
    n_i = \frac{\epsilon_0}e\int_{E_{\rm behind}}^{E_{\rm max}}\bar\alpha(E')dE',
\end{equation}
has been suggested in \cite{babaevaTwodimensionalModellingPositive1996,ebertStreamerPropagationPattern1996} and in the references therein;
here $E_{\rm max}$ is the maximal electric field in the front
and $E_{\rm behind}$ the electric field immediately behind the front.
In the appendix of~\cite{liDeviationsLocalField2007} a more general derivation
of~(\ref{eq:ne}) is given for planar negative fronts without photo-ionization or background ionization: Observe the change of ionization density
and electric field over time at a fixed position in space
while the ionization front is
passing by. Neglecting photoionization, the change of the ion density
is given by $\partial_t n_i = S_e$ (\ref{eq:classical_ni}).
The source term can be written as $S_e=\bar\alpha j/e$,
if the electric current density $j$ is taken
as the drift current density $j = e\mu_e E n_e$ only,
hence neglecting diffusion. The relation between the change
of the electric field and the current density is given by
$\nabla \cdot\left({\bf j}+\epsilon_0\partial_t{\bf E}\right)=0$;
this equation can be derived either as the divergence of Ampere's law,
or from charge conservation~(\ref{eq:conservation})
and Gauss' law~(\ref{eq:Gauss}). If the front is weakly curved
(i.e., if the width of the space charge layer $\ell$ is much smaller
than the electrodynamic streamer radius), and if
the electric field ahead of the front is time independent,
the equation can be integrated through the boundary layer over a length of the order
$\ell$ to the one-dimensional form  $\partial_tE/\epsilon_0+j=0$.
In the resulting system of equations
\begin{eqnarray}
    \partial_t n_i &=& \bar\alpha j/e,\\
    \partial_tE&=&-\epsilon_0 j,
\end{eqnarray}
the time derivative $\partial_t$ can be eliminated, and the integration of
$\partial n_i/\partial E$ results in equation~(\ref{eq:ne}).
In~\cite{liDeviationsLocalField2007}
the approximation (\ref{eq:ne}) was derived for negative streamer fronts
without photo-ionization or background ionization, and it was tested successfully
on particle simulations of planar streamer ionization fronts
in the same paper.

When compared to simulations of positive
curved fronts in air (hence with photo-ionization)~\cite{luqueSpritesVaryingAir2010},
the approximation (\ref{eq:ne}) accounts for approximately half of the ionization density
behind the front. The likely reason is (according to a suggestion by A. Luque),
that the ionization created in the active zone ahead space charge layer
is missing in this approximation.
A further study of this question is needed.

It should be noted that equation (\ref{eq:ne}) is reminiscent of the Meek number (\ref{eq:Meek}), but note
that the integral is performed over the electric field $E$
within the ionization front, rather than over the location $s$
of this field in $\alpha(E(s))$. (We remark that in~\cite{sretenovicIsolatedHeadModel2014} the Meek number was used to estimate the ionization in a streamer, rather than an approximation like (\ref{eq:ne}).)

{\bf Electron density inside the streamer and secondary streamers. }
In electronegative gases such as air, the electron density typically
decreases in the streamer channel, as the electric field is
below the breakdown value and electrons gradually attach ---
though this tendency can be counteracted
by a detachment instability where an inhomogeneous distribution
of electric field and conductivity along the streamer channel
grows further and forms an elongated glow within the
channel~\cite{luqueSpritesVaryingAir2010, liuModelSpriteLuminous2010}.
This mechanism has been suggested as the
cause of afterglow of sprite streamers~\cite{luqueSpritesVaryingAir2010},
of space stems in negative lightning leaders~\cite{malagon-romeroSpontaneousEmergenceSpace2019},
and also of secondary streamers~\cite{Marode1975, Ono2003,Briels2008,Heesch2008}.

\subsection{The stability field or the maximal streamer length}
\label{sec:stabilityfield}

The stability field was originally defined as the
homogeneous electric field where a streamer could propagate in a stable manner~\cite{phelpsFieldenhancedPropagationCorona1971, Gallimberti1979},
i.e., without changing shape or velocity; in modern terms,
one would call this uniformly translating nonlinear object
a coherent structure.
However, nowadays the term `stability field' is used mostly in cases
where the electric field is not homogeneous, but decaying away from
some pointed electrode.
In a geometry with a high-voltage and a grounded electrode separated by a distance $d$, the stability field is the ratio $V/d$, where $V$ is the minimal voltage for streamers to cross the gap. More generally, it denotes the ratio $\Delta V/L$, where $L$ is the maximum length streamers can obtain when the potential difference between their head and tail is $\Delta V$.
Although only rough motivations for this physical concept exist,
experimentally reported values agree remarkably well with each other;
therefore the concept is widely used to
determine the maximum streamer length~\cite{Allen1991,
Allen1995,Babaeva1996,Veldhuizen2002,Qin2014,Seeger2018}
for a given applied voltage. For example, the reported value
of the stability field for positive streamers in ambient air is always
around 5\,kV/cm; and for negative ones, it is 10 to 12 kV/cm.
Further theoretical insight into how this observation is related to conductivity, charge
content and electric field distribution of the streamer will be given
in a future paper by H.~Francisco {\it et al.}

\subsection{Stepped propagation of negative streamers}
\label{sec:stepped-leaders}

Lightning observations show that positive lightning leaders propagate
continuously and negative ones in steps (see e.g. \cite{Dwyer2014} and
references therein); though on smaller scales recently a discontinuous structure
has also been seen in positive leaders~\cite{hareNeedlelikeStructuresDiscovered2019}.
Lightning leaders are based on space charge effects and field
enhancement like streamers, with the addition of heating effects
(cf.~\ref{sec:flowheat}). Why they propagate in a discontinuous manner,
is an open question in lightning physics.

\begin{figure}
  \centering
  \includegraphics[width=8.5 cm]{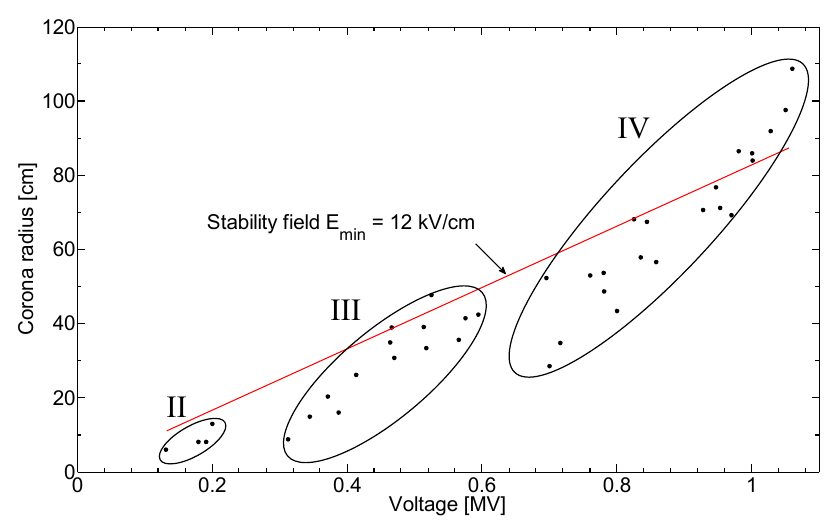}\\
  \caption{Radius of the negative corona as a function of voltage
  in a gap of 127\,cm length in ambient air,  obtained from 39 discharges.
  The voltage increases to 1~MV within 1.2 $\mu$s,
  so the voltage axis corresponds roughly to a time axis.
  The growth of this discharge in ICCD images is shown in
  the lower panel of figure~\ref{fig:kochkin-full-discharge}.
  The so-called stability field of 12\,kV\,cm$^{-1}$ is indicated
  with a red line. The second, third and fourth streamer bursts are indicated with
  II, III, IV and encircled by ellipsoids. Image from~\cite{Kochkin2014}.
  \label{fig:Pavlo-stability-field}}
\end{figure}

Experiments of Kochkin {\it et al.}~\cite{Kochkin2014} have shown
a similar asymmetry between positive or negative
streamers in a 1~m gap in ambient air exposed to a voltage of 1~MV with the so-called
lightning impulse rise time of 1.2~$\mu$s. Images of the evolution of these discharges
are included in figure~\ref{fig:kochkin-full-discharge}.
The negative streamers crossed the gap within 4
consecutive bursts, each one longer than the previous one,
see figure~\ref{fig:Pavlo-stability-field}. The growth of the streamers in each burst
stops when they have reached their maximal length $U/E_{st}$
according to the instantaneous voltage $U(t)$ and the the stability field $E_{st}$.
The final acceleration beyond the stability field line is due
to the proximity of the grounded electrode at 127~cm.

\subsection{Streamer paths}

\begin{figure}
  \centering
  \includegraphics[width=8.5 cm]{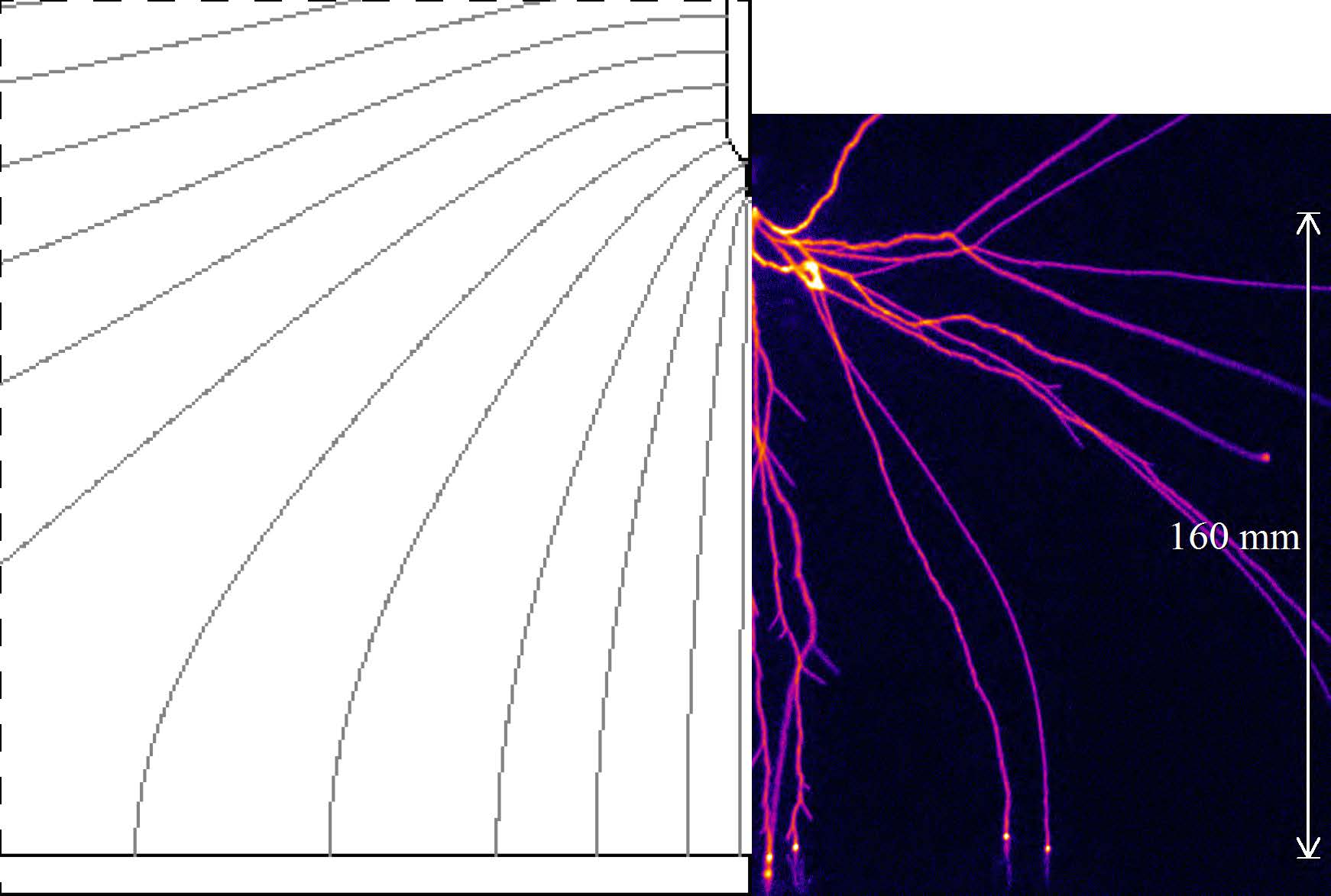}\\
  \caption{Comparison of calculated background
  electric field lines with
  streamer paths in 200\,mbar nitrogen with 0.2\,\% oxygen
  admixture in a 16\,cm point plane-gap with a +11.0\,kV pulse.
  Image adapted from~\cite{NijdamThesis}.
  \label{fig:fieldlinesstreamer}}
\end{figure}

Both positive and negative streamers generally follow electric
field lines, albeit in opposite directions. The origin of this
behaviour is simply that electrons drift opposite to the local
electric field vector and that this electron drift largely
determines the streamer propagation direction. The simple
estimation of a streamer path is therefore a field line of
the background electric field (the field without streamers or any
other free charges). This is illustrated in figure~\ref{fig:fieldlinesstreamer} where it
should be noted that
the streamer image is a 2D projection of the 3D streamer structure,
whereas the field calculation is a radial cross-section produced by an
axisymmetric model.

However, in many cases streamers deviate from these idealized paths.
The most obvious and common cause for this is the charge of
the streamers themselves.
These charges change the electric field distribution and thereby induce a
repelling effect between neighbouring streamers, which is also
visible in figure~\ref{fig:fieldlinesstreamer}, as will be discussed
in more detail below.

Furthermore, positive streamers are very sensitive for changes in
electron density in front of them. In air this hardly affects
the streamer path because of the very high electron density due
to photo-ionization, but in other (pure) gases, the free electron
distribution can almost fully determine both the general streamer paths
as well as the branching behaviour.
In such a case the background electric field plays only a minor role.
This is also illustrated by the avalanche distribution in figures~\ref{fig:streamer-propagation-basics}b-c.

\begin{figure}
  \centering
  \includegraphics[width=8.5 cm]{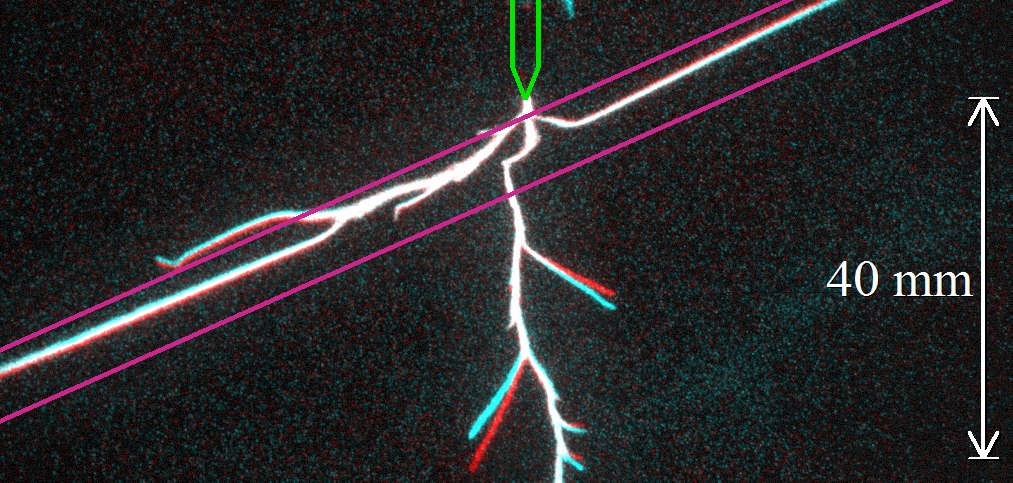}\\
  \caption{Example of streamer guiding by a laser beam. The
  green tip indicates the electrode tip, the two parallel purple
  lines enclose the laser beam position and the cyan/red/white lines
  are stereoscopic images of the propagating streamers. Image made
  in 133\,mbar pure nitrogen with a +5.9\,kV voltage pulse 1.1\,$\mu$s
  after the laser pulse.
  Image adapted from~\cite{Nijdam2014a}.
  \label{fig:laserguiding}}
\end{figure}

In \cite{Nijdam2014a,Nijdam2016} we have shown that
a mildly pre-ionized trail produced by a UV-laser can fully guide
the paths of positive streamers in nitrogen oxygen mixtures with low enough oxygen
concentrations even on a path perpendicular to the electric field.
The ionization density of the trail itself is too
low to have any impact on the global electric field distribution,
so the effect must be fully attributed to the distribution of pre-ionization.
In \cite{Nijdam2016} we compare the vertical offset
of such guided streamers
with the position of the laser beam. We were able to show that
the guiding effect can be explained by free electrons that drift
in the field during the voltage pulse before the streamer arrives.
The vertical offset cannot be explained by drift of other species like
positive or negative ions. Both the guiding by electrons
and the offset due to their drift were confirmed with numerical simulations.

Experiments with more powerful lasers give similar results~\cite{Zvorykin2011,Leonov2012},
although other effects
like gas heating and significantly increased conductivity can play
important roles. In particular, the conductivity can be so high
that it modifies the electric field already before the discharge approaches.

\begin{figure*}
  \centering
  \includegraphics[width=16 cm]{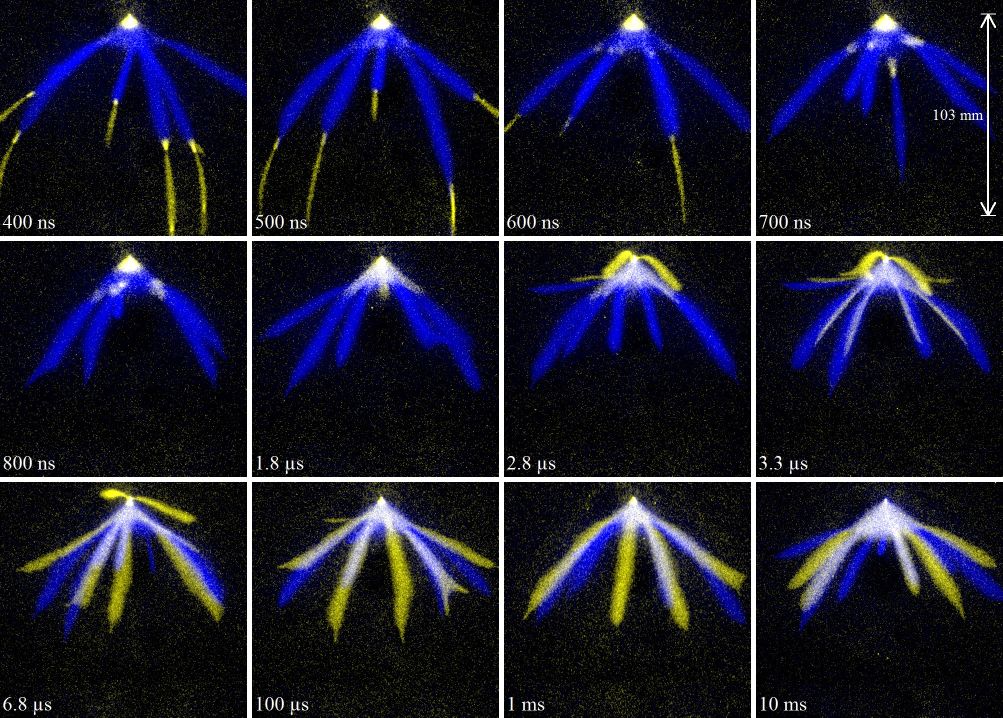}\\
  \caption{Superimposed discharge-pair images for varying pulse-to-pulse delays
   (as indicated in the images).
   Images taken in 133\,mbar artificial air with pulses of 13.6\,kV amplitude and
   200\,ns pulse length in a 103\,mm point plane gap.
   A blue color indicates intensity recorded during the first pulse,
   yellow during the second pulse and white during both pulses.
   Image from~\cite{Nijdam2014}.}
  \label{fig:doublepulse}
\end{figure*}

In~\cite{Nijdam2014} and~\cite{Li2018} we have shown that leftovers
from previous discharges can determine the path of subsequent discharges,
see figure~\ref{fig:doublepulse}.
In these so-called double pulse experiments, streamers follow the
paths of their predecessors at pulse intervals of a few microseconds (in
air) up to tens of milliseconds (in pure nitrogen) at pressures between 50
and 200\,mbar. However, here other effects like metastables or gas heating
cannot be fully excluded as explanation. Similar guiding phenomena
by preceding discharges have been found
in other experiments as well \cite{Nijdam2011}
(see also figure~\ref{fig:Onion})
and are confirmed
by recent modelling results by Babaeva and Naidis~\cite{Babaeva2018}.

A very convincing argument on the role of charged particles in the
guiding of positive streamers comes from recent experiments on pulsed
plasma jets in nitrogen~\cite{vanderSchansThesis}. In these
experiments an electric field was applied perpendicular to the streamer
(or jet) propagation direction during the period between the high
voltage pulses, so between consecutive discharges. It was found that
this electric field causes a displacement of the next discharges,
thereby indicating that the guiding of these discharges
must be due to the memory effect caused by charged particles.
However, both the direction as well
as the magnitude of the displacement are consistent with positive ions
rather than with electrons. The reason for this is not understood at present.

\subsection{Streamer interaction}\label{sec:interaction}

\begin{figure*}
  \centering
  \includegraphics[width=12cm]{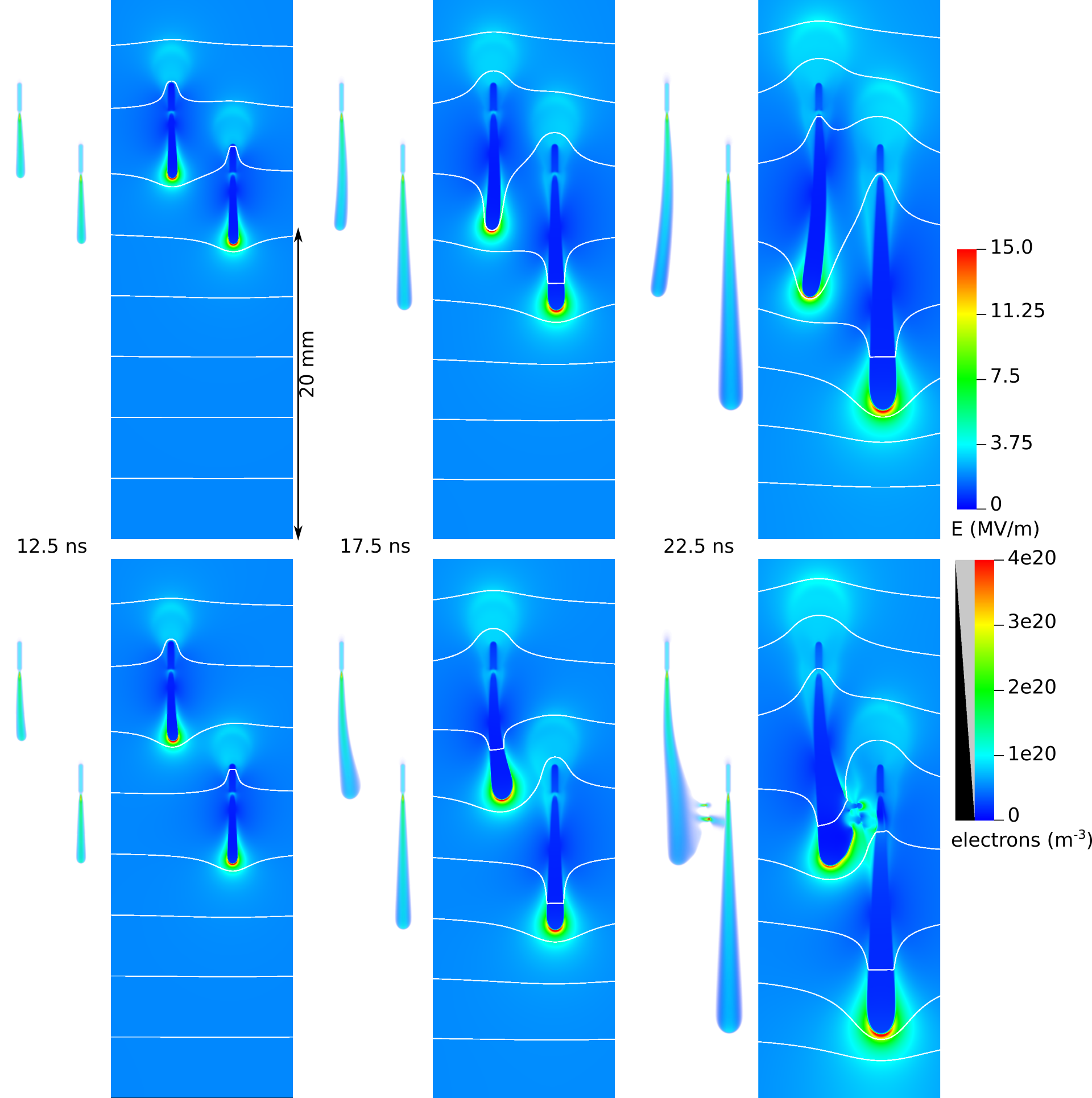}
  \caption{3D Plasma fluid simulations interacting positive streamers in
    atmospheric air. The streamers start from two ionized seeds, which have a
    vertical offset of 4 mm (top row) or 8 mm (bottom row), which leads to
    repulsion and attraction, respectively. The images show the electron density
    (volume rendering) and cross sections of the electric field with
    equipotential lines spaced by 4 kV. Picture taken
    from~\cite{Teunissen_2017}.}
  \label{fig:streamer-path-model}
\end{figure*}

As was mentioned above, an important cause for streamers to deviate
from background electric field lines is the perturbation of this field
by other streamers. Streamers carry a net-charge and thereby perturb
the electric field distribution. Because neighbouring streamers generally
have the same polarity, this effect leads to repulsion between streamers
\cite{Luque2008a,Shi2017}, as shown in the top part of figure \ref{fig:streamer-path-model}.
This also explains why streamers move away from each other after branching.
The repulsion of streamers is not always obvious from camera images,
as the 2D-projection of a branching streamer-tree can lead to apparent
cases of streamers channels connecting to each other. In~\cite{Nijdam2008}
we have shown that 3D-reconstruction of such cases usually reveals
that this is merely an artefact of the projection and no such
connection occurs.

However, under some circumstances, streamers can (re-)connect to other
streamer channels originating from the same polarity electrode.
In~\cite{Nijdam2009} we have shown that this can
happen when one streamer has crossed the discharge gap. Another streamer
can then be attracted to the channel left by the first streamer, likely
due to a change of polarity after crossing. Such behaviour is also
observed in sprites~\cite{Cummer2006,Stenbaek-Nielsen2013} although,
there, no real opposite electrode exists but there is charge polarization
along the sprite streamers. An example of this behaviour is shown
in the bottom part of figure \ref{fig:streamer-path-model}.

In~\cite{Nijdam2009} we also showed that two positive streamers
originating from neighbouring electrode tips can merge to a single
streamer when the distance between these tips is much smaller than
the width of a single streamer, in quantitative agreement with
simulations~\cite{Luque2008}.

\subsection{Streamer branching}\label{sec:branching}

\begin{figure}
\centering
\includegraphics[width=8cm]{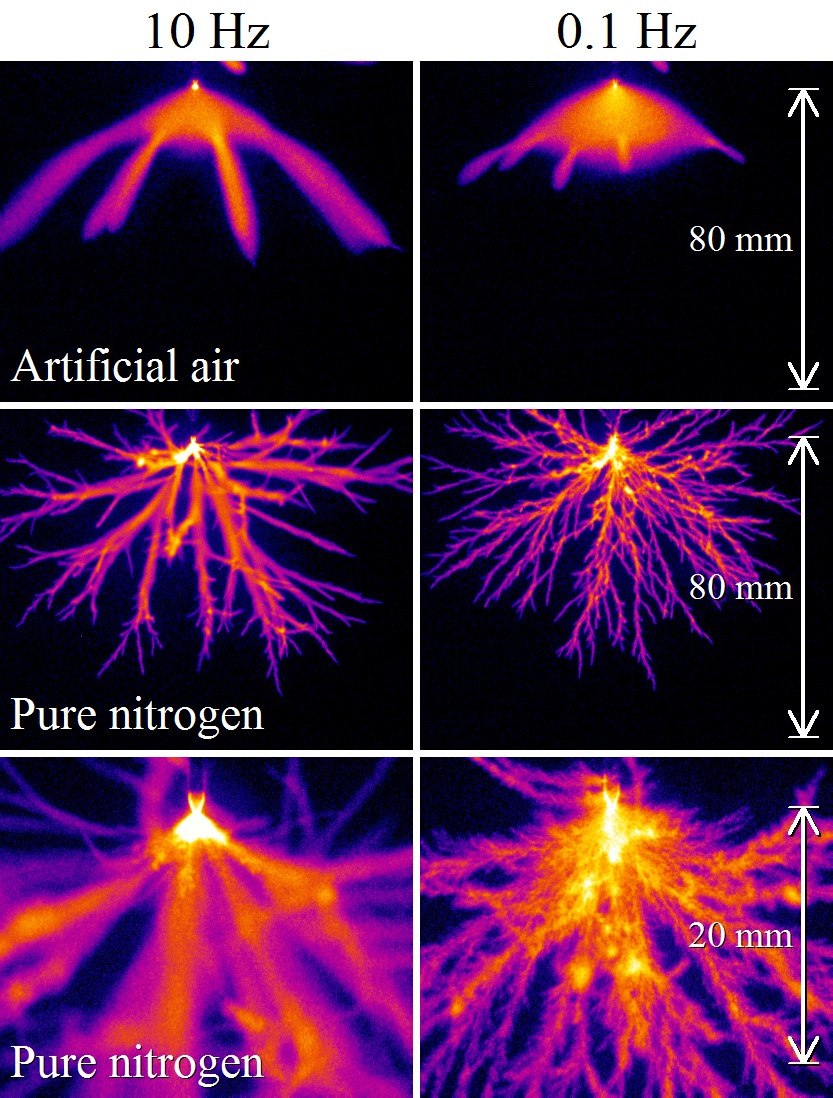}
\caption{\label{fig:Rep-freq-dep} Overview (top and middle row) and zoomed
(bottom row) images of the effects of pulse repetition rate on streamer
morphology at 200\,mbar in air
and nitrogen with 130\,ns, +25\,kV pulses in a 16\,cm
point-plane gap. Image adapted from~\cite{Nijdam2011c}.}
\end{figure}

Sufficiently long and thick streamer discharges frequently split into separate
channels, a process called branching. This can be seen for example in figures
\ref{fig:MultiScale}, \ref{fig:velocity-example}, \ref{fig:gas-pressure-dependence},
\ref{fig:kochkin-full-discharge}, \ref{fig:fieldlinesstreamer}, and
\ref{fig:Rep-freq-dep}.

On the other hand, thin streamers propagating through a spatially decaying electric field do
not branch, but rather they eventually stop propagating. As already discussed above, their diameter approaches a minimal value $d_\mathrm{min}$.

The general questions of when the streamer head is intrinsically unstable and branches, what
the diameters, velocities and directions of the daughter branches are, and when the next
branching takes place, are yet largely unanswered, and we will address them
in future papers.
Here we summarize the state of the literature.

\subsubsection{Experimental results for positive streamer in air. $\quad$}

Quantifying streamer branching is more difficult than one
might expect. The most obvious quantifiable parameters
are branching angles and branching distances, which both
seem straightforward, at least when stereoscopic techniques
are used (see section~\ref{sec:imaging}). However, in many cases
it is very difficult to exactly define a branching event for
smaller branches. There is no fundamental difference between
a small branch and a 'failed' branch. This means that the definition
of a branching event is somewhat arbitrary,
and usually done implicitly. Note that this
issue is not unique for experimental results, but is also
relevant for results of 3D streamer models
\cite{Nijdam2016, Teunissen_2017, Bagheri_2019},
which now are becoming available.
In simulations, streamer paths (like diameters) can be derived from electric
fields, species/charge densities or optical emission, whereas
in experiments generally only the latter is used.

Despite the issues sketched above, there are quite some studies
of streamer branching angles and lengths.
Briels {\it et al.}~\cite{Briels2008b} found that despite the variation of streamer diameters
by more than an order of magnitude for fixed pressure, the ratio $D/d$ of streamer length $D$
(between branching events) over streamer diameter $d$ had an average value of $11\pm4$ for air
and $9\pm3$ for nitrogen, at pressures from 0.1 to 1~bar.
In \cite{Nijdam2008}
we measured branching angles and found an average
branching angle of 43\textdegree$\pm$12\textdegree~for streamers
in a 14\,cm point-plane gap in 0.2~-~1~bar air with a +47\,kV
voltage pulse. These angles were mostly independent of position
and gas pressure. The streamer branching ratio $D/d$ was determined
as 15.
Chen {\it et al.}~\cite{Chen2018a} found similar branching angles in
nitrogen, but larger angles (53\textdegree$\pm$14\textdegree)
in artificial air. The branching ratio they
found was 13 for air and 7 for nitrogen.
They also measured the
ratio between the streamer's cross section before and after branching
$r_{\mathrm{parent}}^2/(r_{\mathrm{branch}\,1}^2+r_{\mathrm{branch}\,2}^2)$,
which was about 0.7 for all conditions.

Streamers generally branch into two new channels, although occasionally
a streamer appears to branch into three new channels
as we reported in 2013~\cite{Heijmans2013a}.
However, such events could also be interpreted as
two subsequent branching events that follow each other
too closely to be distinguished; so this is a matter of definition.
In this study the cross-section ratio
was close to 1 for both branching into two and into three branches.

Note that the above observations have been made for discharges with a modest
number of streamer channels. When the volume is densely filled with streamers,
there is too much overlap in the captured images to properly characterize branching events.

Also note that of the experimental studies mentioned above,
only the ones by Nijdam {\it et al.}\ and by Heijmans {\it et al.}\
use stereoscopic methods. The other studies measure
branching characteristics from 2D images, which can lead to
underestimation of both branching angles and branching ratios.

Good data on streamer branching is an essential ingredient
for streamer tree models like the one described in~\cite{Luque2014}
and in sections~\ref{sec:multiscale} and \ref{sec:simulations}.
These data can come from experiments like the
ones described above or from detailed 3D simulations.

\subsubsection{Theoretical understanding of streamer branching. $\quad$}

As said above, streamers with minimal diameter are not seen to branch.
Generically, the streamer head has to run into an unstable state in order to branch. The
destabilization of an unstable state can be accelerated by noise.
So one needs to identify
\begin{itemize}
    \item[1.] when the streamer head is susceptible to a branching instability, even without noise, and
    \item[2.] which type of noise or fluctuations might accelerate the destabilization.
\end{itemize}

The basic underlying instability is a {\it Laplacian
instability}~\cite{Arrayas2002,Meulenbroek2004,Ebert2011}:
when the space charge layer around a streamer head forms a local protrusion,
the local field is enhanced and the protrusion might grow.
This field enhancement is pronounced only if the thickness of the space charge layer is
smaller than the protrusion, and if the protrusion is smaller than the streamer diameter. This
implies on the other hand that a streamer head filled with space charge as depicted in panel a of
figure~\ref{fig:avalanches-schematic} is intrinsically stable until a thin space charge layer is
formed, as shown in figure~\ref{fig:streamer_sim}.

The Laplacian instability is particularly convenient to analyze for {\it negative streamers
without pronounced photo-ionization}, e.g., in high purity nitrogen. If the electron density
profile decays sufficiently steeply towards the non-ionized region,
the ionization front can be approximated as the 2D surface in 3D space
where the electron density increases steeply. Each part of this surface
propagates essentially with the local electron drift velocity.
The dynamics of such a front is
mathematically similar to viscous fingering in two fluid flow.
In this case strong analytical results can be found as reviewed in~\cite{Ebert2011}. To
summarize them briefly, it can be shown analytically that such a streamer ionization front can
destabilize even in a fully deterministic fluid model. As discussed in \cite{Luque2011}, an infinitesimal perturbation is not sufficient,
but a finite size above a threshold is required to destabilize the streamer head. If the
perturbation is too small, the perturbation is convective, i.e.,
it moves to the side of the streamer and stays behind, before it can grow
to a substantial size, so it cannot determine the
dynamical evolution of the streamer head.

{\it Positive streamers,} on the other hand, require photo-ionization or background ionization
to propagate. This means that the active zone ahead of the space charge layer (where the
electric field is above the breakdown value) is not empty of electrons (in contrast to the
negative streamer case discussed above), but it contributes substantially to the front
dynamics. This extra zone can suppress the growth of protrusions and stabilize the ionization
front by the non-local photo-ionization mechanism. However, there is no analytical stability
analysis available for this case, but only simulation results.
But branching is determined by a
Laplacian instability as well: a protrusion grows due to local field enhancement, also in a
fully deterministic fluid model for a positive streamer with
photo-ionization~\cite{Luque_2011}.

Branching can be accelerated by {\it electron density fluctuations}
in the region with low electron density ahead of a streamer.
This was shown in~\cite{Luque_2011} for positive streamers in air.
For an electron number $N_e$ in a relevant volume, the electron density fluctuations are
proportional to $\sqrt{N_e}$; and these fluctuations matter, e.g.,
in the active zone created by photo-ionization where $N_e$ is small.
The idea that the random photo-ionization events provide the noise for the branching
instability is depicted in schemes like in figure~\ref{fig:avalanches-schematic}a
that show the
photon path and the subsequent ionization avalanches.
However, the number of photo-ionization events
in air is so large, that ionization avalanches cannot be distinguished.
Rather they provide a noisy electron density profile~\cite{Bagheri_2019}, see figure \ref{fig:photoelectrons}.

\subsubsection{Streamer branching in other gases and background-ionizations. $\quad$}\label{sec:branching-other-gases}

In agreement with the discussion above, less photo-ionization would create a
more noisy electron density profile ahead of the space charge layer,
and therefore a larger probability to branch.
And in experiments it is indeed often observed that conditions with low
photo-ionization and background ionization (due to gas density or gas composition or due to
low discharge repetition frequency) exhibit more branching
and a more chaotic or zig-zaggy structure of the
streamers; and streamers with a diameter much larger than the minimal one
(see figure~\ref{fig:TanjaPM}) are not seen.
Streamers in pure gases like nitrogen
and argon branch significantly more than streamers in air
under similar conditions~\cite{Nijdam2010}.
In more extreme cases, low ionization levels can lead to
feather-like structures which may be
interpreted as separate avalanches~\cite{Wormeester2010,Wormeester2011,Nijdam2011c}.

Takahashi \textit{et al.}~\cite{Takahashi2011}
found that they could suppress streamer branching in argon significantly by
illuminating part of the discharge gap with a UV-laser,
thereby increasing the background ionization level
which confirms the theoretical discussion above.
In~\cite{Nijdam2011c} we investigated this effect in pure nitrogen
by varying the streamer pulse repetition
rate (see figure~\ref{fig:Rep-freq-dep})
and by admixing a small quantity
of radioactive $^{85}$Kr in order to increase
the background ionization level. In both cases higher background
ionization levels resulted in suppression of branching and in wider and more stably
propagating streamers.

\subsubsection{Branching due to macroscopic perturbations and peculiar events.
$\quad$}
\begin{figure*}
  \centering
  \includegraphics[width=15cm]{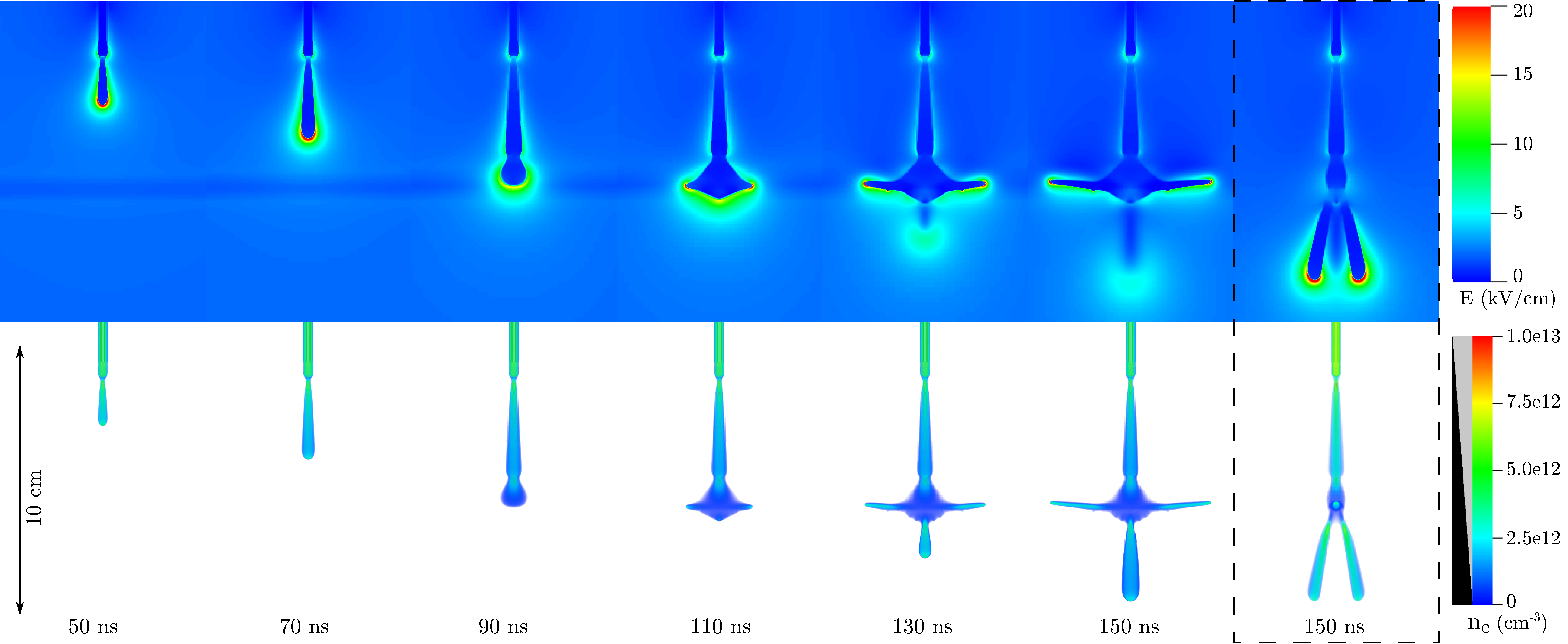}
  \caption{Simulated time evolution of a streamer discharge propagating in
    pure nitrogen and interacting with a
    $10^9 \, \textrm{cm}^{-3}$ preionized trail. Top row: cross sections
    of the electric field, bottom row: volume rendering of the electron density.
    For the rightmost figures, the viewpoint has been rotated by 90$^\circ$,
    revealing that the downwards streamer has branched. Image from~\cite{Nijdam2016}.}
  \label{fig:guiding-time-evolution}
\end{figure*}

Above we have discussed microscopic intrinsic fluctuations that can accelerate a streamer branching instability,
mainly due to low electron densities in the active growth zone
ahead of the streamer tip.
But external macroscopic perturbations can cause streamer branching as well.
An early example is that bubbles in liquids (or in high pressure air)
can influence streamer path and branching, when they are of similar size as
the streamer diameter~\cite{babaevaStreamerBranchingRole2008,
babaevaEffectInhomogeneitiesStreamer2009}.
A hydrodynamic shock front where the gas density is changing suddenly,
can have a similar  effect~\cite{starikovskiyGasdynamicDiodeStreamer2019}.
That localized regions with higher pre-ionization can change the discharge dynamics,
was already discussed above; here the streamer can not only be guided
by laser induced pre-ionization, but it also shows particular branching structures when entering or
leaving a pre-ionised region~\cite{Nijdam2014, Nijdam2016}, see also
figure~\ref{fig:guiding-time-evolution}.

A peculiar branching structure was found by Heijmans \textit{et al.} \cite{Heijmans2015}.
In these point-plane geometry experiments in pure nitrogen,
thick streamers suddenly split into many thin streamers for certain
pulse repetition rates above 1\,Hz.
This occurs, for fixed settings, always at the same distance
from the point electrode. An explanation for this behaviour
was not found, but it suggests that streamer branching in pure
nitrogen could depend on some threshold value of the background
ionization.

\subsection{Interaction with dielectric surfaces}

When a streamer encounters a dielectric surface several types of processes can occur, see section \ref{sec:sim-streamer-dielectric}.
A streamer can deposit charge on a dielectric surface,
thereby affecting the local electric field and suppressing
subsequent discharges or spark formation. This is the working
principle of dielectric barrier discharges
(DBD's) and allows operation of these atmospheric non-thermal discharges
driven by alternating current voltages.
The physics and applications of DBD's were recently extensively reviewed
by Ronny Brandenburg~\cite{Brandenburg2017} and are
outside the scope of this work.

When the field lines are not (nearly) perpendicular to
the dielectric surface, the surface can influence the path of the
streamer. We have observed that streamers can follow dielectric
surfaces even when these are far from parallel to the background field
lines~\cite{Trienekens2014,Dubinova2016}, see also
figure~\ref{fig:strobotrienekens}.
The reason for this attraction is likely
photo-emission from the surface (which requires less energy than
photo-ionization) and field enhancement due to the dielectric itself.
However, the presence of a
dielectric can also repel the discharge by shielding photo-ionization
and avalanches~\cite{Dubinova2016}. In air the effects of photo-emission
will be less prominent than in pure gasses because photo-ionization
makes the streamers insensitive for the electron distribution as
was also seen in the laser-guiding experiments described above.


\section{Streamers in different media and pressures}\label{sec:diff_media}
\subsection{Streamers in different gases}\label{sec:othergasses}

Streamer discharges in gases different from air propagate due to the same mechanisms as
described above, but can have a quite different appearance, see Fig.~\ref{fig:gas-pressure-dependence}. This is mainly due to four gas
properties: photo-ionization, electron attachment, mechanisms of electron energy loss
and visibility.

{\bf Photo-ionization}
\label{sec:photoionization} in air occurs, when an energetic electron in the ionization front
excites a nitrogen molecule to the $\textrm{b}^{1}\Pi$, $\textrm{b}'^1\Sigma^+_u$,
$\textrm{c}^{1}\Pi_u$ or $\textrm{c}'^1\Sigma^+_u$
state~\cite{Zhelezniak1982,Pancheshnyi2015,Stephens_2016,stephensPracticalConsiderationsModeling2018}.
The molecule can then emit a photon with a wavelength in the range
of 98 -- 102.5\,nm
that can ionize an oxygen molecule at some (isotropically distributed) distance. This
distance scales with the inverse oxygen concentration. When the ratio between nitrogen and
oxygen is changed, the availability of free electrons ahead of the ionization front is
changed. In particular, for very low oxygen concentrations, very few electrons are available
ahead of the impact ionization front, and positive streamers attain a characteristically
ragged and zigzagged narrow shape. An
example of this can be seen in figure~\ref{fig:Rep-freq-dep}.

{\bf Electron attachment} occurs in electro-negative gases such as air (due to the presence of
oxygen). In an electro-negative gas, there is a clearly defined break-down value of the
electric field, namely when the growth of electron density due to impact ionization exceeds
the loss of free electrons due to attachment to electro-negative molecules. Without an
attachment reaction, the electron density can slowly grow also in weaker fields, though gas
impurities can never be completely avoided in any experiment \cite{Nijdam2010}.

{\bf Electron energy losses} depend on the gas composition. Noble gases like He or Ar have
no rotational or vibrational excitations and only few electronically excited states. As a
consequence, there are not many energy loss mechanisms for electrons with energy below the
ionization energy of the gas, and electrons are easily accelerated in a given electric
field. On the contrary, in gases consisting of complex poly-atomic molecules like H$_2$O, CO$_2$ or
SF$_6$, there are many inelastic scattering modes for the electrons, and therefore the
electric field has to be higher for the electrons to reach a similar energy.

{\bf Visibility} of the discharge does not influence the physical
processes, but can have a major impact on the diagnostics of the discharge.
Streamers in nitrogen-oxygen mixtures with up to 20\,\%~oxygen and in argon
are generally bright and easy to image. Streamers in
CO$_2$, hydrogen, helium or (especially) oxygen and mixtures of these,
on the other hand,
emit very little radiation in the visible part of the spectrum and
are therefore hard to see by naked eye or to image by (ICCD) camera
\cite{NijdamThesis}.

\subsection{Scaling with gas number density and its range of validity}
\label{sec:scalinglaws}

This subsection contains a short version of the arguments elaborated in the review~\cite{Ebert2011}.

{\bf Townsend scaling.} More than a century ago, Townsend understood that electric
discharges at different gas number densities $N$ can be physically similar. This is the case, if
the dynamics is dominated by electron acceleration in electric fields together with
electron-molecule collisions.
If the product $\ell_{\rm MFP}\cdot E$ of the electron mean free path $\ell_{\rm MFP}$ with
the electric field $E$ stays the same, the electrons gain the same kinetic energy
between collisions, and the discharge evolution is physically similar. As the mean free
path is proportional to the inverse of the gas number density $N$ ($\ell_{\rm MFP}\propto 1/N$), the electric field has to be scaled as $E\propto N$
to show the same physical effects; hence discharges at different gas number density $N$
with the same reduced electric field $E/N$ are physically similar.
(The Townsend unit $1 \, \mathrm{Td} = 10^{-21}$Vm$^2$ has been introduced for $E/N$.)
The mean free path sets the scale for other length scales in the discharge,
therefore they scale with $1/N$ as well.
Characteristic electron velocities are set by the
balance of the kinetic electron energy and the ionization energy of the molecule, hence
they do not vary with $N$. Finally, because velocities don't depend on $N$,
characteristic time scales have to scale in the same manner as the
length scales, hence with $1/N$.

{\bf Nonlinear scaling in streamers.} Streamer discharges are formed by strongly nonlinear
ionization fronts, and the balance between ionization growth and space charge effects leads
to a specific scaling law for streamers according to Gauss' law
(\ref{eq:Gauss}): the charge density
integrated over the width of the ionization front has to screen the electric field ahead of the front. As
fields scale like $N$ and lengths scale like $1/N$, densities of charged particles scale as
$N^2$ \cite{Pasko1998,Rocco2002,Ebert2006a,Pasko2006,Ebert2010}.
Therefore
the degree of ionization $n_e/N$ (where $n_e$ is the electron density) scales like the gas number density $N$,
whereas the total number of electrons in a similar section of a streamer scales like the
total electron density times the relevant volume $N^2/N^3 = 1/N$.

{\bf Limitations of the scaling laws.}
The range of validity of the scaling laws is limited by a number of effects: size of
fluctuations, direct electron electron interactions, three-body reactions and quenching.

\begin{itemize}
    \item {\bf Size of stochastic fluctuations.}
As shown above, the total number of free electrons (and other charge carriers) involved in a physically similar discharge scales with inverse
gas number density. For higher gas number densities, fewer free electrons are
present in a similar discharge, so that stochastic
fluctuations due to the discreteness of electrons increase
and (continuum-based) scaling laws start to lose their validity.
The increased fluctuations can also accelerate streamer branching.

    \item {\bf Electron energy distribution far from thermal equilibrium.}
The degree of ionization $n_e/N$ in a streamer increases linearly with
gas number density $N$. In streamer
modeling, one typically assumes that the free electrons only collide with molecules, and that they
interact with each other only through the collectively generated electric field. Electrons then can gain an energy distribution far from equilibrium, with relatively high
energies. Due to their low mass, they rapidly gain energy from the field, and because the particles they collide with are much heavier, they do not easily lose kinetic energy in elastic collisions.
This extreme electron energy distribution is the key to many applications of streamer
discharges for plasma-chemical processing. In contrast, if the degree of ionization is increased,
electrons can directly scatter on each other and get closer to
thermal equilibrium. This is the case at higher gas densities, and it can
lead to deviations from the scaling laws.

\item {\bf Three-body interactions and quenching.}
Finally, at low gas number density, two-body interactions of charged particles and neutrals
dominate the discharge process. At higher densities, three-body interactions can become
important. Higher gas densities support, e.g., three-body attachment of electrons to oxygen and other three-body plasma
chemical processes, or they suppress the photon emission from excited
states through collisional quenching. Again, this can lead to corrections to the
scaling laws at higher gas densities.
\end{itemize}

\subsection{Discharges in liquid and solids.}

Streamer discharges start to deviate from scaling laws at approximately STP in
air, according to the mechanisms discussed above.
When the medium density increases by
about three orders of magnitude to solid or liquid densities, two additional mechanisms
play a dominant role, namely Ohmic heating and field ionization.

{\bf Ohmic heating.} We recalled above that the degree of ionization $n_e/N$ in a similar streamer
discharge increases linearly with gas number density $N$; therefore the Ohmic heating of the gas by the
electrons $n_e$ becomes more important with growing $N$. At
normal density, one distinguishes between the early stage of a space-charge driven streamer
discharge and a later stage of a heated leader discharge. At solid or liquid density these
stages might overlap much more.

{\bf Field ionization.} With increasing medium density $N$, the mean free path of the electrons
decreases as $1/N$, hence the required electric field for impact ionization increases linearly in $N$.
Eventually, the electric field required to ionize molecules or atoms directly by electric forces can be lower than the field required for impact ionization.
This point was made by Zener in his 1934 paper~\cite{zenerTheoryElectricalBreakdown1934}
for electric breakdown in solids, and field ionization is now known as the Zener mechanism in solid state physics.
Jadidian {\it et al.}~\cite{jadidianEffectsImpulseVoltage2012} used  field
ionization rather than impact ionization in their models of streamers in transformer oil. In such models, streamers still grow due to
local field enhancement at the steamer tip, and they look quite similar to gas streamers.
However, the ionization rate does not depend on the local electron density, but only on the local
electric field.


\section{Other topics}\label{sec:additional}
\subsection{Plasma theory and electrostatic approximation}
\label{sec:plasma-theory}

There exist many types of plasmas, which differ in e.g.~their electron number
density, their degree of ionization and in the temperatures (or energies) of
the plasma species. Compared to most other plasmas, streamer discharges (and
more generally cold atmospheric pressure plasmas) have a low degree of
ionization, a high neutral density and relatively energetic electrons. The
electron-neutral collision frequency $\nu_c$ in streamer discharges is
therefore high; for atmospheric air, it lies in the range $10^{12}$ to
$10^{13} \, \textrm{Hz}$. Due to this high collision frequency, not all
conventional plasma theory is directly applicable to streamer discharges. Another important
difference is that the plasma created by streamer discharges is far from
equilibrium, in particular near their heads.

\textbf{Plasma oscillations. } In most plasmas, there are high-frequency
electron density fluctuations described by the plasma frequency, which in
simple cases is given by $\omega_\mathrm{pe} = \sqrt{n_e e^2 / (m_e
\varepsilon_0)}$. The underlying mechanism is that a fluctuation in the
electron density gives rise to an electric field, which acts as a restoring
force. However, plasma oscillations are not relevant for streamer discharges,
as we typically have $\omega_\mathrm{pe} < \nu_c$. This means oscillations
are almost immediately damped by electron-neutral collisions.

\textbf{Debye length. } If the potential inside an equilibrium collisionless
plasma is locally perturbed, a characteristic length scale for electric
screening is the Debye length $\lambda_D = v_\mathrm{th}/\omega_\mathrm{pe}$,
where $v_\mathrm{th} = \sqrt{k_B T_e/m_e}$ is the thermal velocity of
electrons. This length scale is determined by the competition between thermal
motion and electrostatic forces. It is hard to define $\lambda_D$ for
developing streamer discharges, since electric fields and collisions lead to
a strongly non-thermal electron velocity distribution. If we consider a
stationary streamer channel (for example, after the voltage has been turned
off) in which electrons have been thermalized, then $\lambda_D$ is typically
smaller than all other length scales of interest. For example, an electron
density of $10^{20}/{\rm m}^3$ and a thermal energy corresponding to room
temperature give a Debye length of about 70~nm.

\textbf{Ionization length $1/\bar\alpha$. }
Relevant length scales
in a streamer discharge are the distances between neutrals and between
charged particles (see section~\ref{sec:Conductivity}) and the mean free path
of electrons between collisions with neutrals.
But the most relevant length
scale that is characteristic for the nonlinear streamer dynamics, is the
ionization length $1/\bar\alpha(E)$; it depends on the local electric field. The width
of the space charge layer (see section~\ref{sec:prop-theory}), the electron density
(\ref{eq:ne}) behind the front and the Meek criterion for
the avalanche length until a streamer is formed (see section~\ref{sec:inception})
are all functions of $\bar\alpha(E)$.
The region ahead of the space charge layer where $\bar\alpha>0$
is the active zone; here the electron density grows on the spatial scale
$1/\bar\alpha(E)$.
Therefore it is essential to resolve the local length scale $1/\bar\alpha(E)$
in numerical simulations (see section~\ref{sec:numerical-methods}).



\textbf{Electron gyrofrequency. } In the absence of collisions, electrons
gyrate around magnetic field lines with a frequency $\omega_\mathrm{ce} = e
B/m_e$, where $B$ is the magnetic field strength and $c$ the speed of light.
Collisions disturb the electron gyration, and the ratio
$\omega_\mathrm{ce}/\nu_c$ indicates the \emph{magnetization} of the plasma.
For a streamer discharge with $\nu_c \sim 10^{12} \, \mathrm{Hz}$, a magnetic
field of more than $5 \, \textrm{T}$ is required to have a ratio
$\omega_\mathrm{ce}/\nu_c \sim 1$. The effect of magnetic fields is therefore
usually negligible, except under the conditions of a high magnetic field 
lab~\cite{mandersPropagationStreamerDischarge2008}.

Since $\nu_c$ scales with the neutral gas number density, so does the magnetic field
required to magnetize a plasma. The effect of the geomagnetic field on
streamer-like discharges at different altitudes in the atmosphere (hence at
different air densities) is elaborated
in~\cite{Ebert2010,kohnCalculationBeamsPositrons2015}. 

\textbf{Induced magnetic field. } 
The strength of the magnetic field along a circular line with radius $r$ around an enclosed current $I$ is $B(r)=\mu_0 I/(2\pi r)$.
Therefore the magnetic field $B(r)$ increases approximately linearly with $r$ inside the streamer and decreases like $1/r$
outside; hence it is maximal precisely on the streamer radius
and at the streamer head.
According to section~\ref{sec:currents}, streamer currents of
up to 25~A have been measured in ambient air. This yields a magnetic field of $1.7\cdot 10^{-3}$~T~(3~mm/$r$) outside the streamer, hence if the streamer has 3~mm radius,
the maximal magnetic field is about $1.7\cdot 10^{-3}$~T. According to the
estimate on the streamer magnetization $\omega_\mathrm{ce}/\nu_c$ above, this
field has no influence on the electron motion in the streamer.
Finally, we remark that using the estimate for $I_\mathrm{max}$ from section \ref{sec:currents-theory} gives the following estimate for the streamer's maximal magnetic field: $B_\mathrm{max} \approx v \, E_\mathrm{max} / c^2$,
where $v$ is the streamer's velocity, $E_\mathrm{max}$ its maximal electric field and $c$ the speed of light.


\textbf{Electrostatic
approximation. } In general, electric fields can have two components, one
determined by equation (\ref{eq:Gauss}) and one by
\begin{equation}
  \label{eq:maxwell-faraday}
  \nabla \times \vec{E} = -\partial_t \vec{B}.
\end{equation}
In the electrostatic approximation, only equation (\ref{eq:Gauss}) is taken
into account. The electric field can then be computed as $\vec{E} = -\nabla
\phi$, where the electric potential $\phi$ is obtained by solving equation
(\ref{eq:poisson-phi}).

To show the validity of the electrostatic approximation, we estimate the
magnitude of the right-hand sides of equations (\ref{eq:Gauss}) and
(\ref{eq:maxwell-faraday}). The charge density $\rho$ at the
streamer head is typically in the range $0.1 e n_i$ -- $0.3 e n_i$, where $e$ is the elementary charge and $n_i$ the ionization density in the streamer head. The numeric factor takes into account that the degree of ionization is still increasing in the charge layer, and that there is partial charge neutrality.
Using the densities of section~\ref{sec:Conductivity},
the value of $|\rho/\varepsilon_0|$
(which is the right hand side of (\ref{eq:Gauss}))
is in the range of $10^{10}$ to $10^{13}$~V/m$^2$
in atmospheric air.
The right hand side of (\ref{eq:maxwell-faraday}) is
$|-\partial_t \vec{B}|$. By multiplying the maximal magnetic field
at the streamer head with the streamer velocity over the
streamer radius ($v/R$), a rough estimate for $|-\partial_t \vec{B}|$ is
obtained. For the fast and wide streamer already considered above, with maximal magnetic field $B=1.7\cdot 10^{-3}$~T, radius $R=3$~mm, and velocity $v=10^6$~m/s,
the maximal $|-\partial_t \vec{B}|$ is approximately $1.7\cdot10^6$~V/m$^2$. From these
estimates, it follows that the contribution of equation
(\ref{eq:maxwell-faraday}) to the electric field is much smaller than that of
equation (\ref{eq:Gauss}), so that the electrostatic approximation is valid.

\subsection{Basic streamer plasma chemistry}
\label{sec:chemistry}
In plasma applications, chemical activity is usually the main purpose
of using streamer-like discharges. Below, we briefly describe some of the key reactions occurring during and after
streamer propagation in dry air.
In other gases or in wet air, many variations
on these reactions are possible, although the general mechanisms are
always the same.

Generally, higher applied voltages lead to thicker streamers,
which carry more current
(see also section~\ref{sec:VeloRadField}),
but these are also chemically more
active, as was shown by van Heesch \textit{et al.}.~\cite{Heesch2008}.

The two essential reactions for a propagating streamer in air
(or any other nitrogen-oxygen mixture) are the electron impact
ionization reactions:
\begin{eqnarray}
\mathrm{O_{2}+e} & \rightarrow & \mathrm{O_{2}^{+}+2e;}\label{eq:Oxygen-impact}\\
\mathrm{N_{2}+e} & \rightarrow & \mathrm{N_{2}^{+}+2e}.\label{eq:Nitrogen-impact}
\end{eqnarray}
These reactions create the electrons and positive ions that separate
in an external field and cause the field enhancement at the streamer tip
that in turn enables streamer propagation.

Free electrons created in this way, however, do not remain available forever, but are
lost by a multitude of reactions. One of the most notable reactions,
especially in air, is attachment to oxygen, either by two body attachment
\cite{Kossyi1992,Morrow1997,Liu2004,Dujko2011}
\begin{eqnarray}
{\rm e+O_{2}\to O+O^{-},}\label{eq:Two-body-attachment}
\end{eqnarray}
predominantly at lower air density or higher electron energy,
or by three-body attachment
\begin{eqnarray}
{\rm e+O_{2}+M\to O_{2}^{-}+M}\label{eq:Three-body-attachmetn}
\end{eqnarray}
where M is an arbitrary other molecule. Three-body attachment is
more important at higher air densities, and does not require the dissociation energy for the ${\rm O_2}$ molecule.

Alternatively, electrons can be lost by recombination with positive ions.
In air, the most likely positive ion to recombine with is O$_{4}^{+}$ because
N$_{2}^{+}$ and O$_{2}^{+}$ are quickly converted according to the following
scheme~\cite{Aleksandrov1999}:
\begin{eqnarray}
\mathrm{N_{2}^{+}\dashrightarrow N_{4}^{+}\dashrightarrow O_{2}^{+}\dashrightarrow O_{4}^{+}}.\label{eq:Positive-ions-evolution}
\end{eqnarray}
In pure nitrogen, this scheme stops at N$_{4}^{+}$, while
in mixtures with low oxygen concentrations it stops at O$_{2}^{+}$, as was
shown by us in~\cite{Nijdam2014}.

Other reactions will lead to the formation of radical species like O,
N, NO and O$_3$. In wet air, these are accompanied or replaced by OH,
We have described the formation processes of these
species in~\cite{Nijdam2012} while a more elaborate overview
can be found in~\cite{Kim2004}.

\subsection{Interaction with gas flow and heat}
\label{sec:flowheat}

Streamers can both cause gas heating and/or flow, but they are also
affected by both phenomena which can lead to complex interactions.
Below we will start with the interaction of gas heat with streamers,
followed by the interaction with gas flow.

\subsubsection{Streamers in hot gases. $\quad$}
In recent years, three groups have investigated the effects of elevated
gas temperature
on streamer discharges in air~\cite{Huiskamp2013,Ono2016a,Ono2018,Komuro2018}.
Huiskamp \textit{et al.}~\cite{Huiskamp2013} studied the effect of temperatures up
to 773\,K on positive streamers at constant gas density; they changed
temperature and pressure simultaneously
in order to keep the gas density fixed. They found that the dissipated
plasma energy as well as the propagation velocity increase with temperature
which suggest the existence of a specific temperature effect.
They suggest that this may be due to a higher streamer conductivity at higher
temperature.

A similar experiment was performed more recently by Ono and Ishikawa~\cite{Ono2018} but for
a much larger temperature range, up to 1438\,K. They confirmed the trends
observed by Huiskamp \textit{et al.} and noted that at 1438\,K the pulse energy was
approximately 30 times larger than at room temperature. They also showed
that temperature affects the shape of the discharge for temperatures
exceeding 900\,K where streamers near the anode became thinner.

Komuro \textit{et al.}~\cite{Komuro2018} show through their models and simulations that
temperature-dependent changes in the recombination and attachment rates can explain
most of the effects of elevated gas temperatures on streamer properties.

\subsubsection{Gas heating by streamers and the transition to leaders. $\quad$}
\label{sec:gas-heating}
Electrons and ions moving a distance $\vec{d}$ in an electric field $\vec{E}$
gain the energy $q \vec{E} \cdot \vec{d}$ where $q$ is the particle charge.
As the kinetic energy of the particles on average
does not change along the path due to the balance of field acceleration
and energy losses in inelastic collisions, their energy $q \vec{E} \cdot \vec{d}$
is deposited in the gas, in the form of translational, rotational,
vibrational and electronic excitations of the molecules,
and of ionization, dissociation etc.
According to the argument above, the energy density deposited per time
is $\vec{j} \cdot \vec{E}$ for an electric current density $\vec{j}$
due to the drift of electrons and ions.
Initially the energy distribution over these degrees of freedom is
very far from equilibrium and one cannot define a temperature;
this is just the mode of operation of non-equilibrium pulsed discharges
to deliver energy to particular excitations for plasma-chemical
applications. But eventually, at least a fraction of the energy is
available in the translational modes of the molecules and ions.
The energy in these modes determines the pressure of the gas,
and an increased pressure within a discharge channel will drive
a gas expansion wave into the surrounding colder gas.
When the gas density has decreased in the hot channel,
the discharge is called a leader in lightning physics and
high voltage technology. The reduced gas density $N$ within a leader
channel leads to a higher reduced electric field $E/N$,
and hence to an easier maintenance of the discharge than in the surrounding
colder gas.

The energy transfer between electrons and the gas can be greatly
accelerated by a few processes collectively called fast heating~\cite{Aleksandrov_2010,Popov_2011}.
Dominant reactions are electron impact dissociation of O$_2$ and quenching
of electronically excited N$_2$ by O$_2$ and of excited O atoms by nitrogen.
For high fields (over 400 Td), electron impact dissociation of N$_2$ become
dominant.

\subsubsection{Gas flow induced by streamers and the corona wind. $\quad$}
It is well-know that corona-discharges can cause air flow. This
phenomenon is commonly known as corona or ion wind and was first
reported in 1709~\cite{Robinson1962}. In many cases, (part of) the
air flow is induced by gas heating, but often the main process is
directed momentum exchange between charged particles (electrons and
ions) and the neutral background gas. In particular, the momentum change
$e(n_+ - n_- -n_e)\vec{E}$ of the charged particles in the electric field
is transferred to the gas molecules. Including also the diffusion of the charged
particles, the force $\vec{F}$ on the gas is~\cite{Boeuf2005}:
\begin{equation}\label{eq:IonWind}
  \vec{F} = e(n_+ - n_- -n_e)\vec{E}-k(T_g\nabla n_+ + T_g\nabla n_- + T_e\nabla n_e),
\end{equation}
where $n_{+,-,e}$ is the density of positive ions, negative ions or electrons,
$e$ the elementary charge, $\vec{E}$ the electric field vector, $k$ the Boltzmann
constant and $T_{g,e}$ the gas or electron temperature.
In most cases this equation is dominated by its first term, although
many authors (mistakenly) neglect the contribution from
electrons~\cite{Moreau2007,Rickard2006},
attributing
this to their small mass, even though equation (\ref{eq:IonWind})
does not contain the particle masses.

Actually, equation~(\ref{eq:IonWind}) nicely shows that in a quasi-neutral plasma
without any large gradients, the body-force is negligible. Therefore,
ion wind can only be generated by a streamer discharge zone at the
streamer tips (where large gradients as well as space charges occur)
or outside the streamer
area in a so-called ion-drift zone due to the net charge there.

In \cite{Chen2017} we have shown in numerical simulations that drift
of negative ions is dominant in the production of corona wind from
a negative DC discharge in air in a pin-ring geometry. Electrons are
quickly attached to oxygen molecules and therefore play only a minor
role in the drift region. The calculated Trichel pulse frequency
and amplitude and flow
patterns match our experimental results remarkably well.

In \cite{Chen2018} we have improved these simulations by self-consistently
adding the effects of gas heating on both the flow and the discharge.
Gas heating can have a detrimental effect in applications where corona
wind is used for cooling purposes. In this work we found that for the same
geometry as used in~\cite{Chen2017}, at voltages above 10--15\,kV, positive
DC coronas provide more gas flow than negative DC coronas. In both
cases, gas heating plays an important role, which was confirmed
experimentally by Schlieren photography.

\begin{figure}
  \centering
  \includegraphics[width=8.5 cm]{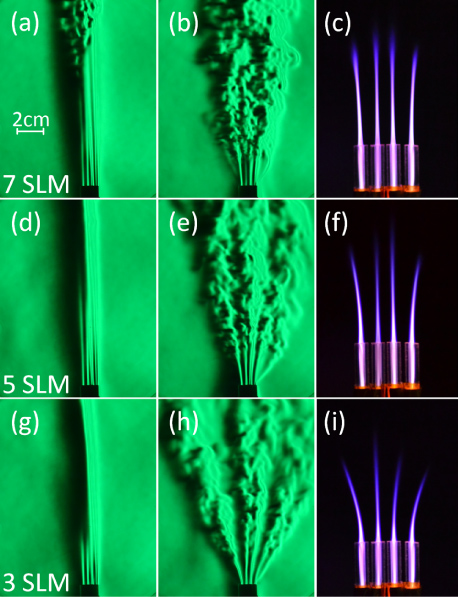}\\
  \caption{Image from~\cite{Ghasemi2013} showing a helium jet array operating
  at gas flow rates of 7, 5 or 3\,SLM in the upper (a-c), middle (d-f)
  and lower (g-i) row. The left column (a, d, g) shows Schlieren images
  of the gas flow
  without applied electric field, and the middle column (b, e, h)
  with the plasma voltage switched on.
  The right column (c, f, i) shows camera images of the plasma
  plume trajectory for the case when the voltage is on.
  }
  \label{fig:walshmultiplejets}
\end{figure}

Flow also plays a crucial role in plasma jets, which will be discussed in
more detail in section~\ref{sec:jets}. One striking example of the
feedback between streamer-like discharges and flow is given by
Ghasemi \textit{et al.}~\cite{Ghasemi2013}. They image an array of four plasma
jets using Schlieren imaging and direct photography, see
figure~\ref{fig:walshmultiplejets}. Here, it can be seen that 
the plasma accelerates the onset of turbulence, but it also 
leads to a repulsion between the
four flow paths; the interaction
between discharge and flow is indeed very complex.
\subsection{High-energy phenomena} \label{sec:fastelectrons}

\subsubsection{Electron runaway. }
The reaction-drift-diffusion model for the electron density as discussed up
to now in this paper, is based on the assumption that all electrons undergo a
drift motion with a similar velocity and energy in a given electric field.
However, a second ensemble of electrons with relativistic energies
(i.e., with energies larger than $511~{\rm keV= m_e c^2}$
where ${\rm m_e}$ is the rest mass of the electron) can exist
even in a field below the classical breakdown value (which is
$\approx$~3\,MV/m in STP air). Electrons that initially have energies
in the eV range, can reach these high energies, if the electric field
is above the so-called runaway threshold of
about 26\,MV/m in STP air, and if the field is sufficiently extended in
space and time.
The classical argument~\cite{wilsonElectricFieldThundercloud1924,
Gurevich1960} for electron runaway is based on a friction curve: the energy
losses due to inelastic and ionizing collisions are maximal for an electron
energy of about 200\,eV in air. If an electron can reach this energy
in a given field, it can accelerate further and ''run away'',
because the friction is then decreasing with energy.
The electron will continue to
accelerate up to energies where friction again becomes large, mainly due to
Bremsstrahlung radiation, or when it reaches conditions with lower
electric fields.

We remark that this intuitive friction picture has two shortcomings:
first, the friction is not deterministic, but due to stochastic collisions,
which allows the electrons to "tunnel" to high energy states even
if the electric field in a deterministic interpretation would be too low.
And second, the dynamics does not concern a fixed number of electrons,
but in electric fields
above the classical breakdown value, there is also a continuously growing
reservoir of low energy electrons to run away. Both aspects and their consequences
for the electron energy distribution are elaborated in~\cite{dinizColdElectronRun2018}.

\subsubsection{X- and $\gamma$-rays, anisotropy and discharge polarity. }
When electrons run away to high energies, Bremsstrahlung becomes an important
process in electron-molecule collisions; it converts fractions of the
electron energy into photon energy. This is how pulsed discharges in nature
and technology can generate X-rays and gamma-rays. Runaway electrons
propagate predominantly in the direction against the electric field, and
bremsstrahlung photons created by relativistic electrons follow essentially
the electron beam. On the other hand, for electron energies below 100~keV,
the emission of Bremsstrahlung photons in different directions varies
typically by not more than an order of
magnitude~\cite{kohnAngularDistributionBremsstrahlung2014}. Both X-rays and
gamma-rays can travel much larger distances without scattering or energy
losses than runaway electrons.

Runaway electrons in a high electric field can leave a trace of lower energy
electrons behind, and therefore they can determine the shape of pulsed
negative discharges~\cite{kohnCalculationBeamsPositrons2015}. The same
applies to the directed motion of $\gamma$-rays, although they interact much
less with matter than relativistic electrons. X-rays, on the other hand, are
emitted somewhat more isotropically. For this reason, it has been suggested
that they could replace photoionization in positive
discharges~\cite{tarasenkoRunawayElectronsDiffuse2019}. However, in recent
simulations~\cite{kohnInfluenceBremsstrahlungElectric2016a} bremsstrahlung
photons cannot support the typical propagation of a positive streamer
in high purity
nitrogen, and they are not relevant in air since photo-ionization is dominating.

\subsubsection{High energy phenomena in pulsed discharges.}
The existence of runaway electrons and consecutive X-rays and $\gamma$-rays
is now well established in thunderstorm
physics~\cite{dwyerHighEnergyAtmosphericPhysics2012},
and they are a topic of much current research in the geosciences.
They appear in terrestrial gamma-ray
flashes observed from space~\cite{fishmanDiscoveryIntenseGammaRay1994a, neubertTerrestrialGammarayFlash2019}, in lightning leaders approaching ground~\cite{mooreEnergeticRadiationAssociated2001} or in
long lasting gamma-ray glows measured above
thunderclouds~\cite{mccarthyFurtherObservationsXrays1985}. The $\gamma$-rays
in turn can create electron-positron beams~\cite{briggsElectronpositronBeamsTerrestrial2011a}, and photo-nuclear reactions~\cite{enotoPhotonuclearReactionsTriggered2017, rutjesTGFAfterglowsNew2017}.

All these emissions (except for gamma-ray glows) are correlated
with the impulsive ``stepped'' propagation of negative lightning leaders
that involve streamer coronas in their dynamics.
While simplified calculations with stationary
fields near a leader tip show that electrons can run away,
accelerate to relativistic motion and create gamma-rays
through bremsstrahlung on air molecules~\cite{xuSourceAltitudesTerrestrial2012, kohnCalculationBeamsPositrons2015}, a detailed model
of the dynamics of leader stepping and of the related electron
acceleration is currently missing.

A joint feature is the pulsed nature
of the discharge that accelerates the electrons. Clearly,
in a pulsed discharge, the electric field can reach high values
before plasma formation and electric screening set in.
If the discharge is negative, electrons could even surf on an ionization wave
with local field enhancement and gain more energy than is available
in a static electric field created by the same voltage; this interesting
physical concept has been suggested by different
authors~\cite{babichModelElectricField2015,Luque_PRL_2014}.

Electron runaway has also been found in pulsed lab discharges. Experiments
where 1~m of ambient air was exposed to positive or negative voltages of 1~MV
(with voltage rise times of 1.2~$\mu$s), showed the formation of meter long
streamers that emitted X-rays with characteristic energies of about 200~keV.
The discharge evolution is shown in figure~\ref{fig:kochkin-full-discharge}.
The upper panel shows positive streamers propagating downward, and a pulse of
X-rays occurs when these streamers encounter the negative upward propagating
counter-streamers in panel (h)~\cite{Kochkin2012}. The lower panel shows
negative streamers propagating downward~\cite{Kochkin2014, Kochkin2015};
these streamers propagate in 3 to 4 pulses downward, see Fig.~\ref{fig:Pavlo-stability-field}, and they emit X-rays
from close to the upper electrode, independently of whether positive
counter-leaders propagate upward from the grounded electrode.

Electron runaway from negative streamers has been modeled in simulations~\cite{mossMonteCarloModel2006, li3DHybridComputations2009, chanrionProductionRunawayElectrons2010a} as well, though the electric
potential available at the streamer head limited the electron energies to a few keV.

Runaway electrons are also suggested as a relevant mechanism in other
laboratory discharges, like fast ionization waves, or so-called diffuse
discharges~\cite{naidisSubnanosecondBreakdownHighpressure2018,
tarasenkoRunawayElectronsDiffuse2019}, but note that in section~\ref{sec:inception_cloud},
we have suggested to identify diffuse discharges with inception clouds that
do not require electron runaway.

\subsection{Plasma jets}\label{sec:jets}

A plasma jet (often called nonequilibrium atmospheric pressure plasma jet,
N-APPJ)
is a repetitive discharge in a stream of Argon, Nitrogen or other gas
that usually flows into ambient air. Plasma jets were first reported by
Teschke \textit{et al.}~\cite{Teschke2005} and Lu and Laroussi~\cite{Lu2006}.
In the past one-and-a-half decade, a multitude of plasma jet
designs has emerged.
In many of these, the actual discharge is (almost) identical
to a traditional streamer discharge with the only exception that
the streamers are guided by the flow itself or by the gas composition
distribution it induces.
Due to its reproducibility and the fact that propagating streamers
emit light only from their tips, the discharges of such
plasma jets are often called `plasma bullets' or `guided streamers'.

The mechanism of this guidance depends on the medium; in nitrogen
the guiding is primarily due to leftovers from the previous discharge
carried by the gas flow. The dominant leftover species for this process is
probably free electrons, although some authors also mention
other species like negative ions or metastables.
Two recent reviews on the guiding mechanism are given
by Lu and co-authors in~\cite{Lu2014,Lu2018}.
In~\cite{Lu2018} they conclude that electrons are the
main factor for the streamer guidance in plasma jets (in all relevant gases)
although this seems to disregard the mechanism sketched below.

For plasma jets in helium flowing in air, the boundaries
of the helium channel also play a major role in the
guiding mechanism, as can be observed from the light emission
of the discharge, which is ring-like~\cite{Lu2014,Mericam2009}.
This is generally
attributed to Penning ionization in the air-helium mixing
layer and thereby differs from the purely
electron-density driven guiding observed in for example
nitrogen plasma jets~\cite{vanderSchansThesis}. Nevertheless,
leftover species are still essential for the inception
of such helium-jets,
evidenced by their requirement for a minimum pulse repetition frequency.

Besides the applications of plasma jets in plasma medicine and
industrial surface treatment, they also provide something for the lab that
most other streamer discharges cannot: a very reproducible discharge
both in time and space. This makes them ideally suited for a range
of plasma diagnostics like optical emission spectroscopy and
laser diagnostics which cannot easily be performed on traditional,
stochastic branching streamer discharges.

\subsection{Sprite discharges in the upper atmosphere}\label{sec:sprites}

\begin{figure}
  \centering
  \includegraphics[width=8.5 cm]{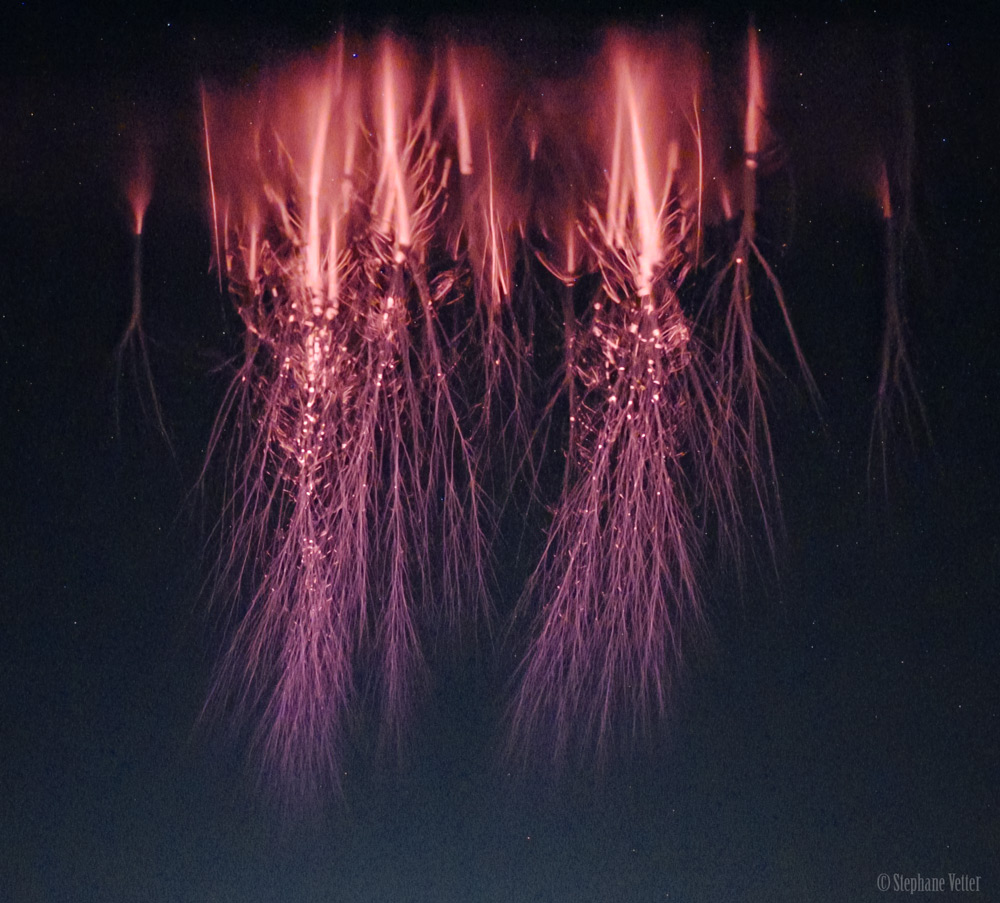}
  \caption{Colour image of a sprite discharge in the upper atmosphere.
  Image is a frame from a
  video (Sony A7s - Nikkor 105mm/1,4@1,4 - Atomos Shogun - 1/25s - 64000 ISO).
  Location \& Time: Peninsula Orbetello, Italy - 10 September 2019 2h UTC.
  Image courtesy of Stephane Vetter and reproduced with permission.
  \label{fig:sprites}}
\end{figure}

Sprite discharges are the largest streamer discharges on our planet,
crossing altitudes of 40 to 90\,km in the night-time atmosphere,
see figure~\ref{fig:sprites}.
They are initiated by the fast changes in electric field induced
by cloud-to-ground lightning and start at altitudes where the resulting
electric field exceeds the local breakdown field.
The relation between streamers and sprites is given by the scaling laws
discussed in section~\ref{sec:scalinglaws}: According to the
US Standard Atmosphere, air density at 83\,km altitude is a factor
of $10^{-5}$ lower than at sea level. This means that length and time scales
are a factor $10^5$ larger (centimeters scale to kilometers, and nanoseconds
to 0.1\,milliseconds), ionization degrees $n_e/N$ a factor $10^{-5}$ smaller
etc. Indeed the similarity laws have been shown to be a good
first approximation, and in that sense sprites are the first lightning-related discharge that we understand from first principles
\cite{Pasko2007,Ebert2008a,Ebert2010,Qin_2015}.

\section{Recent advances in streamer simulations}\label{sec:simulations}

Numerical simulations can be a powerful tool to study the physics of streamer
discharges. In simulations, the electric field and all species densities are known,
both in time and in space. Furthermore, physical mechanisms can be turned off or
artificially increased, the discharge conditions can easily be modified, and
simulations can be performed in simplified geometries. For these reasons,
simulations are increasingly used to help explain experimental results, see for
example~\cite{Boeuf_2013,Nijdam2016}.

The first streamer simulations were performed around thirty years
ago~\cite{Dhali1987,Wu_1988}. A short review of plasma fluid models for streamer discharges has been
presented in~\cite{Luque2011}. Here we also introduce
other types of models, with a focus on developments in the last decade. For a
detailed description of the foundations of the different models (although not aimed at streamers) we refer to the
recent review of Alves \textit{et al.}.~\cite{Alves_2018}.

Modeling streamer discharges can be challenging. Time-dependent
simulations have to be performed in at least two (and often three) spatial
dimensions. Due to the strong electric fields, steep density gradients and thin
space charge layers that are typical for streamers, simulations require a high
temporal and spatial resolution. The streamer dynamics are strongly non-linear,
due to the coupling between electric field, electron transport and source terms.
Because of this non-linearity, a lack of resolution can significantly change the
outcome of a simulation, making low-resolution approximations generally
infeasible. In atmospheric air, features of a few $\muup$m have to be resolved,
whereas a typical streamer is centimeters long. This \emph{multiscale} aspect is
especially challenging for three-dimensional simulations. For this reason, most
simulations have been performed in either Cartesian 2D or axisymmetric
geometries. Even then, simulations of a single streamer can take several hours
to a day~\cite{Bagheri_2018}.

In streamer simulations, the electric field $\vec{E}$ is computed in the
electrostatic approximation as $\vec{E} = -\nabla \phi$, where $\phi$ is the
electric potential. Arguments for the validity of the electrostatic
approximation can be found in section \ref{sec:additional}. Numerical solvers to compute the electric potential are discussed in section \ref{sec:field-solvers}.

We have already briefly introduced particle and fluid models in section
\ref{sec:intro-models}. Below, these models are described in more detail,
after which hybrid models and reduced macroscopic models are also discussed.

\subsection{Particle (PIC-MCC) models}
\label{sec:intro-particle}

Streamer discharges can be simulated with particle-in-cell (PIC) codes
\cite{Hockney_1988,Birdsall_1991a} coupled with a Monte Carlo collision (MCC)
scheme~\cite{Nanbu_2000}. As already discussed in section \ref{sec:first_view_PIC}, electrons and sometimes also ions are tracked as
discrete simulation particles. Each simulation particle has a position
$\vec{x}$, velocity $\vec{v}$, acceleration $\vec{a}$ and a weight $w$, which
determines how many physical electrons it represents.

Compared to fluid models, the main advantages of particle simulations are that
relatively few approximations have to be made, that the electron distribution
function $f(\vec{x}, \vec{v}, t)$ is directly approximated, and that
fluctuations in regions with few particles can be captured. The main drawback of
particle simulations is their high computational cost.

A direct evaluation of particle-particle forces requires $O(N^2)$ operations,
where $N$ is the number of particles. PIC codes therefore compute the forces
between particles using a numerical grid. A typical cycle for a PIC streamer
simulation consists of the following steps:
\begin{enumerate}
  \item \label{pic-step-1}Map the charged particles to a charge density $\rho$
  on a numerical grid.
  \item Compute the electrostatic potential by solving Poisson's equation
  $\nabla \cdot (\varepsilon \nabla \phi) = \rho$ and determine the electric
  field as $\vec{E} = -\nabla \phi$, see section \ref{sec:field-solvers}.
  \item Interpolate the electric field to particles to update their acceleration.
  \item Advance the particles over $\Delta t$ with a \emph{particle mover} and perform
  collisions using a Monte Carlo procedure.
  \item If required, adjust the weights of simulation particles, and go back to
  step \ref{pic-step-1}.
\end{enumerate}
For streamer simulations, electrons are typically tracked as particles whereas
the slower ions can be tracked as densities on a grid. In some applications it can be relevant to model ions as particles, see e.g.~\cite{Babaeva_2011}.

\textbf{Collisions} Typically, only electron-neutral collisions are considered.
Neutral gas molecules are included as a background that electrons can
stochastically collide with. Such collisions can be divided in four categories:
elastic, excitation (rotational, vibrational, electronic), electron-impact
ionization and electron attachment, see e.g.~\cite{Nanbu_2000}. The probability
of a collision per unit time is given by the collision rate
$\nu_i = N_0 v \sigma_i$, where $N_0$ is the number density of the neutral
species, $v$ is the electron velocity and $\sigma_i$ is the energy-dependent
cross section for the collision. Cross sections are often obtained through the
LXCAT website at \href{https://www.lxcat.net}{www.lxcat.net}~\cite{TheLXCatTeam_}. The Monte Carlo
sampling of collision times is usually performed with the null-collision method
\cite{Koura_1986}. With this procedure, an artificial dummy collision is added
to make the total collision rate energy-independent, which greatly simplifies
the sampling procedure.

Because forces are computed via a grid, close range interactions between
electrons are not accurately captured, but this is a good approximation as long
as the discharge is weakly ionized. It is possible to include electron-electron
Coulomb collisions as a separate process~\cite{Nanbu_2000}.

\textbf{Stochastic fluctuations} The combination of discrete simulation
particles with a Monte Carlo collision procedure naturally leads to stochastic
fluctuations. When the simulation particles have a weight of one, these
fluctuations can be regarded as physical. However, in practice the number of
electrons in a streamer discharge is much too large to individually simulate
them. Therefore super-particles with weights $w \gg 1$ are used, which increase
the stochastic fluctuations beyond the physical level. The relative magnitude of
these fluctuations is roughly given by $1/\sqrt{k}$, where $k$ denotes the
number of particles in a region. For example, in a cell with 100 simulation
particles, fluctuations in the particle density would typically be around
$10\%$. Fluctuations are therefore increased by a factor $\sqrt{w}$ compared to
their physical level.

The amount of noise can be controlled by specifying that there should be
$N_\mathrm{cell}$ particles per cell (if there are at least that many physical
particles). For streamer simulations this means that adaptive particle weights
have to be used, since the electron density greatly varies inside and outside
the streamer channel. The use of an adaptively refined mesh, see section
\ref{sec:numerical-methods}, is another reason why weights have to be adjusted
dynamically. Approaches for adaptively setting weights are described in e.g.
\cite{Teunissen2014,Lapenta_1994,Vranic_2015}. Finally, we remark that certain
stochastic fluctuations can also be included in fluid models~\cite{Luque_2011}.

\textbf{Computational cost} The computational cost of a PIC simulation depends
on the number of particles per cell $N_\mathrm{cell}$, the number of cells in
and around the streamer, the gas pressure etc. In atmospheric air, a typical
collision time is $10^{-13} \, \textrm{s}$. If $N_\mathrm{cell} = 100$, and
there are about $10^6$ cells in and around the streamer, this means that about
$10^{12}$ collisions have to be evaluated to advance the simulation over
$1 \, \textrm{ns}$. The costs of the field solver, particle mover, the
adjustment of weights and other model components have to be added. Because
electron-neutral collisions largely determine the electron dynamics, streamer
simulations generally do not need to resolve the Debye length and the plasma
frequency. Parallelization helps to speed simulations up, but due to the
adaptive weights, adaptive refinement and the cost of the field solver, it is
nontrivial to scale simulations to a very large number of processors.

\textbf{Examples of recent work} An PIC-MCC model for streamer simulations in
axisymmetric geometries was presented in~\cite{Chanrion_2008} to investigate
the production of runaway electrons from negative streamers. A 3D PIC-MCC
model with adaptive particle weights and adaptive mesh refinement (AMR) was
presented in \cite{Teunissen_2016}, and it was used to study discharge
inception in nitrogen-oxygen mixtures. The same model was also used to show
the difference between discharges above and below the breakdown threshold
in~\cite{Sun_2014a}. A 3D PIC-MCC model with AMR was presented
in~\cite{Kolobov_2016}, as part of a larger flow simulation
package~\cite{Kolobov_2012}.



\subsection{Fluid models}
\label{sec:intro-fluid}

Most streamer simulations are performed with plasma fluid models, which were already briefly introduced in section \ref{sec:intro-fluid-model}. In a fluid model, the
evolution of several densities is described with partial differential equations
(PDEs). Such models can be derived based on phenomenological arguments and
conservation laws, or from velocity moments of Boltzmann's
equation~\cite{Dujko_2013}. In the simplest case, just the electron and ion
density are considered, but equations for e.g. the electron momentum and/or
energy density can also be included.

Most streamer simulations have been performed with fluid models of the
drift-diffusion-reaction type~\cite{Luque2011}. The equations in this model are
of the form
\begin{equation}
  \label{eq:drift-diffusion}
  \partial_t n_j + \nabla \cdot \left(\pm n_j \mu_j \vec{E} - D_j \nabla n_j\right) = S_j,
\end{equation}
where $n_j$ is the density of species $j$, $\pm$ is the sign of the species'
charge, $\mu_j$ is the mobility, $D_j$ is the diffusion coefficient and $S_j$ is
the source term. Note that the term in parentheses is the flux of the species,
so that equation (\ref{eq:drift-diffusion}) is a conservation law with a source
term. In the simplest case only electrons $n_e$ and a single immobile positive
ion species $n_i$ are considered, so that the equations can be written as
\begin{eqnarray*}
  \partial_t n_e &= \nabla \cdot \left(n_e \mu_e \vec{E} + D_e \nabla n_e\right) + \bar S_e,\\
  \partial_t n_i &= \bar S_i,
\end{eqnarray*}
and we must have $\bar S_e = \bar S_i$ due to charge conservation.
This model is usually called the \emph{classical fluid model}.
We remark that $\bar S_e$ here denotes the sum of all electron generation processes, such as
impact ionization $S_e$ and photo-ionization $S_{ph}$. More complex models can
include several positive and negative ion species, and also keep track of
e.g.~excited molecules to describe light emission or the first stage of the
plasma-chemical conversion processes. The transport coefficients $\mu_j$ and
$D_j$ are often determined using the \emph{local field approximation}, which
assumes that the velocity distribution of electrons (or ions) is relaxed to the
local electric field. Alternatively, transport coefficients can be determined based on the mean electron
energy, see section \ref{sec:fluid-model-comparison}. 

\subsubsection{Transport and reaction coefficients}
Transport coefficients for fluid models can be computed (and tabulated) using a
Boltzmann solver such as Bolsig+~\cite{Hagelaar2005}, which takes
electron-neutral cross sections as input, with the assumption of isotropic scattering. Bolsig+ uses a two-term expansion, i.e,
a first order expansion about an isotropic velocity distribution,
which can be sufficient depending on the gas and the required accuracy
\cite{White_2003}. However, the approximation
can become problematic, for example at high electron energies. More accurate multi-term Boltzmann solvers have been
developed by several authors, e.g.~\cite{Dujko2011,Stephens_2018a,Tejero-del-Caz_2019}.
Monte Carlo swarm simulations can also be used to obtain transport coefficients
\cite{Biagi_1999,Rabie_2016a}. Such Monte Carlo simulations are more expensive,
but e.g. anisotropic scattering or magnetic fields at arbitrary angles can
relatively easily be included. (We remark that accurate anisotropic cross sections can be hard to obtain, and that proper rescaling is important when they are based on their isotropic counterparts.)

There are so-called \emph{bulk}
and \emph{flux} transport coefficients, see for example
\cite{Petrovic_2009,Dujko_2013}. Bulk coefficients describe average properties
of a group of electrons, taking ionization and attachment into account, whereas
flux properties are averages for ``individual'' electrons. These coefficients differ
when there is strong impact ionization or attachment. Fluid models typically use flux coefficients, but in some cases the use of bulk coefficients can be beneficial; this depends on the type of model used and the quantities of interest, see e.g.~\cite{Dujko_2013}.

\subsubsection{Source terms}
The electron source term $\bar S_e$ can contain several components. The most
important is electron impact ionization (reactions (\ref{eq:Oxygen-impact}) and (\ref{eq:Nitrogen-impact})
in air),
often written as $\alpha \mu E n_e$,
where $\alpha$ is the ionization coefficient. Like $\mu_e$ and $D_e$, $\alpha$
can be tabulated using a Boltzmann solver. In electro-negative gases, such as air,
electrons can be lost in attachment reactions. This can be described with a sink $-\eta \mu E n_e$, where $\eta$ is the attachment coefficient. Depending on the gas number
density and the electron energy, two-body or three-body attachment reactions are dominant. Another
important source term in air is photo-ionization~\cite{Pancheshnyi2015}, which is
discussed in more detail in section \ref{sec:simulations-photoionization}. The
detachment of electrons from negative ions can also be important, especially at
longer time scales~\cite{Naidis_1999,Pancheshnyi_2013}; for a further discussion of electron detachment and related phenomena, we refer to section~\ref{sec:Conductivity}. Electrons can also be
generated from conducting or dielectric boundaries through e.g. secondary
emission, but this is typically incorporated in the fluxes near those
boundaries~\cite{Hagelaar_2000}.

\subsubsection{Comparison of fluid models for streamer discharges}
\label{sec:fluid-model-comparison}

For simulations of for example RF discharges, fluid models based on a local
energy equation are often more accurate~\cite{Grubert_2009} than those based on
the local field approximation. Models with an energy equation can also have
advantages when applied to streamer
discharges~\cite{Eichwald_2005,Markosyan_2015}, but they also have drawbacks, which are discussed below. Several types of second order
models have been constructed~\cite{Guo_1993,Becker_2013b,Kanzari_1998}, as well
as higher-order models~\cite{Becker_2013a,Dujko_2013}.

The drift-diffusion-reaction model combined with the local field approximation
is probably the most popular model for streamer simulations. One of the
underlying approximations is that the electron velocity distribution is relaxed
to the local electric field. This approximation can be inaccurate when electric
fields change rapidly, for example near the streamer head. It can also be
inaccurate when momentum or energy relaxation of electrons is relatively slow,
for example in a noble gas~\cite{Markosyan_2015}. Another shortcoming is that
the model cannot capture kinetic and non-local effects~\cite{Petrovic_2007}. For
example, even in a uniform electric field, there can be a gradient in the
electron energy~\cite{Li2010}. To correct for this, an extra source term based
on the gradient of the electron density was introduced in \cite{Li2010}.

Despite the potential inaccuracies outlined above, the use of the local field
approximation has some advantages. For electrons, only a single
drift-diffusion-reaction equation has to be solved, and for most gases, input
data can readily be generated with a Boltzmann solver or is already available.
Furthermore, the electric field is a relatively `safe' and smooth variable to
base transport coefficients on. When one uses for example the mean energy, a
division by the electron density is required, which is problematic when the electron density goes to zero.

\subsubsection{Time stepping}
\label{sec:fluid-time-stepping}

The fluid equations can be solved implicitly or explicitly in time. With a
typical explicit approach, the state $Q(t + \Delta t)$ can directly be
constructed from the past state $Q(t)$ as
\begin{equation}
  Q(t + \Delta t) = f(Q(t), t),
\end{equation}
where $f$ is a function to advance the solution in time. This function includes
a field solver, see section \ref{sec:field-solvers} below, but it is otherwise
computationally cheap to evaluate. Conversely, with an implicit approach, the
new state is defined implicitly as
\begin{equation}
  Q(t + \Delta t) = f(Q(t + \Delta t), Q(t), t).
\end{equation}
Solving such an implicit equation can be quite costly, and it typically requires
an iterative procedure. One reason for this is that the new state in a grid cell
depends on the new states in neighboring cells, which again depend on their
neighbors etc.

A drawback of the explicit approach is that the time step $\Delta t$ is limited
by several constraints to ensure stability of the numerical method, where
stability means that errors should not blow up in time. Perhaps the most
important of these is the well-known \emph{CFL condition}, which for a simple 1D
problem reads
\begin{equation}
  \Delta t < C \frac{\Delta x}{v},
\end{equation}
where $v$ and $\Delta x$ denote the velocity and grid spacing and $C$ is a
number of order unity. The dielectric relaxation time~\cite{Barnes_1987}
$\tau = \varepsilon_0 / \sigma$, where $\sigma$ is the conductivity of the
plasma, is another common constraint; however, this constraint can be avoided with semi-implicit methods~\cite{Ventzek_1993,Hagelaar_2000a} or by limiting the electric current~\cite{Teunissen_2020}. Time step constraints from fast
chemical reactions or the diffusion of species can also be avoided by solving
for these terms implicitly.

Both implicit and explicit approaches are used for streamer simulations
\cite{Bagheri_2018}. A drawback of the implicit approach is that small time
steps are still required to capture the strongly non-linear evolution of a
streamer discharge, making them typically more costly than explicit methods.
This is particularly relevant for 3D simulations, which require high
computational efficiency.

\subsubsection{Spatial discretization}
\label{sec:fluid-spatial-discretization}

Finite volume and finite element methods have been used to simulate streamer
discharges. With a finite volume approach the fluxes between grid cells are
first computed, after which they are used to update the solution. This approach
naturally ensures that quantities such as charge are conserved. For streamer
simulations, finite volume methods are often used in combination with explicit
time stepping, in which case fluxes have to be computed with a suitable scheme
\cite{LeVeque_2002,Toro_2009} to ensure that errors and oscillations do not blow
up in time.

The use of second order flux/slope limiters is common in recent
work, see e.g.~\cite{Montijn2006a,Luque_2008,Eichwald_2012,Bessieres_2007,Pancheshnyi_2008,Teunissen_2017,Plewa_2018}.
A high-order method has been tested in~\cite{Zhuang_2014}. Because of the
steep density gradients around a streamer head, which are approximately a shock
in the solution, even high-order schemes need a high resolution in this region.
Other methods that have been used for streamer simulations are the
Scharfetter-Gummel scheme~\cite{Kulikovsky_1995,Liu2004,Bourdon2010} and the
flux-corrected transport method~\cite{Georghiou_2000}.

For streamer simulations in complex geometries, it can be attractive to use a
finite element method on an unstructured grid, see section
\ref{sec:numerical-methods}. Examples of finite-element based streamer
simulations can be found
in~\cite{Papageorgiou_2011,Ducasse_2007,OSullivan_2008,Zakari_2015}.

\subsection{Hybrid models}
\label{sec:hybrid-models}

Hybrid models aim to combine the strong points of different simulation
models. Fluid models are computationally cheap, but they are often inaccurate
near dielectrics and electrodes, where sheaths form, and they cannot capture
phenomena like electron runaway. Furthermore, the continuum approximation
breaks down for very low densities. By using a particle description in these
regions, a more accurate description of the discharge can in principle be
obtained.

In~\cite{Li2010, Li2012}, particle and fluid models were used in different regions in space.
The model was used to simulate electron runaway from negative
streamers~\cite{li3DHybridComputations2009}, with the fluid model describing
the low-field region behind and in the streamer, and a particle model
describing the high-field region ahead of it with fewer and energetic
electrons. With such a spatial separation a buffer region is required to
describe particles moving back and forth across the boundary. At the
interface, particle fluxes have to be converted to fluid fluxes and vice
versa, although the latter operation is not always
required~\cite{li3DHybridComputations2009}.

Another approach is to separate models in energy space. This was done in
\cite{Chanrion2014} to study the link between streamers, runaway electrons
and terrestrial gamma-ray flashes (TGFs). The authors used a PIC-MCC model
for electrons with energies above $100 \, \textrm{eV}$ and a fluid model for
the other electrons. The authors computed special transport coefficients so
that the fluid model could take only the low-energy electrons into account.
Another example of a hybrid model was presented in~\cite{Babaeva_2016} to
study the interaction between streamers and dielectrics. The authors used a
Monte Carlo particle model to simulate energetic secondary electrons coming
from a dielectric surface. The other electrons were described by a
conventional fluid model. Different computational grids and time steps were
used to combine the two models.

Hybrid modeling can offer significant advantages by combining the strengths of
different models. However, from a practical point of view, implementing two
models together with a consistent coupling between them can be quite
challenging. In~\cite{Kushner_2009}, some of these practical considerations are
discussed for hybrid plasma models aimed at equipment design.

\subsection{Macroscopic models}
\label{sec:macroscopic-models}

Although particle and fluid models already contain various approximations, we
consider them to be \emph{microscopic}. These models simulate the (drift) motion
of electrons, they include fast processes such as electron impact ionization,
and they resolve the thin charge layers around a streamer, so that a high
spatial and temporal resolution is required. If one wants to simulate a
discharge containing tens to hundreds of streamers (like those in
figure~\ref{fig:Rep-freq-dep}), a
microscopic description therefore becomes computationally unfeasible.
A solution could be to use a more \emph{macroscopic} model that describes the evolution of streamer channels as a whole, without resolving thin space charge
layers and the microscopic electron dynamics.
Although such models have already been proposed in the 1980s, their development is still in an early stage.

A model to describe a streamer based on the velocity, radius, electron density,
potential and current at its tip can be found
in~\cite{Raizer_2006,Bazelyan_spark_discharge_1998}.
In~\cite{Niemeyer_1984}, the phenomenological
\emph{dielectric breakdown model} was introduced to investigate the fractal
nature of planar discharges such as Lichtenberg figures. In this model, the
discharge is evolved on a numerical grid in which each cell is either part of
the discharge or not. At each iteration, the electric potential on the grid is
computed assuming that streamer channels are perfect conductors. The probability
of extending the discharge in a particular direction then depends on the
potential difference in that direction. The model was applied to explain the
`fractal structure' of sprite discharges in \cite{Pasko2000}. (Note that streamers are not true 
fractals, as they have a minimum diameter, and as they eventually do not branch anymore, see Figs.~\ref{fig:velocity-example}-\ref{fig:kochkin-full-discharge}.)

In~\cite{Akyuz2003}, an alternative macroscopic model was presented, in which
streamers are modeled as multiple segments of perfectly conducting cylinders,
capped with spherical heads. In this phenomenological model, a new segment is
added to an existing streamer when the ionization integral (see equation
(\ref{eq:Meek})) ahead of it is
sufficiently large, and branching occurs according to heuristic rules.

In reality, streamers are no perfect conductors. They carry electric currents
that vary in space and time, due to changes in the external circuit, their own
growth and the growth of nearby streamers. Some authors have included a fixed
internal field along each channel in the dielectric breakdown
model~\cite{Pasko2000}. However, this approach does not conserve charge and it
cannot capture the dynamic currents carried by the streamers.

\begin{figure}
  \centering
  \includegraphics[width=7cm]{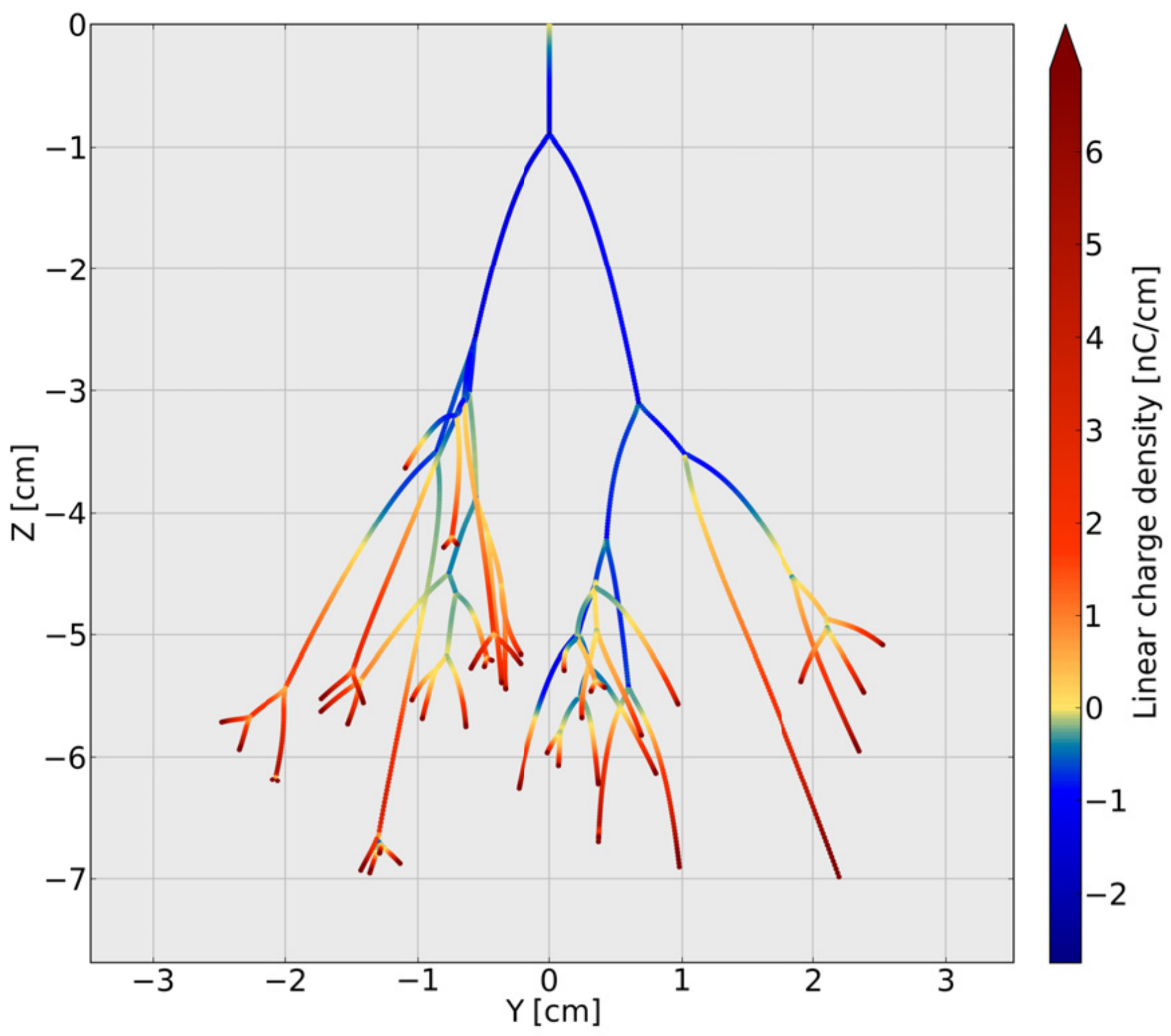}
  \caption{Charge distribution in a streamer simulation using a tree model.
    Picture taken from~\cite{Luque2014}. The color scale is truncated and does
    not show the charge density at the streamer tips, as they would dominate the
    plot.}
  \label{fig:tree-charge-luque}
\end{figure}

A more realistic
\emph{tree model} was presented in~\cite{Luque2014}. In this model, the
streamers are represented as growing linear channels, see figure \ref{fig:tree-charge-luque}. Along these channels, the
streamer radius, charge density, conductivity and electric current are evolved.
The authors implement a simple version of this model, where the radius and
conductivity are kept fixed and equal for all channels,
the streamer velocity is linear in the local
electric field, and branching is implemented as a Poisson process. Numerically,
the channels are represented by a series of finely spaced points with a
regularized kernel to avoid singularities in the electric potential.

Even the simple tree model incorporates charge conservation and transport.
Therefore, the streamer channels are polarized, 
with positive line charge in the growing tips 
(colored in red in Fig.~\ref{fig:tree-charge-luque}),
and negative line charge at the channel backs (colored in blue). 
The electric field configuration inside the channels in the same simulation
is shown in panel d in Fig.~\ref{fig:MultiScale}. Clearly, the electric field
varies along the channels, especially behind branching points,
and it is even inverted in the channels colored in blue. 
This is in remarkable contrast to the assumption of a constant interior field
that is sometimes used to motivate the concept or a stability field, see section~\ref{sec:stabilityfield}.

\textbf{Outlook} Even with a continuing increase in computational power,
macroscopic models will be required to study systems with tens or hundreds of
streamer discharges. The approach presented in~\cite{Luque2014} can give
physically realistic results if the evolution of the linear channels is
described accurately. For that, we need to better understand the dynamics of
streamers: how their velocity and branching statistics
depends on radius, channel conductivity and generated electric field profile,
as well as background or photo-ionization, and how tip radius and channel
conductivity develop in time. Partial answers to these questions can be
found in section~\ref{sec:propagation}. A general approach for future model reductions
is outlined in section~2 of~\cite{Luque2014}.


\subsection{Numerical grid and adaptive refinement}
\label{sec:numerical-methods}

\begin{figure}
  \centering
  \includegraphics[width=7cm]{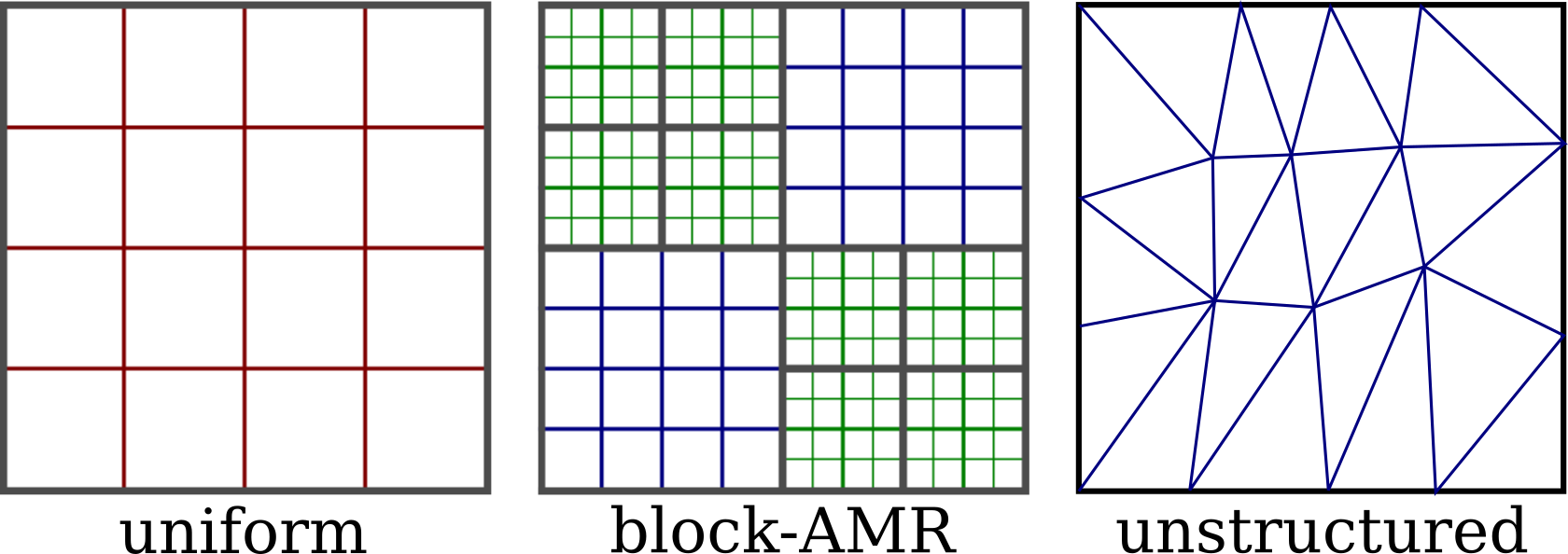}
  \caption{Schematic illustration of three types of numerical grids in 2D.
    Uniform grids are the simplest to work with. Adaptive mesh refinement (AMR)
    allows for a varying resolution in the domain, and unstructured grids are
    the most flexible.}
  \label{fig:grid-examples}
\end{figure}

Computational efficiency is important for streamer simulations, as they can be
quite costly. An important factor for
the performance is the type of numerical grid that is used. This is not only
true for fluid simulations, in which all quantities are defined on this grid,
but also for particle simulations, in which the grid is used to keep track of
particle densities and to compute electric fields (see section
\ref{sec:field-solvers}). Most macroscopic models also make use of a numerical grid to compute electric fields.
Below, we briefly discuss the most common types of grids, which are illustrated in figure \ref{fig:grid-examples}.

\paragraph{Structured grids}
Uniform grids are simple to work with, and they allow for efficient computations
per grid cell. However, the total number of grid cells rapidly grows with the
domain size due to the fine mesh that is required to capture the streamer
dynamics. This can be avoided by using adaptive mesh refinement (AMR), because a
fine mesh is usually only required around the streamer head. With AMR, the
resolution in a simulation can change in space and in time. This is usually done
by constructing the full mesh from smaller blocks that are locally rectangular.
More details about structured AMR and AMR framework can be found in
e.g.~\cite{Dubey_2014,Teunissen2017a}, and streamer codes with AMR have been
presented
in~\cite{Montijn2006a,Pancheshnyi_2008,Kolobov_2012,Duarte_2012,Teunissen_2017}.
Different refinement criteria have been used, based on for example density
gradients, local error estimators and the ionization length $1/\bar\alpha$
(see section~\ref{sec:plasma-theory}), where
$\bar\alpha$ is the ionization coefficient. However, an ideal criterion suitable for
both positive and negative streamers has not been established.

\begin{figure}
  \centering
  \includegraphics[width=7cm]{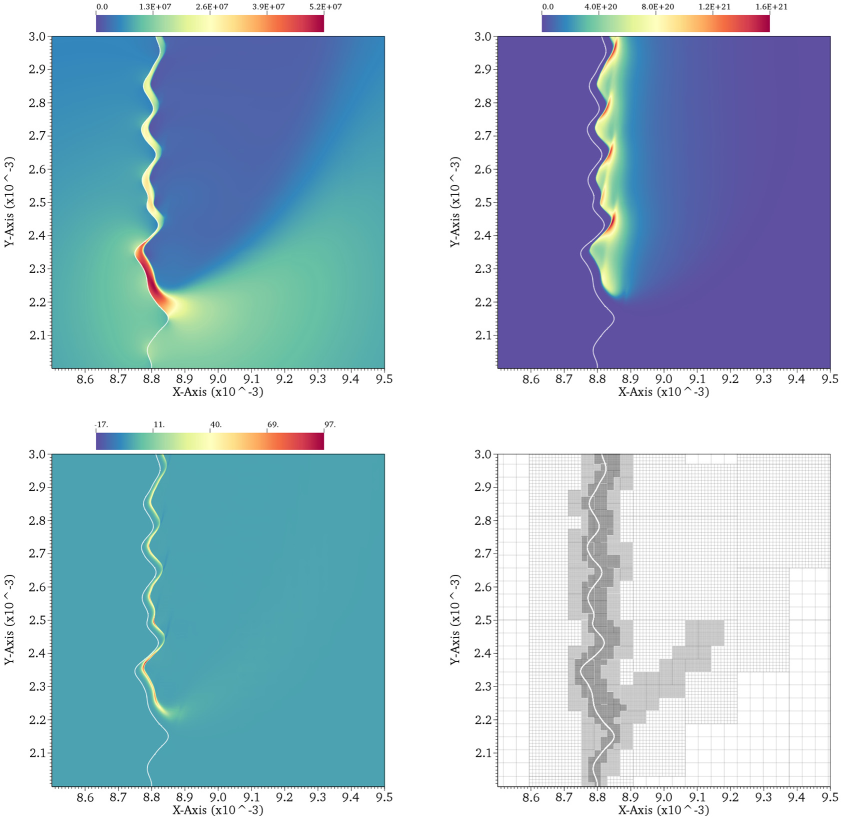}
  \caption{2D simulation with adaptive mesh refinement of a positive streamer
    propagating over a rough dielectric surface. Figure taken
    from~\cite{Marskar_2019a}
}
  \label{fig:embedded-boundary-example}
\end{figure}

Modeling curved electrodes and dielectrics in a structured grid requires special
interpolation procedures, such as the ghost fluid method~\cite{Celestin_2009a}.
It is quite challenging to combine such methods with AMR, but significant
progress in this direction has recently been made in
\cite{Marskar_2019,Marskar_2019a}, see figure \ref{fig:embedded-boundary-example}.

\paragraph{Unstructured grids}

Operations on an unstructured grid are generally more costly than those on a
structured grid. For each cell, the shape and the connectivity to other cells
has to be stored. However, unstructured grids have an important advantage: the
cells can be aligned with complex geometries, such as curved electrodes and
dielectrics, see e.g.~\cite{Zakari_2015}. There exist several frameworks for
finite-element and finite-volume simulations in such geometries, for example
Comsol, Openfoam and Fluent. Unstructured grids are also used in
nonPDPSIM~\cite{Xiong_2010,Xiong_2012} and they are currently being incorporated into
Plasimo~\cite{vanDijk_2009}.

Streamer discharge simulations with unstructured grids have for example been
performed in~\cite{Georghiou_2000,Babaeva_2015a,Zakari_2015,Bagheri_2018}. The
cost of such simulations is usually higher than for finite-volume simulations
with structured AMR, so that they are not ideal for 3D simulations. There are
several reasons for this. Operations on unstructured grids are more expensive,
it is more costly to adapt unstructured grids in time, and unstructured grids
typically require some type of implicit time stepping, see section
\ref{sec:fluid-time-stepping}. With an explicit method, the presence of a single
small cell would severely restrict the time step.

\subsection{Field solvers}
\label{sec:field-solvers}

Streamer simulations are usually modeled using the electrostatic approximation,
so that the electric field can be determined as $\vec{E} = -\nabla \phi$.
The electrostatic potential $\phi$ can be obtained by solving
a Poisson equation, see equation (\ref{eq:poisson-phi}).
A new electric field has to be computed at least once per time
step. An $n$\textsuperscript{th} order accurate time integrator typically
requires $n$ evaluations of the electric field.
It is therefore important to
use a fast Poisson solver, but solving a Poisson equation efficiently (and in
parallel) is not trivial due to its non-local nature.

There exist many numerical methods to solve elliptic partial differential
equations like (\ref{eq:poisson-phi}). Which solvers can be used depends on the
simulation geometry and mesh type, the boundary conditions and the variation of
$\varepsilon$ in the domain. Solvers can generally be classified in two groups:
direct methods, which do not need an initial guess, and iterative methods, which
improve an initial guess. A somewhat dated overview of classical methods can be
found in~\cite{Hockney_1988}, and a more recent comparison of solvers suitable
for high-performance computing can be found in~\cite{Gholami_2016}. Poisson
solvers have also been compared specifically for streamer simulations in
\cite{Kacem_2012,Duarte_2015}. Below, we briefly list some of the most efficient
solvers for different mesh types.

\subsubsection{Field solvers for uniform grids}
\label{sec:field-uniform-grids}

Almost all Poisson solvers can be used on a rectangular grid with standard
boundary conditions (Dirichlet or Neumann) and a constant $\varepsilon$.
Efficient direct methods are for example based on the fast Fourier transform
(FFT), potentially combined with cycling reduction~\cite{Swarztrauber_1979}. The
computational cost of these methods scales as $O(N \log N)$, where $N$ is the
total number of unknowns, and they can be used in parallel. Geometric multigrid
methods~\cite{Trottenberg_2000,Brandt_2011} can achieve ideal $O(N)$ scaling in
the number of unknowns. Implementations for uniform grids can be found in e.g.
\cite{Adams_1989, Falgout_2002}. They are discussed in more detail below in
section \ref{sec:field-solvers-amr}.

The use of free space boundary conditions, i.e. $\phi(r) \to 0$ for
$r \to \infty$, can be attractive for streamer simulations. Such boundary
conditions can be implemented in several ways. In~\cite{Malagon-Romero_2018}, a
procedure is described to apply such conditions in the radial direction in
axisymmetric simulations. There also exist special FFT-based solvers that
implement these boundary conditions, see e.g.~\cite{Genovese_2006,Teunissen_2019_gmg}. We remark
that with fast multipole methods (FMMs)~\cite{Greengard_1997} free space
boundary conditions are naturally imposed, but such solvers are more suitable
for isolated point sources than for grid-based computations.

A uniform grid requires special methods when curved dielectrics or curved
electrodes are present. The numerical discretization of Poisson's equation
around such objects can be modified, using for example the ghost fluid method
\cite{Celestin_2009a}. Methods that can solve the resulting equations are
discussed below in section \ref{sec:field-unstructured-grids}.

We remark that classic SOR (successive over-relaxation) is still used by some
authors~\cite{Plewa_2018}. Although it offers benefits in terms of simplicity,
SOR is typically much less efficient than the fastest methods discussed here.

\subsubsection{Field solvers for structured grids with refinement}
\label{sec:field-solvers-amr}

FFT-based methods cannot directly be used with mesh refinement (see section
\ref{sec:numerical-methods}), because they operate on a single uniform grid.
There exist strategies to still employ such solvers with mesh refinement, see
e.g.~\cite{Ricker_2008,Teunissen_2016}, but this has drawbacks in terms of
solution accuracy, parallelization or the flexibility of the refinement
procedure.

Geometric multigrid methods~\cite{Trottenberg_2000,Briggs_2000,Brandt_2011} are
naturally suited for meshes with refinement. The basic idea is to apply a simple
relaxation method that locally smooths the error. By doing this on grids with
different resolutions, different `wavelenghts' of the error can efficiently be
damped. The operations in geometric multigrid methods are defined by the mesh
and no matrix has to be stored. This is an advantage in streamer simulations
where the mesh frequently changes to track the streamer head. Geometric
multigrid methods have recently been used in several streamer simulations, see
e.g.~\cite{Kolobov_2012,Komuro_2014,Teunissen_2017}.

\subsubsection{Field solvers on unstructured grids}
\label{sec:field-unstructured-grids}

For unstructured grids it is common to work with more general sparse matrix
methods, which can be direct or iterative. First, a matrix $L$ corresponding to
the discretized Poisson's equation
$$L x = b$$
is stored in a sparse format, and analyzed by the solver package, which does
some pre-computation. Afterwards, the solution $x$ can be determined for one or
more right-hand sides $b$. The cost of the sparse solver not only depends on the
type of solver that is used, but also on the sparsity and the structure of the matrix $L$.

Examples of direct sparse solvers are MUMPS~\cite{Amestoy_2001},
SuperLU~\cite{Li_2005} and UMFPACK~\cite{Davis_2004}. In general, such direct
methods are more robust than iterative approaches, but their parallel scaling is
usually worse, and their memory and computational cost can become prohibitive
for 3D problems~\cite{Benzi_2002}.

There exist several types of iterative sparse methods. So-called Krylov methods
such as GMRES~\cite{Saad_1986} or the conjugate gradient method do not converge
very rapidly by themselves. However, when used in combination with a suitable
\emph{preconditioner}~\cite{Benzi_2002}, large sparse systems can be solved
efficiently. For unstructured grids, algebraic multigrid~\cite{Henson_2002} can
for example be used as a preconditioner. Algebraic multigrid is a generalization
of geometric multigrid that directly works with a matrix instead of a mesh.
Experimenting with different solvers and preconditioners can conveniently be
done using the PETSc framework~\cite{petsc-user-ref}.

\subsection{Computational approaches for photo-ionization}
\label{sec:simulations-photoionization}

In air, photo-ionization is usually an important source of free electrons
ahead of a streamer discharge, see section \ref{sec:photoionization}. This is
particularly important for positive streamers, because they propagate against
the electron drift velocity. Zheleznyak's model~\cite{Zhelezniak1982} has
frequently been used to describe photo-ionization in air. Photo-Ionization in
air, N$_2$, O$_2$ and CO$_2$ has been analyzed in detail
in~\cite{Pancheshnyi2015}. In~\cite{Stephens_2016}, the precise excited
states and transitions responsible for photo-ionization in air have recently
been investigated.

The approaches for photo-ionization in streamer simulations can be divided in two
categories: continuum methods and Monte Carlo methods. With a continuum method,
the photo-ionization rate (ionizations per unit volume per unit time) is computed
from the photon production rate (photons produced per unit volume per unit
time). This is commonly done using the so-called \emph{Helmholtz approximation},
in which the absorption function is approximated~\cite{Bourdon2007,Luque2007}
to obtain multiple Helmholtz equations of the form
\begin{equation}
  \label{eq:photoi-helmholtz}
  (\nabla^2 - \lambda_j) S_j = f,
\end{equation}
see for example Appendix A of~\cite{Bagheri_2018} for details. A comparison of
different continuum approximations for photo-ionization can be found
in~\cite{Bourdon2007}, which also emphasizes the need for proper boundary
conditions for equation (\ref{eq:photoi-helmholtz}).
It can also be important to adjust the Helmholtz coefficients when changing the gas composition or pressure. Scaling with the gas number density can only be done over a limited range, because the absorption function is fitted with functions that have a different algebraic decay~\cite{Luque2007}. This was not taken into account in~\cite{Wormeester2010}.

A Monte Carlo approach for photo-ionization in streamer discharges was first
presented in~\cite{Chanrion_2008}. The idea is to use random numbers to generate
discrete photons. Their direction and absorption distance are also determined
by random numbers. Some authors have also included the life time of the excited
states, see e.g.~\cite{Jiang_2018}. Depending on the number of photons that is
produced, a Monte Carlo method can be computationally cheaper or more expensive
than the Helmholtz method described above.

With a Monte Carlo approach, the photo-ionization rate will be stochastic,
whereas a continuum approach will lead to a smooth profile. Both approaches have
recently been compared using 3D simulations in~\cite{Bagheri_2018}. It was found
that stochastic fluctuations can play an important role in the branching of
positive streamers.

\subsection{Modeling streamer chemistry and heating}


Typical time scales for streamer propagation at atmospheric pressure are nanoseconds to
microseconds. Several fast chemical reactions occur within such time scales, see section 
\ref{sec:chemistry}.
Slower reactions can also play an important role, for example when studying repetitive discharges, or when
the long-lived chemical species produced by a discharge are of interest.
An extensive list of chemical reactions in nitrogen/oxygen mixtures can be found
in~\cite{Kossyi1992}. Basic research in this direction is still ongoing, see 
e.g.~\cite{Hosl_2017,Haefliger_2019}.
Below, we highlight several studies with extensive chemical modeling.

In~\cite{Sathiamoorthy_1999}, NO$_{\rm x}$ removal was modeled in a
pulsed streamer reactor, using different reactions sets when the voltage was on
or off. In~\cite{Sentman2008} and~\cite{Gordillo-Vazquez2008}, the chemistry
and emission of sprite streamers was studied with extensive chemical models
containing hundreds of reactions. In~\cite{Levko_2017}, the chemistry in 2D
fluid and 0D global simulations was compared in an air/methane mixture,
obtaining quite good agreement for typical species concentrations.

There are many more computational studies that include a large number of
reactions, but of great importance will be the development of standardized and
validated chemical datasets~\cite{Tennyson_2017}.
For large chemical datasets it
is often beneficial to use some type of dimensionality reduction. Examples of
such methods can be found in~\cite{Peerenboom_2015,Markosyan_2014}.

The interaction between streamers and gas heating and gas flow has already been
discussed in section \ref{sec:flowheat}. Here we mention some of the numerical
studies that have been performed, all of them in air. In~\cite{Tholin_2013},
streamers between two pointed electrodes were simulated with an axisymmetric
fluid model. Assuming a certain fraction of the discharge energy was deposited
into fast gas heating~\cite{Aleksandrov_2010,Popov_2011}, the effect of the
streamers on the gas dynamics was studied. Fast gas heating by streamer
discharges was also studied in~\cite{Komuro_2014}, using an axisymmetric plasma
fluid model directly coupled to Euler's equations. Temperatures up to
$2000 \, \textrm{K}$ were observed directly below the tip of a positive needle
electrode.



\subsection{Simulating streamers interacting with surfaces}
\label{sec:sim-streamer-dielectric}

For many applications, the interaction of streamer discharges with dielectric or
conducting surfaces is of importance. This includes the interaction with liquids
and tissue in plasma medicine, see e.g.~\cite{Tian_2014,Norberg_2014}. Some of
the mechanisms that can play a role are electrostatic attraction, secondary
electron emission due to ion impact, photo-emission, the charging of surfaces,
field emission, and the transport of species across an interface. Below, we only
refer to a few of the computational studies from this broad field.

Between a positive streamer and a dielectric, a narrow sheath with a high
electric field can form. In~\cite{Babaeva_2011}, it was shown that this sheath
can accelerate positive ions to energies of tens of eVs. The effect of different
boundary conditions was investigated in~\cite{Yan_2014}. The differences between
positive and negative streamers near dielectrics have also been investigated in
\cite{Babaeva_2016}.

Discharge propagation in capillary tubes, relevant for the plasma jets described
in section \ref{sec:jets}, was simulated in~\cite{Jansky_2010}.
Plasma jets
touching different dielectric and metallic surfaces were simulated in
\cite{Norberg_2015}. The dynamics of surface streamers on a dielectric bead have
recently been investigated in~\cite{Kang_2018}, using both experiments and
simulations. Experiments and simulations were also used to study the effect of
dielectric charging on streamer propagation in~\cite{Celestin_2009}.

\subsection{Validation and verification in discharge simulations}

Streamer physics is complex, and in many cases just the demonstration
of a nonlinear physical mechanism is very valuable to develop understanding,
even in different parameter regimes, or in two rather than three spatial dimensions.
However, the quest for quantitative models, and hence for the
validation and verification (V\&V) of simulation codes~\cite{Roache_1998} is
becoming more important in our community. Code \emph{verification} is verifying
that a model is correctly implemented, whereas code \emph{validation} is
validating the results against experiments. Closely related to this are code
benchmarks, in which the results of several simulation codes are compared on a
set of test problems.

Let us first discuss some general work, not directly aimed at streamer
simulations. In~\cite{Kim_2005}, particle, fluid and hybrid models were compared
for capacitively and inductively coupled plasmas and plasma display panels. The
results were also compared with experimental data. A benchmark of
particle-in-cell codes for the 1D simulation of capacitively coupled discharges
was presented in~\cite{Turner_2013}. Just as important as the models and their
implementation is the input data that is used. This was illustrated for a
complex plasma chemistry in~\cite{Turner_2015}.

There have been several model comparisons for streamer simulations.
In~\cite{Ducasse_2007}, a finite-element and a finite-volume code were compared
for a positive streamer in an axisymmetric geometry.
In~\cite{Li_2012}, 3D particle, fluid and hybrid models were compared
for a short negative streamer in an overvolted gap. However, we should point out that the classic fluid model was not implemented correctly in this paper. In~\cite{Markosyan_2015},
three fluid models were compared to PIC results for a 1D streamer discharge in
different gases. The first extensive comparison of streamer simulation codes
from different groups was presented in~\cite{Bagheri_2018}. Six groups compared
their plasma fluid models for a positive, axisymmetric streamer discharge under
different conditions. Three of the groups did a convergence study with higher
spatial and temporal resolutions than are commonly used. On the finest grids,
these groups observed relatively good agreement in their results, with
differences of a few percent in the maximal electric field. On coarser grids,
differences were significantly larger.

For streamer simulations, no complete validation studies have been performed.
However, there have been several studies in which experimental results and
simulations were compared. A few examples are listed below.
In~\cite{Pancheshnyi_2005}, velocity, diameter and current of positive streamer discharges were compared between simulations and experiments, finding agreement within 30-35\%.
In~\cite{Nijdam2016}, the experimentally observed guiding of streamers by weak laser preionization could be reproduced and explained with 3D simulations.
The influence of surface charge on dielectric barrier discharges was investigated using simulations and experiments in \cite{Celestin_2009}. Finally, the interaction between streamers and a dielectric rod was studied both experimentally and with axisymmetric simulations in \cite{Dubinova2016}.



\section{Modern streamer diagnostics}\label{sec:diagnostics}
The earliest diagnostics of streamers are of course the observations of
corona discharges by human eyes and ears. In the dark a corona discharge is
visible as a faint purple glow and it can often be heard as a hissing sound.
These earliest observations were followed by electrical diagnostics and
later by more and more advanced imaging techniques \cite{Creyghton1994a}.

Most techniques that are used to study streamer-like discharges rely on
electromagnetic radiation,
primarily in the visible part of the spectrum
or in ranges close to it. Such methods either use the light emitted by
the discharge itself or use light that has been modified (scattered, absorbed,
etc.) by the discharge or its remnants. In all cases, the intensity of the
light that has to be measured is generally very low. Furthermore, the
highly transient and often stochastic nature of a streamer discharge can require single shot
measurements with high spatial and temporal accuracy and resolution.
Together this makes optical diagnostics on streamer discharges
more challenging than on most other laboratory plasmas. Reviews
of optical diagnostics relevant for streamers were published
by Ono~\cite{Ono2016}, {\v{S}}imek~\cite{Simek2014}
and Laux \textit{et al.}~\cite{Laux2003}; all provide far more detail than we can
do here.

\subsection{Electrical diagnostics}
Electrical diagnostics are still essential for characterizing streamer
discharges. It is hard to find experimental papers that do not mention or
show the voltage and/or current waveforms of the measured discharges. Both are
required in order to understand some of the basic properties of a discharge.
Recent advances for this diagnostic aspect are the
use of faster and more advanced oscilloscopes and probes.

One possible issue with such measurements is their synchronization. When one
wants to measure the power dissipated by a discharge, it is of utmost
importance that the current and voltage measurements are properly
synchronized. When these are out of sync by as much as a fraction of a
nanosecond, this can lead to errors in the measured power. Such a
small mis-synchronization can be easily produced by for example differences in
cable lengths.
The measurement location is also important because (stray) capacitances,
cable losses and build-up of space charge can greatly influence the results.
It is therefore
good practice to measure currents on both electrodes instead of only
on one side of the discharge gap.

For many applications, such power or energy measurements are very important
because one is often interested in the energy efficiency of the discharge.
One technique that is often employed is to use a Lissajous figure (a
Voltage-Charge plot) to measure the dissipated power \cite{Manley1943,
Kriegseis2011, Hofmann2011, Gerling2017}. However, this method is only
useful for repetitive discharges with high repetition rates, like those
driven by radio frequent power sources. For other discharges, a simple
multiplication of current and voltage is often the best approach
\cite{Briels2008}. In this case, one should subtract the capacitive current
from the current signal. This capacitive current can for example be measured
by applying the voltage pulse to an evacuated vessel instead of a gas-filled
vessel so that no discharge can occur.

In all cases great care should be taken to insure that the results are not
influenced by stray capacitances, impedance (mis)matching and dissipation in
other places like matching networks or cables.

\subsection{Optical imaging techniques}\label{sec:imaging}

Streamer discharges often have a complex morphology and imaging this
morphology can tell a lot about the actual discharge, although the
applicability of such techniques depends on the gas (see
section~\ref{sec:othergasses}). The most common method
to image streamers is by Intensified CCD (ICCD) camera. Such a camera is
capable of capturing the dim streamer discharges, and, more importantly, can
be gated such that it captures a very well-defined period of time. Timing
accuracy is often better than nanoseconds while the shortest exposure
times range from about 100 picoseconds to a few nanoseconds so that fast
phenomena like those in figure~\ref{fig:velocity-example} can be imaged.
For short exposures,
synchronization of high voltage (pulses), electrical measurements and
camera measurements is essential. This may be achieved by good
understanding of the complete measurement system, including measurements
of each cable and instrument delay. Alternatively
one can employ a very fast LED to synchronize camera and voltage measurements.

\begin{figure}
  \centering
  \includegraphics[width=8.5 cm]{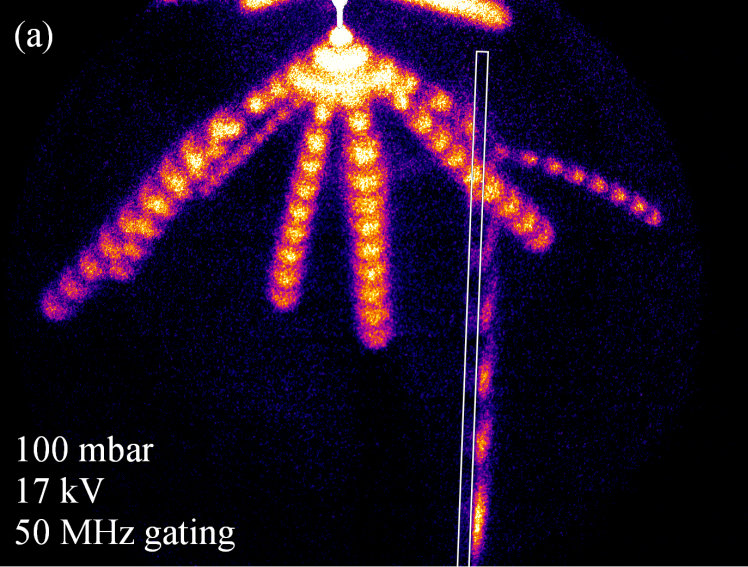}\\
  \caption{Stroboscopic ICCD image of a streamer discharge in artificial air
  in the vicinity of a
  dielectric rod. Image from~\cite{Trienekens2014}.
  }
  \label{fig:strobotrienekens}
\end{figure}

When an ICCD camera is capable of switching its gate on and off at
frequencies well above 10\,MHz, it becomes possible to use so-called
stroboscopic imaging. With this method, the gate is switched on and off so
fast that the resulting image will not show continuous streamer channels but
instead strings of beads or lines that are separated in time with the gating
frequency of the camera. This is possible because only the streamer head
(ionization front) emits light. Note that this depends on the decay
time of the upper levels of the dominant emission lines or systems. When this
decay is too slow, as is for example the case in argon, this method
will not work. Stroboscopic imaging allows one to see both the streamer
development, as well as its propagation velocity in one single image. This
technique was first demonstrated by Pancheshnyi {\it et
al}~\cite{Pancheshnyi_2005} and later improved by Trienekens {\it et
al}~\cite{Trienekens2014}, see also figure~\ref{fig:strobotrienekens}.

Another addition to standard streamer photography is the use of stereoscopic
techniques. Such techniques make it possible to make a 3D-reconstruction of
the entire discharge morphology. This is necessary when one wants to understand
essential properties of the discharge like branching angles and propagation
velocities. In a 2D single camera projection such quantities can be easily
underestimated due to the projection. The simple use of some mirrors and/or
prisms makes it possible to do stereoscopic measurements with one single camera
\cite{Nijdam2008,Nijdam2009,Ichiki2011,Ichiki2012,Heijmans2013a,Yu2016}.
Processing of such data is mostly done (semi-)manually, but currently
significant progress towards automatic processing is made by Dijcks~\cite{Dijcks2019}.

A variation on the ICCD camera is the streak camera. This camera can image
fast phenomena better than an ICCD camera because it can show sub-nanosecond
dynamics of a single event. The disadvantage of streak cameras though, is
that they only image a one-dimensional strip, which severely limits their
usefulness to image objects like branching streamers. Therefore, they can
only be used in situations where the discharge will follow a predictable
line like in (DBD-)discharges in short gaps \cite{Brandenburg2013}.

\subsubsection{Measuring diameters and velocities}\label{sec:diamvelodiagnostics}

Propagation velocity and streamer diameter are two basic properties which
are obvious and relatively easy to measure. However, both are non-trivial,
as will be explained below.

{\bf Measuring diameters. }
The diameter of a streamer is usually measured from an (ICCD) image.
This requires a definition of diameter, as the
longer exposure images of streamers have a roughly Gaussian intensity profile.
A commonly used definition for streamer diameter is the full width at
half maximum (FWHM) of the emission profile~\cite{Briels2008,Nijdam2010,Kanmae2012,Chen2013}.
Other definitions can also be used, like the width of a
certain pixel count threshold~\cite{Ono2003,Pancheshnyi_2005}.

Because ICCD images of streamers are usually quite noisy (due to the low
light intensity), averaging along the length of the channel may
be required in order to get a reliable measurement. When channels are
not straight, this can be quite difficult and more advanced algorithms
may be needed.
To be fully correct, one should first
perform an Abel inversion of the measured channel. Luckily, the Abel
inversion of a Gaussian curve is exactly the same curve, so as long
as the profiles are close enough to this shape, such an operation can be avoided.

Furthermore, due to limitations in actual resolution, streamers
should be wide enough for a reliable diameter measurement. In~\cite{Briels2006}
we used a minimum reliable width of about 6 camera pixels, while later we
increased this threshold to 10 pixels~\cite{Nijdam2010}.

Finally, the diameter obtained in this way is only one definition of streamer
diameter. It could depend on the spectral sensitivity of the camera (other
wavelength regions may give other diameters) and is surely different from
diameters calculated from from the electric field or the electron density
distribution in models, see also section~\ref{sec:VeloRadField}.
Therefore, when one wants to compare measured
streamer diameters with simulated results, one should try to process the
model results in such a way that the real emission profile is shown.

{\bf Measuring velocities. }
The definition of propagation velocity is simpler than that of streamer
diameter. Here, one can also debate which exact definition to use, but
this will hardly affect the obtained values because each defined front
edge propagates with the same velocity. This also makes it easier to
quantitatively compare with models.
Measuring it, however, can be more difficult than measuring streamer diameter
for several reasons. Firstly, because the propagation velocities
are very high ($10^5 - 10^7\,$m/s), one needs fast equipment to measure
velocities in the lab. Measuring velocities in very
large discharges like sprites is easier because the velocities do
not scale with the gas number density $N$, whereas the lengths and widths do scale
with $1/N$ (see section~\ref{sec:scalinglaws}).
Secondly, depending on gas species and
density, optical emission can last relatively long on these timescales.
For example, in atmospheric air, lifetimes of (bright) excited nitrogen states
are on the order of a few nanoseconds \cite{Stancu2010},
but other gasses like argon have much
longer radiative lifetimes \cite{Niermann2012}.
Due to these issues, a variety of measurement
methods has evolved.

The simplest and probably cheapest measurement technique uses
the current profile of the discharge to detect when it has
crossed the gap and determines an average velocity from this~\cite{Dawson1965}.

Another relatively simple technique employs
multiple (at least two) photomultiplier tubes (PMTs) pointing at different locations
along the expected streamer path. The time between signals divided
by the distance between locations now directly gives a propagation
velocity~\cite{Dawson1965,Allen1995,Grange1995,Pritchard2002,Meng2015}.
The main disadvantage of this method
is that it gives information on just the average velocity between
only a few points and that streamers can potentially be missed.

Most other velocity measurements on streamers use ICCD cameras.
Often, a short exposure is used, which gives an image with
a short line for each propagating streamer during exposure. When the
gate time is long compared to the radiative lifetime
and the line length is long compared to the diameter,
the velocity can simply be approached by dividing
the measured line length by the exposure time (see also
figure~\ref{fig:velocity-example}).
In other cases, corrections may have to be applied
for lifetime or streamer head size.
This method of measuring streamer velocity from short exposure images
is often employed and can give very good results
\cite{Veldhuizen2002,Namihira2003,Ono2003,Briels2008,Briels2008b,Nijdam2010}, although it
only gives the velocity for part of the propagation, so
multiple images are needed to get an overview of velocity
development.
Alternatively, one can use a stroboscopically
gated ICCD camera and measure the distances between subsequent
maxima as is explained above.
Another point of attention with this technique
is that the 2D projection makes lines which are out of plane appear
shorter, so only the longest ones
can be used. This can be remedied by
stereoscopic techniques.

An alternative velocity measurement method is to
measure the distance to the high-voltage electrode of the
longest streamers as function of time by looking at multiple
images with different exposure end-times with respect to the
discharge start~\cite{Winands2008a}. Because multi-frame imaging is
generally impossible during
a discharge, this requires the use of images from multiple
discharges and can therefore only be used when discharge jitter
is either very low or when the discharge inception time can be measured
in other ways (from e.g. a current or photomultiplier signal).
This method is able to measure velocities for discharges with
long emission lifetimes because it only uses the front of
the propagating streamer. The method can be automated to quickly
process hundreds to thousands of images and thereby get a
very good overview of velocity development trends~\cite{Heijmans2015,Chen2015}.

Other available methods to measure streamer velocities
are streak cameras~\cite{Yi2002,Tardiveau2002,Teramoto2011,Brandenburg2013},
although only
for known roughly straight channels (see above) and multi-frame
or extremely high frame-rate cameras~\cite{Zeng2013,Takahashi2011a}.

\subsection{Optical Emission Spectroscopy} Optical emission spectroscopy
or OES is probably the most versatile technique used to investigate
plasmas \cite{Laux2003,Machala2007,Horst2012,Dilecce2014,Simek2014,Ono2016}. It can be used
for simple purposes like recognizing specific species
in a discharge up to complex tasks like determining ionization degrees or
rotational, vibrational and excitational temperatures. Generally speaking,
measuring a spectrum is quite straightforward, although signals can be low,
but the processing and interpretation of spectra requires models or
complex fit routines.

One commonly used OES-technique is the determination of the electric field
strength by the ratio of specific emission lines from singly ionized
and neutral nitrogen
molecules~\cite{Kozlov2001,Paris2004,Paris2005,Shcherbakov2007,Schans2016}.
In this technique
the intensity of a line of the second positive system (SPS) of N$_2$
at 337.1\,nm is
compared to a line of the first negative system (FNS) of N$_2^+$
at 391.4\,nm.
A detailed description of this technique and other
line-ratio techniques in nitrogen and argon plasmas is given in~\cite{Zhu2010}.
Disadvantages of all such techniques are that they firstly are usually
based on line-of-sight averaging (and are usually also temporally averaged)
and secondly that they assume some state of equilibrium in the plasma
which is not always the case in a streamer head.

\begin{figure}
  \centering
  \includegraphics[width=8.5 cm]{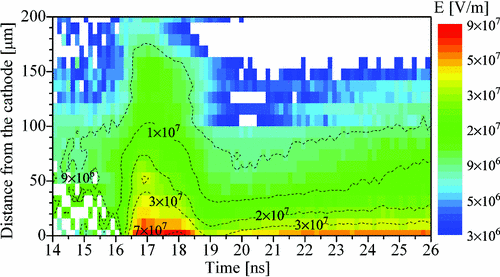}\\
  \caption{Electric-field strength distribution in repetitive
   Trichel pulse in atmospheric-pressure air measured with
   cross-correlation spectroscopy using the nitrogen line
   ratio method.
  Image from~\cite{Hoder2012}. 
  }
  \label{fig:CCSLineRatio}
\end{figure}

For highly reproducible discharges a technique like cross-correlation
spectroscopy developed in Greifswald \cite{Kozlov2001,Brandenburg2013,Hoder2012}
can give highly accurate phase-resolved results as can be
seen in the example in figure~\ref{fig:CCSLineRatio}.
When one is interested in complex spatial structures in discharges
a tomography technique like shown in~\cite{Schans2016} can be used.
In both cases many spectra are measured, which together give
great insight in the discharges.

\subsection{Laser diagnostics}

Active laser-based diagnostic techniques can be very powerful tools for
quantification of a large range of plasma parameters. They can measure
species densities, kinetics, temperatures and even electric fields.
However, they generally require a significant investment in experimental
equipment and in time to set up and align an experiment.

A large disadvantage of using laser diagnostics is that such diagnostics
generally require many laser shots in order to get reliable results.
H\"{u}bner \textit{et al.}~\cite{Hubner2017} show that in a practical laser scattering set-up,
only a fraction of about $10^{-19}$ of the incident laser photons is detected
as scattered photons,
which makes single-shot operation nearly impossible
with laser intensities that do not influence the plasma.
The stochastic nature of most streamer-like discharges
prevents good laser diagnostics;
laser diagnostics can only be applied to discharges that are highly
reproducible, like plasma jets or pin-to-pin discharges with
short gaps.\\

The most straightforward laser diagnostic is Rayleigh scattering
\cite{Bates1984,VanGessel2012} where the intensity of scattered laser light is
proportional to the gas number density and therefore, through
the ideal gas law, to the inverse of the gas temperature.
However, the scattering cross sections for different atomic and molecular
species vary a lot. This means that the proportionality between signal and number density is only
valid for constant gas composition and that quantitative measurements require exact
knowledge of the gas composition.

A next step is Thomson scattering \cite{Gregori1999,Hubner2014,Carbone2015a,Hubner2017}
in which
the light scattered on free electrons is measured.
It employs their large Doppler shift caused by
their high velocities compared to the heavy species. With this technique
one can measure both the electron density and temperature.

Raman scattering uses the inelastic scattering of laser light on molecules
to measure molecular densities as well as rotational temperatures
\cite{Belostotskiy2006,VanGessel2012}. In many cases the Raman and
Thomson signals have to be disentangled \cite{VanGessel2012}.
Raman scattering on known gas mixtures is often
employed as a calibration tool for other techniques.

The laser techniques that can best target specific species are Laser Induced
Fluorescence or LIF and variations on it like two-photon atomic LIF (TA-LIF).
LIF employs a laser tuned to an excitation wavelength
of a species while emission at another wavelength is monitored.
This allows one to probe densities of very specific species like the
vibrational levels of N$_2$(A$^3\Sigma_u^+$) metastable species
\cite{Simek2015,Simek2017} or OH-radicals
\cite{Ono2003a,Choi2011,Verreycken2012,Ouaras2018}.

A related technique uses the optical absorption of the light instead
of scattered or re-emitted light. In atmospheric air discharges this
is mostly used to determine ozone concentrations \cite{Ono2003a}.

Finally, it is possible to measure the electric field using non-linear optical
properties of gases. One way to do this is by using four-wave mixing Coherent
Anti-Stokes Raman Scattering (CARS)
\cite{Schans2017} which uses two colinear laser beams of different wavelengths
as well as a few detectors. A simpler alternative is
electric field induced second harmonic generation (EFISH) which has recently
become popular~\cite{Goldberg2018,Orr2020,Adamovich2020}.

\subsection{Other diagnostics}

In quite a few cases, especially at atmospheric and higher pressures,
streamer-like discharges can lead to gas heating and/or
gas flow (see section~\ref{sec:flowheat}). Two methods to visualize this are Schlieren
photography \cite{Ono2004,Xu2011,Papadopoulos2014,Chen2018} and
shadowgraphy \cite{Tetsuo2009,Ono2010}. Both these methods employ the effects of density
gradients
on the refractive index $n$. Schlieren visualizes $\partial n / \partial y$
while shadowgraphy visualizes $\partial^2 n / \partial y^2$
with $y$ a spatial coordinate perpendicular to the light path.

\begin{figure}
  \centering
  \includegraphics[width=8.5 cm]{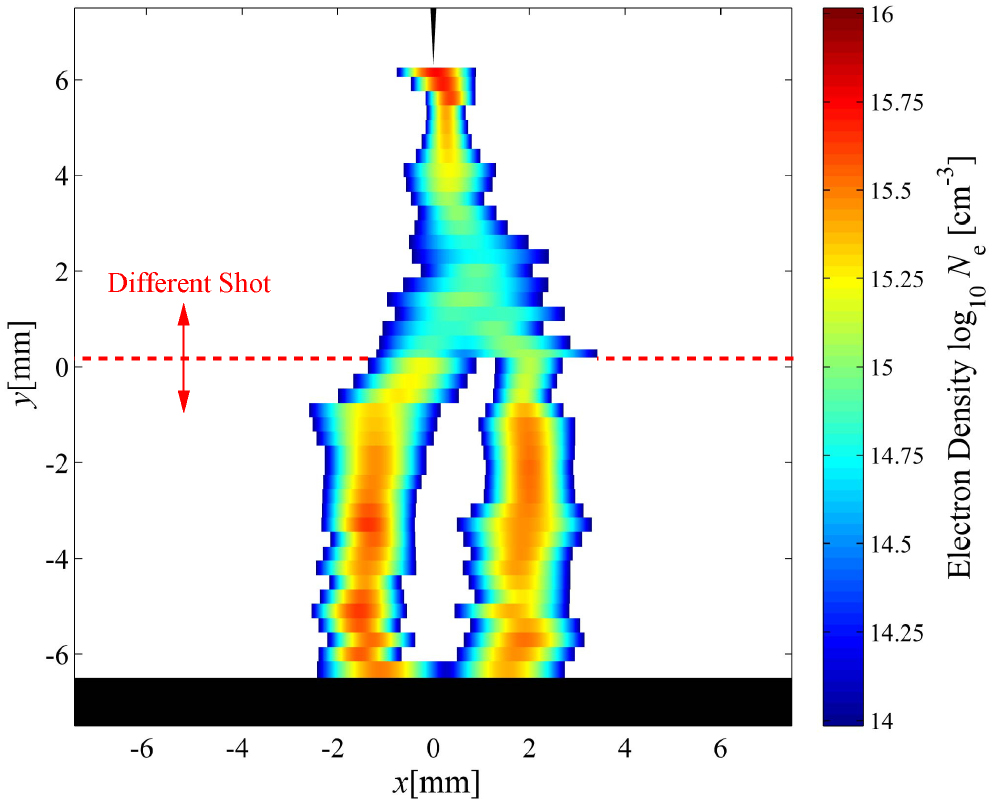}\\
  \caption{Two dimensional electron density distribution acquired by measuring
  refractive index variations. Image from~\cite{Inada2017}.}
  \label{fig:inadaelectrondensity}
\end{figure}

A variation on these techniques gives an elegant way to measure
the electron density in a streamer discharge in a
single shot. It employs the decrease of the refractive index with electron density.
Inada \textit{et al.}~\cite{Inada2017} have shown that with this method they can acquire a
two-dimensional image of the electron density with a 2~ns temporal
resolution. For this they use two Shack-Hartmann type laser wavefront
sensors illuminated by laser light of two distinct wavelengths (blue and
red) to distinguish gas density and electron density effects.
A resulting electron density distribution is shown in
figure~\ref{fig:inadaelectrondensity}. A disadvantage of this method is that
essentially electron density is integrated over a line-of-sight. Therefore,
an Abel inversion on the data is required to get the full information.
However, this requires the distribution to be cylindrically symmetric,
which is often not the case.

\newpage
\section{Outlook and open questions}\label{sec:outlook}

We have reviewed our present understanding of streamer discharges, addressing
basic physical mechanisms and observed phenomena.
Our emphasis has been on positive streamers in air under lab conditions, whose properties (e.g.,~velocity, radius, maximal field) already span a wide parameter space due to the progress of fast pulsed high-voltage sources.

We think that advances in diagnostic techniques and numerical modeling have brought a quantitative understanding of all streamer-related phenomena within reach.
However, there are still many open questions, even when only considering positive streamers in air. Below, we have collected several of the most important ones.
Our goal is to answer these questions, and to develop quantitative theories for streamer discharges that can be generalized to other gases or polarities.


\subsection{Discharge inception:}

\begin{itemize}
    \item Which are the dominant electron sources for streamer inception
    in different parameter regimes, and for different polarities?
    What is the role of surfaces in this? Sections~\ref{sec:incept-sources} and ~\ref{sec:incept-surf}.
    \item Which quantitative criteria for discharge inception can be developed beyond the classical
    Meek number criterion? Section~\ref{sec:avalanche-to-streamer}.
    \item Does our proposed mechanism for the inception cloud also
    explain seemingly similar phenomena like ''diffuse discharges''?
    Section~\ref{sec:inception_cloud}.
   \item What determines the break-up of the inception cloud into streamers?
   Section~\ref{sec:inception_cloud}.
\end{itemize}

\subsection{Streamer evolution}
A particular problem is that most experiments show a burst of many branching streamers,
whereas fluid or particle simulations become computationally very expensive when more than one streamer is present.
Therefore, the overall discharge evolution cannot easily be compared, except if the experiment
produces a single streamer, or if the microscopic models can be reduced to quantitative macroscopic
tree models. To achieve the latter, the following questions need to be answered.
\begin{itemize}
    \item Can we understand the large range of velocities, and radii of streamers in air for both
    polarities? How are they related
    to the electric field and other conditions? Can this be described analytically
    from basic physics or only empirically from simulations and experimental results?
    Section~\ref{sec:VeloRadField}.
    \item Can we understand the charge distribution in a streamer tree?
    What is the physical background of the experimentally
  often reported `stability field'? Section~\ref{sec:stabilityfield}, ~\ref{sec:interaction}
  and~\ref{sec:macroscopic-models}.
    \item What are the mechanisms causing streamer branching
   under varying conditions? 
   Can we identify the distribution of branching lengths and angles?
   Section~\ref{sec:branching}.
   \item Why do negative streamers and leaders propagate in a step-like fashion while
   positive ones seem to propagate continuously?
   Section~\ref{sec:stepped-leaders}.
\end{itemize}

\subsection{Further evolution after passage of ionization front}
The reactions occurring after the passage of a streamer front depend strongly on
the gas composition, so answers on the questions below can and will also
vary with gas mixture.

\begin{itemize}
    \item Which plasma chemistry is precisely triggered by streamers?
    How does it depend on velocity, electric field and radius of the streamer head?
    How does this influence discharge evolution and how can it best be used in applications?
    Section~\ref{sec:chemistry}.
    \item What is the interaction between a streamer corona and a leader
    and what is the role of gas heating?
    Section~\ref{sec:gas-heating}.
    \item How is conductivity in a streamer channel maintained?
    What are the roles of heating, plasma chemistry and the detachment instability?
    Are there nonlinear self-enforcing mechanisms in the streamer channel?
    Section~\ref{sec:Conductivity}.
\end{itemize}

\subsection{Particular physical mechanisms}

\begin{itemize}
    \item In which parameter regime of negative discharges
    do electron runaway and bremsstrahlung become important?
    Can they play a role in positive discharges as well?
    Section~\ref{sec:fastelectrons}.
    \item Can electrons be
    accelerated in streamer discharges to energies that are possibly far higher than electrostatic
    acceleration would predict? Section~\ref{sec:fastelectrons}.
    \item Do runaway electrons have a significant
    influence on streamers and, if so, how? Section~\ref{sec:fastelectrons}.
    \item Is there photo-ionization or are there other substitutes for it
    in gases different from air? Why do we see widely varying streamer diameters
    for positive streamers in air but not in gases like pure nitrogen?
    Section~\ref{sec:photoionization}.
\end{itemize}


\section*{References}

\bibliographystyle{unsrt}
\bibliography{References_combined_limited}

\end{document}